\documentclass[12pt]{article}
\usepackage[english]{babel}
\usepackage[latin1]{inputenc}
\usepackage{amsfonts,amssymb,amsmath, epsfig}
\usepackage{color,graphicx,graphics,psfrag}
\usepackage{amsmath,amstext,amssymb,amsfonts, amscd}
\usepackage{hyperref}        
\usepackage{subfigure}

\def\Box{\leavevmode\vbox{\hrule
     \hbox{\vrule\kern4pt\vbox{\kern4pt}%
           \vrule}\hrule}}
\def\blackbox{\leavevmode\vrule height 5pt width 4pt depth 0pt\relax}
\def\endproof{\null\hfill {$\blackbox$}\bigskip}

\newcounter{appendix}
\def\appendix{\advance\c@appendix by 1
   \def\thesection{\Alph{section}}
   \ifnum\c@appendix=1 \setcounter{section}{-1} \fi
   \@startsection {section}{1}{\z@}{-3.5ex plus -1ex minus 
   -.2ex}{2.3ex plus .2ex}{\Large\bf}}

\def\paragraph#1{{\bf #1\ }}

\newtheorem{lemma}{Lemma}[section]

\newtheorem{proposition}[lemma]{Proposition}

\newtheorem{remark}{Remark}[section]

\newcommand{\norm}[1]{\lVert#1\rVert}

\title{Time-delayed Follow-the-Leader model \\for pedestrians walking in line} 
\author{J\'er\^ome Fehrenbach$^{1,2}$,  Jacek Narski$^{1,2}$, Jiale Hua$^3$,\\ 
Samuel Lemercier$^4$, Asja Jeli{\'c}$^{5,6}$, C\'ecile Appert-Rolland$^{7,8}$,\\ St\'ephane Donikian$^9$, Julien Pettr\'e$^4$ and Pierre Degond$^{10}$} 
\date{} 
\begin{document}

\maketitle


{\footnotesize
\begin{center}
1. Universit\'e de Toulouse; UPS, INSA, UT1, UTM ; \\
Institut de Math\'ematiques de Toulouse ; \\
F-31062 Toulouse, France\\
email: jerome.fehrenbach@math.univ-toulouse.fr, \\jacek.narski@math.univ-toulouse.fr
\end{center}

\begin{center}
2. CNRS; Institut de Math\'ematiques de Toulouse UMR 5219 ;\\
F-31062 Toulouse, France.
\end{center}

\begin{center}
3. Donghua University, 1882 Yan'an Road West, \\
Changning district, Shanghai 200051, China\\
email: jiale.hua@dhu.edu.cn
\end{center}

\begin{center}
4. INRIA Rennes - Bretagne Atlantique, \\
Campus de Beaulieu, 35042 Rennes, France \\
email: samuelemercier@hotmail.com, \\
julien.pettre@irisa.fr
\end{center}

\begin{center}
5. Istituto Sistemi Complessi, Consiglio Nazionale delle Ricerche, \\
UOS Sapienza, 00185 Rome, Italy.\\
email: asja.jelic@gmail.com
\end{center}

\begin{center}
6. Dipartimento di Fisica, Universit\`a Sapienza, \\
00185 Rome, Italy
\end{center}

\begin{center}
7. Univ. Paris-Sud, Laboratoire de Physique Th\'eorique, \\
B\^at. 210, F-91405 Orsay Cedex, France \\
email: Cecile.Appert-Rolland@th.u-psud.fr
\end{center}

\begin{center}
8. CNRS, LPT, UMR 8627, B\^at 210, 
F-91405 Orsay Cedex, France
\end{center}

\begin{center}
9. Golaem S.A.S., B\^atiment Germanium, \\
80 avenue des Buttes de Co\"esmes, 35 700 Rennes, France\\
email: stephane.donikian@golaem.com
\end{center}

\begin{center}
10. Imperial College London, South Kensington Campus, \\
London SW7 2AZ, United Kingdom \\
email: pdegond@imperial.ac.uk
\end{center}
}

\vspace{0.5 cm}
\begin{abstract}
We use the results of a pedestrian tracking experiment to identify a follow-the-leader model for pedestrians walking-in-line. We demonstrate the existence of a time-delay between a subject's response and the  predecessor's corresponding behavior. This time-delay induces an instability which can be damped out by a suitable relaxation. By comparisons with the experimental data, we show that the model reproduces well the emergence of large-scale structures such as congestions waves. The resulting model can be used either for modeling pedestrian queuing behavior or can be incorporated into bi-dimensional models of pedestrian traffic. 
\end{abstract}

\medskip
\noindent
{\bf Acknowledgements:} This work has been supported by the french 'Agence Nationale pour la Recherche (ANR)' in the frame of the contract ``Pedigree'' (ANR-08-SYSC-015-01). JH acknowledges support of the ANR and the Institut de Math\'e\-matiques de Toulouse, where he conducted this research. AJ acknowledges support of the ANR and of the Laboratoire de physique théorique in Orsay where she conducted this research. PD is on leave from CNRS, Institut de Mathématiques de Toulouse, France.


\vskip 0.4cm

\section{Introduction}

The need for accurate predictions of pedestrian behavior is rapidly growing, due to the constant increase of urban populations worldwide and to the  strengthening of safety regulations imposed to buildings and public areas. Yet there is no consensus about what model of pedestrian behavior is the most appropriate. This is probably due to the difficulty of precisely assessing the validity of models in an unambiguous way. Indeed, in natural conditions, many factors which are difficult to entangle may influence pedestrian behavior, such as the environment, the topology of the premises, the social and psychological state of the pedestrians, etc. On the other hand, the design of experiments in fully controlled situations is costly, time-consuming and strongly impeded by experimental constraints.  

Many models of pedestrian dynamics have been proposed in the literature. A recent review on crowd modeling can be found in \cite{Bellomo_Dogbe_SIAMRev11}. A vast majority of the models are based on Individual-Based Models (IBM), which describe the behavior of each pedestrian and its interactions with the neighboring pedestrians individually.  Such models can be roughly categorized as follows. There are models based on e.g. heuristic rules \cite{Reynolds_ProcGameDev99, Moussaid_etal_PNAS11}, mechanical models \cite{Helbing_BehavSci91, Helbing_Molnar_PRE95, Helbing_Molnar_SelfOrganization97},  optimal control theory models, \cite{Hoogendoorn_Bovy_OptControlApplMeth03}, Cellular-Automata  \cite{Burstedde_etal_PhysicaA01, Nishinari_etal_IEICETranspInfSyst04} and Vision-Based models \cite{Guy_etal_Siggraph09, Huang_etal_RoboticsAutonomSyst06, Pettre_etal_Siggraph09, Vandenberg_Overmars_IntJRoboticsRes08}. All these models partially reproduce the behavior of real crowds, each of them having its pros and cons. 

One of the difficulties in reproducing actual pedestrian dynamics comes from its two-dimensional nature and the fact that the transversal and longitudinal dynamics (with respect to the walking direction) combine in a complex way which is hard to desantangle. Collision avoidance manoeuvres by pedestrians involve controls on both their velocity direction and amplitude \cite{Ondrej_etal_Siggraph10, Paris_etal_Eurographics07}, a feature that has already been implemented in some models \cite{Degond_etal_JSP13, Degond_etal_KRM13, Lemercier_etal_CompGraphForum12, Moussaid_etal_PNAS11}. In order to better understand the role of each of these controls, in this paper we consider a one-dimensional configuration where pedestrians walk on line without being able to pass each other. This situation is a paradigm for narrow corridors \cite{Chraibi_etal_PRE10} or dense situations where lane form and pedestrians tend to follow a predecessor walking in the same direction \cite{Moussaid_etal_PLOSCB12, Zhang_etal_JStatMech12}. Moreover, following behavior as observed in one-dimensional experiments can be used for analyzing or modeling two-dimensional situations \cite{Johansson_PRE09, Seyfried_etal_PhysicaA06}. The study of one-dimensional pedestrian following behavior has also triggered interest for its own sake \cite{Jelic_etal_PRE12-1, Jelic_etal_PRE12-2, Jezbera_etal_JStatMech10, Seyfried_etal_JStatMech05}.

In this paper, we rely on experimental results using motion capture techniques reported in \cite{Lemercier_etal_LNCS11} to calibrate the parameters of a follow-the-leader model inspired from car traffic \cite{Chandler_etal_OperRes58, Gazis_etal_OperRes61}. However, on the basis of the experimental data, we demonstrate that it is necessary to consider a non-zero time delay in the model, i.e. that the acceleration of a pedestrian at a given time is determined by the relative position and velocity of this pedestrian with respect to his predecessor at an earlier time. This results in a system of delay differential equations. It is well known that such delay differential systems may not be stable depending of the choice of the model parameters \cite{Bellman_Cooke_AcadPress63}.  In this paper, we show that, within the range of parameters found from the calibration of the model, the model is actually not stable. This leads us to introduce some dissipation mechanism in the form of a relaxation of the pedestrian velocity to the average velocity of a certain number of his predecessors. Thanks to the introduction of this additional mechanism, the model becomes well-posed. We numerically show that, with a physically consistent choice of the parameters of the relaxation operator, the model does indeed provide extremely good agreement with the experiments for large pedestrian densities. For lower density however, it seems that additional mechanisms which are not included in the model are at play.  

The paper is organized as follows. In Section \ref{sec:data} we describe the experiments and the filtering method that is applied to the data. In Section  \ref{sec:model} we discuss the model and show that a delay term must be added. We perform a stability analysis of the resulting delay-differential system and show that a relaxation term must be added to obtain a stable model. We then estimate the parameters of the model from the experimental data. In Section \ref{sec:results} we present the result of the so-obtained calibrated model and we assess the quality of the calibration by looking at macroscopic observables such as the statistics of jams. We show that the model is able to reproduce the experimental data in a very satisfactorily manner. The paper is concluded by a discusion in Section \ref{sec:discussion}. A technical annex explores how the results of the calibration depends on the data processing parameters and shows that apart from the choice of the cutoff frequency in the data filtering step, they are insensitive to this choice. The cutoff frequency is chosen fo filter out the pedestrian stepping frequency while keeping all phenomena occuring at lower frequencies.

\setcounter{equation}{0}
\section{Materials and methods}
\label{sec:data}

\subsection{Experiments}
\label{subsec:experiment}

The description of the experimental setup can be found in \cite{Jelic_etal_PRE12-1}. We recall it briefly for the sake of completeness. In this experiments, subjects were instructed to walk in line on a circular path without passing each other. The trajectory of each pedestrian was recorded using  high precision motion capture technology \cite{Moussaid_etal_PLOSCB12}. Experiments took place in a ring-shaped arena. The imposed circular path was chosen close to either the inner or the outer boundary of the arena. This provided a way to modify the density of pedestrians (another way being by changing the number of subjects enrolled in the experiment). The choice of circular paths was made in order to avoid spurious clogging effects arising at the ends of a rectilinear path when the subjects enter or exit the path. The trajectories of the pedestrians were reconstructed using data processing methods described in \cite{Lemercier_etal_LNCS11}.

Up to $28$ subjects were enrolled in the experiments. They were volonteers and uninformed of the purpose of the experiment. They were instructed to walk at their natural pace and forbidden to talk to each other.  The inner and outer radii of the circular arena were respectively $2$ and $4.5$ m. The observed average radius of the pedestrians' circular trajectory in the experiments using the inner radius was $2.4$ m (implying an average perimeter of $15.08$ m). For the experiments using the outer radius, the respective figures were $4.1$ m and $25.76$ m. 
Table I of \cite{Jelic_etal_PRE12-1} gives a summary of the experimental parameters (number of subjects, use of inner or outer circle, pedestrian density and number of replications). The typical duration of each replication was $1$ minute. Each subject was equipped with 4 markers, one on the left shoulder, two on the right shoulder, and one on the forehead. The ring-shaped arena was surrounded by $12$ infra-red cameras which detected the markers. A dedicated software converted this information into the three-dimensional coordinates of each marker. After some processing of the data described in  \cite{Lemercier_etal_LNCS11}, the planar two-dimensional coordinates of the barycenter of the four markers of each subject was reconstructed with a frequency of $120$ Hz.

\subsection{Experimental data}
\label{subsec:expe_data}

The experimental data keep records of the planar positions $(x(t_n),y(t_n))$ of each pedestrian at sampling times $t_n$, with a sampling frequency of $120$ Hz
(i.e. two consecutive sampling times are separated by $\Delta t = 1/120s$). To exploit the fact that the pedestrians are moving on a circle, we transform the cartesian coordinates $(x(t_n),y(t_n))$ into polar coordinates $(r(t_n),\theta(t_n))$ relative to the center of the circle and to some reference axis. We can then estimate the angular velocity $\omega$:
\begin{gather}
 \omega= \dot\theta=\frac{1}{r}(-\sin \theta \, \dot{x}+\cos \theta \, \dot{y}),    
\end{gather}
where dots denote time derivatives and $\dot x$ and $\dot y$ are approximated by finite differences. 

The experimental data are perturbed by quasi-periodic oscillations due to the pedestrian stepping behavior. We will discard the effect of steps in the models as we are interested in longer time scales where they have no predominant influence. For this reason, we strongly reduce the amplitude of these oscillations by applying a linear fourth order filter to the position data. In the frequency domain $\nu$, and for any quantity $u(t)$ whose Fourier transform in time is denoted by $\hat u(\nu)$, the resulting filtered quantity $u_f(t)$ has Fourier transform $\widehat{u_f}(\nu)$given by:
  \begin{gather}
    \widehat{u_f}(\nu)=\frac{1}{1+c \, \nu^4} \, \widehat{u}(\nu) \quad
    \text{ for some constant }c>0.\label{eq:7}
  \end{gather}  
Due to its high order, this filter also reduces the amplitude of the oscillations of the first and second order time-derivatives (i.e. the velocity and the acceleration of the pedestrians). The  cutoff frequency associated to the filter \eqref{eq:7} is $f_c=\frac{2\pi}{\omega_c}$ where $\omega_c$ satisfies
\begin{gather}
 \frac{1}{1+c\omega_c^4}=\frac{1}{\sqrt{2}}.\label{eq:6}
\end{gather}
The cutoff frequency is chosen in such a way that as much information
in the data is kept as possible. From the experiments, the step period is about $2$ s. Consequently, the cutoff frequency $0.5$ Hz has been chosen. A study of the influence of the cut-off frequency on the results supports this conclusion (see section \ref{subsec:parameterfc}).

\setcounter{equation}{0}
\section{Theory: the Follow-The-Leader model}
\label{sec:model}

\subsection{The model}
\label{subsec:model}

The Follow-The-Leader (FTL) model is a microscopic model which describes the individual behavior of each agent and his interactions with his neighbors.  It has been first used in the context of car traffic \cite{Chandler_etal_OperRes58, Gazis_etal_OperRes61}. There are many variants of the FTL model but a fairly common one is written as follows (adopting notations of polar geometry which are best suited to the experimental setup). We consider $N$ pedestrians  on the 1D circle with angular positions $\theta_i(t)$, and angular velocities $\omega_i(t)$, for $i=1,\cdots, N$ functions of time $t$. These quantities evolve according to the following system of delay-differential equations, written for $i=1,\cdots, N$:
\begin{eqnarray}
&&\hspace{-0.5cm}  \dot{\theta}_i(t) =\omega_i(t),\label{eq:1.0}\\
&&\hspace{-0.5cm}  \dot{\omega}_i (t+\tau)=C\frac{(\omega_{i+1}-\omega_i)(t)}{|\theta_{i+1}-\theta_i|^{1+\gamma}(t)}.\label{eq:2.0}
\end{eqnarray}
Here $C>0$, $\tau >0$ and $\gamma\geq -1$ are modeling constants to be calibrated on the data. The $(i+1)$-th pedestrian is the leader of (i.e. the one exactly before) the $i$-th pedestrian. This model can be interpreted as follows. There is a first phase where the pedestrians observes his leader, acquires the knowledge of the quantities $\omega_{i+1}(t)$ and $\theta_{i+1}(t)$ and makes a decision about what reaction should be implemented in response to these observations. This decision-making is represented by the right-hand side of (\ref{eq:2.0}). The second phase is the action phase where the pedestrian acts on his own velocity to comply with his decision-making rule. It is represented by the left-hand side of (\ref{eq:2.0}). The time delay $\tau$ corresponds to the time needed between the decision and its translation into action. The decision-making rule itself describes how the pedestrian adjusts his velocity to that of his leader. To this aim, he decelerates if he is faster than his leader and accelerates if he is slower, in proportion to the speed difference, as shown at the numerator of the right-hand side of (\ref{eq:2.0}). This adjustment is modulated by the proximity of the pedestrian to his leader, a short distance inducing a stronger reaction, as expressed by the denominator of the right-hand side of (\ref{eq:2.0}). The constant $C$ quantifies the intensity of the reaction and $\gamma$ its dependence upon the distance to the leader. Note that $(\theta_{i+1}-\theta_i)(t)$ is always positive if the pedestrians are moving counterclockwise and always negative if they are moving clockwise. Since angles are quantities defined up to the addition of a multiple of $2 \pi$, we take for $(\theta_{i+1}-\theta_i)(t)$ the smallest positive (respectively largest negative) quantity among all such possible values.

\begin{figure}
  \centering
  \includegraphics[scale=0.34]{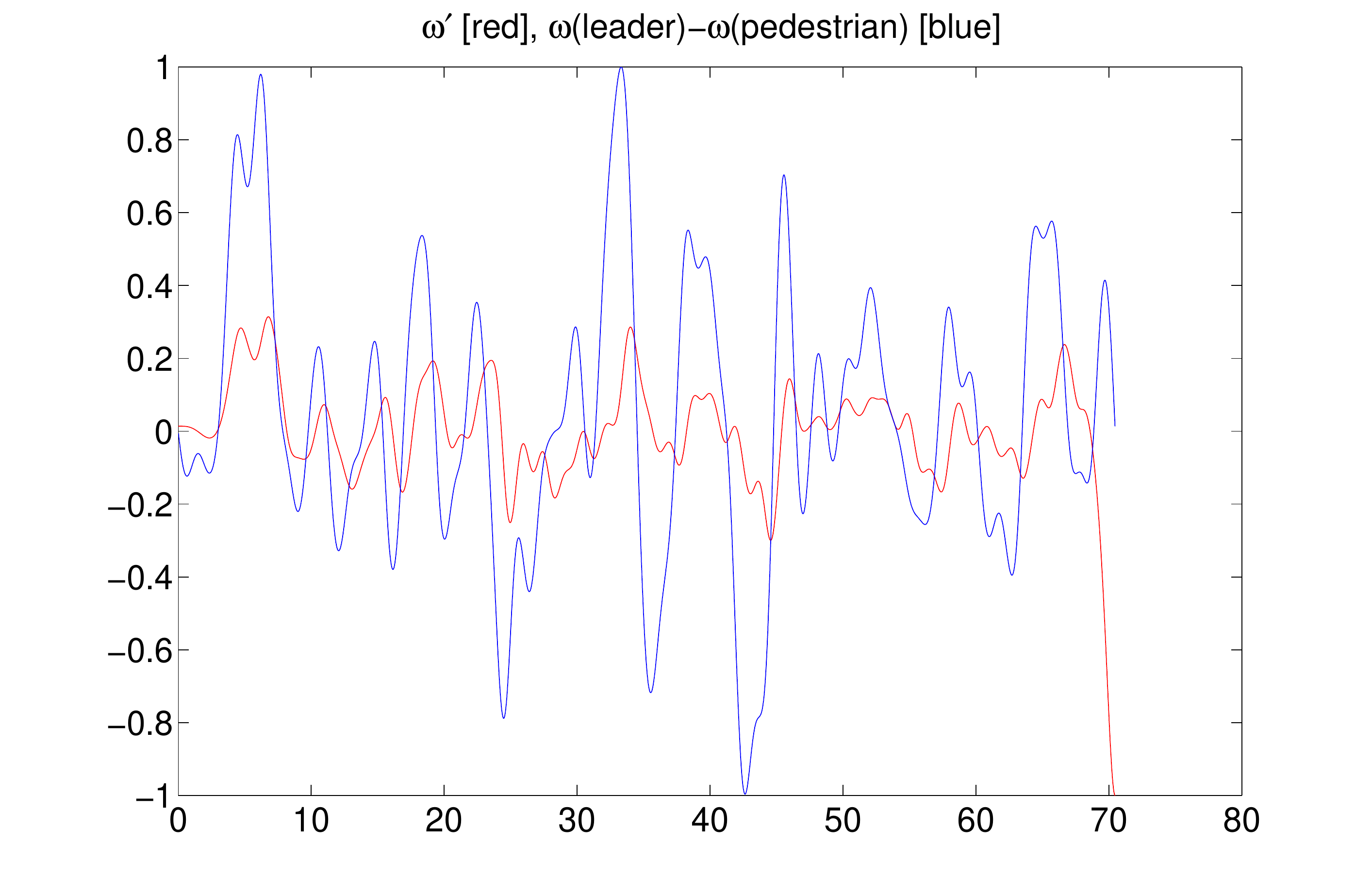}
  \caption{$(\omega_{i+1} - \omega_i)(t)$ (in blue) and $\dot{\omega}_i(t)$ (in red) as functions
    of time $t$ for a typical pedestrian trajectory.  For the sake of a better representation, both functions are normalized by their maximal value. We notice that $\dot{\omega}_i(t)$ has roughly the same features as $(\omega_{i+1} - \omega_i)(t)$ with some time delay.}
  \label{fig:70}
\end{figure}

While the general FTL model used in traffic incorporates a time-delay \cite{Chandler_etal_OperRes58, Gazis_etal_OperRes61}, it has often been neglected in the literature (see e.g. \cite{Aw_etal_SIAP02}). Here, the experimental data suggest that a non-zero time-delay should be used. Indeed, Fig. \ref{fig:70} shows the quantities $(\omega_{i+1}-\omega_i)(t)$ and $\dot{\omega}_i(t) $ as functions of time, for a typical pedestrian trajectory. From this figure, it appears clearly that these quantities are correlated but with some time shift. This observation is generic: it applies to a large proportion of our experimental data. Then, for the sake of simplicity, we restrict ourselves to $\gamma=-1$ and will consider the following model, written for $i=1,\cdots, N$: 
\begin{eqnarray}
&&\hspace{-0.5cm}  \dot{\theta}_i(t)=\omega_i(t), \label{eq:1}\\
&&\hspace{-0.5cm}  \dot{\omega}_i(t+\tau)=C (\omega_{i+1}-\omega_i)(t). \label{eq:modelcst}
\end{eqnarray}
Indeed, later on, it will prove interesting to consider even more general models in which the constants $C$ and $\tau$ are density-dependent, which contains the previous model as a particular case. 

Unfortunately, time-delay differential equations are not always linearly stable. Linear stability means that if a steady-state solution (here for instance, a solution where all velocities $\omega_i$ are equal) is slightly perturbed, the linearized system has bounded solutions.  The stability analysis of the FTL model (\ref{eq:1}), (\ref{eq:modelcst}) has been performed in  \cite{Chandler_etal_OperRes58}. It is shown that the model is stable if $C\tau<1/2$. With the values of $C$ and $\tau$ calibrated from the experiments, we find that this condition is not always fulfilled and that the FTL model can be ustable. After the development of the instability, it is observed that the solution does not fit with the observed trajectories. In particular, we observe particle crossings that are forbidden in the experiments. Therefore, in order to be usable, the FTL model has to be stabilized. 

Here, the stabilization consists in adding a relaxation term which describes the relaxation of the subject's velocity to an averaged velocity over a certain number of neighbors. It is intended to model the fact that a given pedestrian may perceive other subjects than just his leader and take them into account in the decision-making process. Since it is not precisely known how the informations on the various neighbors are combined, the choice of a relaxation model, being the simplest possible one, is the most reasonable. Additionally, by tuning how the average velocity is computed, it allows some flexibility and calibration by comparisons with the data. Therefore, the considered relaxed form of the FTL model is as follows, for $i=1, \ldots, N$: 
\begin{eqnarray}
&&\hspace{-0.5cm}  \dot{\theta}_i(t)=\omega_i(t), \label{eq:1.1}\\
&&\hspace{-0.5cm}  \dot{\omega}_i(t+\tau)=(1-\alpha) C
 (\omega_{i+1}-\omega_i)(t) + \alpha C (\hat{\omega}_i-\omega_i)(t),\label{eq:generalODE}
\end{eqnarray}
where the most general form of the weighted average velocity $\hat{\omega}_i$ is
\begin{gather}\label{eq:8}
  \hat{\omega}_i=\sum_{\ell=0}^{N-1}b_\ell \omega_{(i+\ell)|_N}, \quad
  (i+\ell)|_N=(i+\ell) \text{ modulo } N,
\end{gather}
for positive $b_\ell$ such that $\sum_{\ell=0}^{N-1}b_\ell=1$. The quantity
$\alpha\in[0,1]$ gives the balance between the follow-the-leader term
and the relaxation term. When $b_\ell=\frac{1}{N}$, $\hat{\omega}_i = \hat \omega$ is independent of $i$ and corresponds to the global average velocity. When $b_\ell$ is more strongly peaked about $\ell = 0$, the average velocity becomes more local. In (\ref{eq:generalODE}), we have chosen to write $\dot{\omega}_i(t+\tau)$ as a convex combination of the two terms (respectively corresponding to the relaxation towards the leader's velocity and towards the neighbors' average velocity) rather than writing them as a sum. The reason for this is that the leader also appears in the neighbors' average velocity and that the actual measured reaction rate $C$ should be distributed among these two terms. It is also the choice which provides the best fit with the experimental data as we will see below. 

\begin{remark} The model \eqref{eq:generalODE} is equivalent to the model
\begin{equation*}
    \dot{\omega}_i(t+\tau)= C (\widetilde{\omega}_i-\omega_i),\qquad i=1,\cdots,N,
 \end{equation*}
where the average relaxation speed $\widetilde{\omega}_i$ is defined by
\begin{gather*}
  \widetilde{\omega}_i=\sum_{\ell=0}^{N-1}\widetilde{b_\ell}\omega_{(i+\ell)|_N}, 
\end{gather*}
and is associated with averaging coefficients $\widetilde{b_1}=(1-\alpha)+\alpha b_1$ and $\widetilde{b_\ell}=\alpha b_\ell$ for $\ell\ge 2$.  The coefficients $\widetilde{b_\ell}$ are positive and sum up to 1. In other words, one can restrict to the case $\alpha=1$ for the analysis. However  the parameter $\alpha$ provides a tuning between the FTL model and the relaxation-to-the-mean model, and the interpretation of this parameter is intuitive. We thus propose an analysis depending on $\alpha$ (and the parameters $b_\ell$ will remain fixed).
\label{rem:relaxation}
\end{remark}

We now perform a stability analysis of model (\ref{eq:1.1}), (\ref{eq:generalODE}). In the most general setting considered in the present paper, the constants $C$ and $\tau$ depend on the local density, i.e. $C = C(\rho_i)$ and $\tau = \tau(\rho_i)$ where $\rho_i$ is computed by enumerating the number of agents in a neigborhood of $\theta_i$. In this case, Eqs. (\ref{eq:1.1}), (\ref{eq:generalODE}) are coupled, since $\rho_i$ depends on the values of $\theta_j$ for all $j$. Furthermore, this coupling makes the problem nonlinear. In the special case where $C$ and $\tau$ are constants, this coupling disappears. Indeed, eq. (\ref{eq:generalODE}) can be solved for $\omega_i(t)$, $i=1,\ldots,N$ without knowing the values of $\theta_j$. Once the functions $\omega_i(t)$ are determined, Eq. (\ref{eq:1.1}) can be integrated and the values of the functions $\theta_j(t)$, determined. Additionally, the delay differential equation (\ref{eq:generalODE}) becomes linear. This will be the case considered in the stability analysis performed in the section below.

\subsection{Stability analysis}
\label{subsec:stab_ana}

We write the model \eqref{eq:generalODE} in vector form:
\begin{gather}
  \dot{\boldsymbol{\omega}}(t+\tau)=CA\boldsymbol{\omega}(t),
\end{gather}
where $\boldsymbol{\omega}$ is the vector 
$$ \boldsymbol{\omega} = (\omega_1, \ldots, \omega_N)^T, $$
and the exponent $T$ denotes the transpose of a vector or a matrix. The matrix $A$ is such that the vector $A\boldsymbol{\omega}$ has entries $(A\boldsymbol{\omega})_i$ given by 
$$(A\boldsymbol{\omega})_i = (1-\alpha) (\omega_{i+1}-\omega_i) + \alpha (\sum_{\ell=0}^{N-1}b_\ell \omega_{(i+\ell)|_N}-\omega_i). $$
Concerning this matrix, we have the following

\begin{proposition}\label{prop:eigenvalues}
All the eigenvalues of the matrix $A$ have a non-positive real part. More precisely, 0 is an eigenvalue, and all the other eigenvalues have a negative real part. The non zero eigenvalues lie in the closed disk of center -1 and radius 1.
\end{proposition}

\medskip
\noindent 
{\bf Proof.} $A$ is a circulant matrix. If we denote by $K$ the matrix of entries $k_{ij}$ such that $k_{ij}=1$ if $j=i+1$, $k_{N1}=1$ and $k_{ij}=0$ otherwise, we can write:
$$A=- \mbox{Id} +(1-\alpha) K+ \alpha\sum_{\ell=0}^{N-1} b_\ell \, K^\ell.$$
Now, we introduce $\nu=\exp(2i\pi/N)$. The eigenvalues of $K$ are $\nu^k$ for $k=0, \ldots, N-1$. Hence the eigenvalues of $A$, which will be denoted by $\beta_k$ are given by:
\begin{align}
\beta_k&=-1+(1-\alpha)\nu^k+\alpha\sum_{\ell=0}^{N-1} b_\ell \, \nu^{k\ell} \label{eq:stab2}\\
		& = (1-\alpha)(-1+\nu^k)+\alpha\sum_{\ell=0}^{N-1} b_\ell \, (-1+\nu^{k\ell}). \nonumber 
\end{align}
Therefore if $k=0$ then $\beta_0=0$. Let us now consider the case $k\neq 0$. Then, $-1+\nu^k$ has a negative real part and $-1+\nu^{k\ell}$ has a non-positive real part. 
Then, since $\beta_k$ is a convex combination of all these terms, it has a non-positive real part. More precisely, $\beta_k$ is a barycenter of $-1+\nu^k$ which lies on the circle centered at $-1$ of radius 1, and of  $\sum_{\ell =0}^{N-1} b_\ell (-1+\nu^{k\ell})$ wich is a barycenter of points of the same circle, and hence, which lies in the disk centered at $-1$ of radius 1. Therefore, by convexity, $\beta_k$ also belongs to the disk of center -1 and radius 1.
This ends the proof. \endproof

\begin{figure}
  \centering
  \includegraphics[scale=0.45]{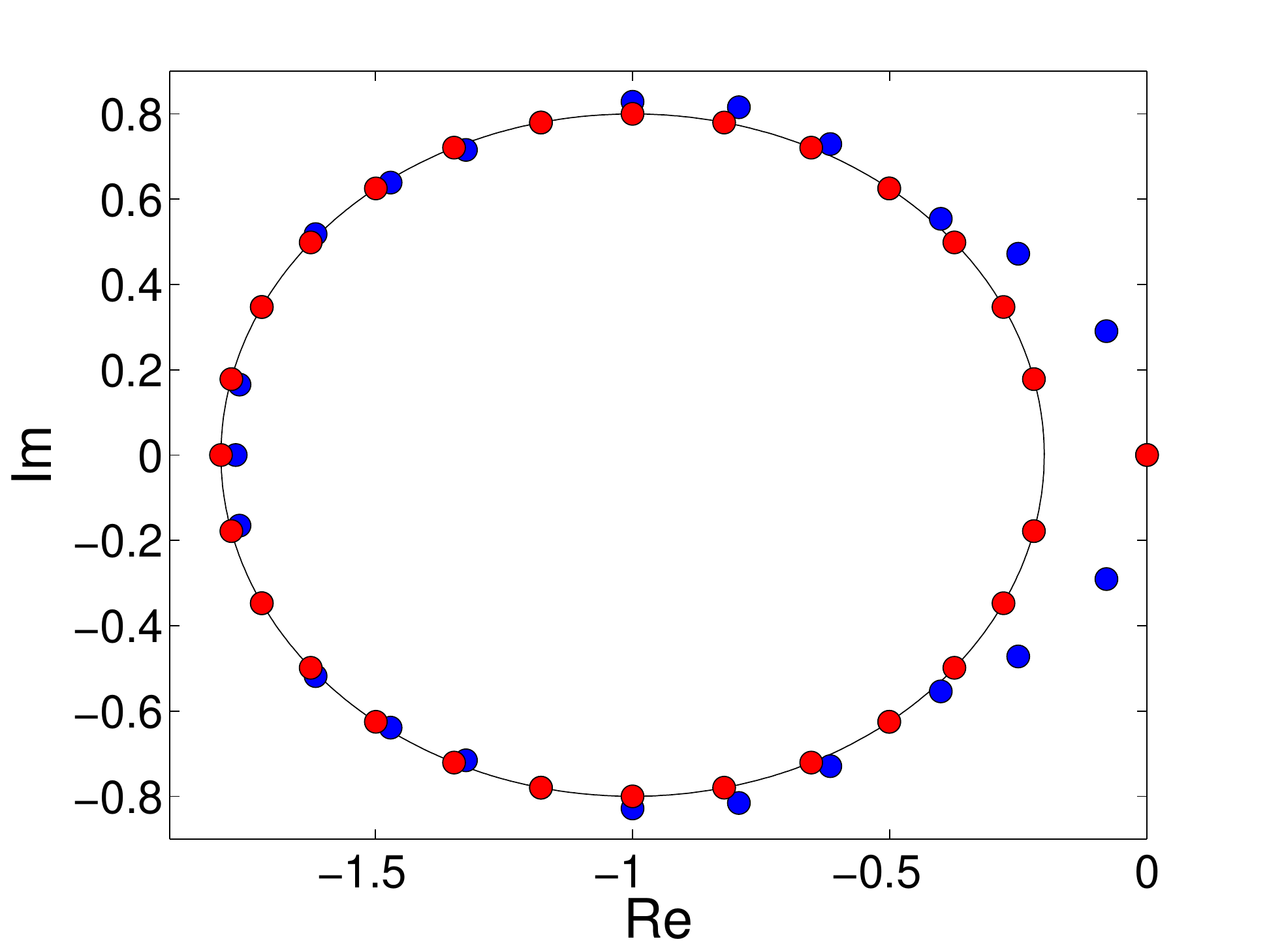}
 \caption{Eigenvalues of the system \eqref{eq:generalODE} for $N=28$ with relaxation to
   the global mean velocity in red ($b_\ell=1/28$ for all $\ell$) and  relaxation to
   the local  mean velocity in blue ($b_\ell=1/7$ for $\ell=1..7$ and $0$ otherwise), with $\alpha=0.2$. The black circle is centered at -1 and has radius $1-\alpha=0.8$.}
  \label{fig:42}
\end{figure}

\medskip
As an illustration, in Figure \ref{fig:42}, we represent  the eigenvalues of \eqref{eq:generalODE} in the complex plane in the case $N=28$, $\alpha=0.2$ and for two examples of relaxation operator (i.e. two different choices for the coefficients $b_k$). The case of a relaxation to the global average (i.e. all the $b_k$ being equal and summing up to $1$) is represented in red and that of a relaxation to a local average computed on the seven closest neighbors in front (with equal weight) is represented in blue. In both cases, we see that the eigenvalues of $A$ have non-positive real part, and consequently, the linear ODE  $\dot{\boldsymbol{\omega}}(t) = C A \boldsymbol{\omega}$ is stable for any value of $C \geq 0$. 

\medskip
The following stability analysis for the ODE with delay \eqref{eq:generalODE} follows  standard works on delay ODEs \cite{Pontrjagin_AMSTrans55, Bellman_Cooke_AcadPress63}. The characteristic equation, i.e. the equation that $\lambda \in {\mathbb C}$ must satisfy for the existence of a solution of the form $\boldsymbol{\omega}(t) = \boldsymbol{\omega}_0 e^{-\lambda t}$ with $ \boldsymbol{\omega}_0 \not = 0$, reads:
\begin{equation}
\mbox{det}(-\lambda \, \mbox{Id}+CAe^{-\tau \lambda})=0.
\label{eq:chara}
\end{equation}
The value $\lambda = 0$ is always a solution to \eqref{eq:chara} (corresponding to the eigenvalue $\beta_0 = 0$ of $A$), whatever the value of $\tau \geq 0$ is, and corresponds to a state where all the agents have the same constant velocity. The system without delay is stable, as stated in the previous proposition. Since the eigenvalues depend continuously on the delay $\tau$, the system with delay is stable for delays $\tau$ such that $\tau<\tau^\star$, 
where the critical delay $\tau^\star$ is the smallest delay $\tau$ such that the associated eigenequation (\ref{eq:chara}) admits a non-zero pure imaginary solution.

\begin{proposition}\label{prop:delay1}
The critical delay $\tau^\star$ for the model \eqref{eq:generalODE} satisfies
$$\tau^\star\ge \dfrac{1}{2C}.$$
\end{proposition}

\medskip
\noindent
{\bf Proof.} The critical delay $\tau^\star$ is the smallest value of $\tau$ such that the characteristic equation \eqref{eq:chara} admits a non-zero pure imaginary solution, i.e. the smallest $\tau$ such that there exists $\omega_0\in{\mathbb R}$ and some $k\in\{1,\ldots, N-1\}$ satisfying 
$$-i\omega_0+C\beta_ke^{-i\tau\omega_0}=0,$$
where we recall that $\beta_k$, $k\in\{0,\ldots, N\}$, denote the eigenvalues of $A$. Decomposing $\beta_k=\rho e^{i\theta}$, we get:
\begin{equation}
-i\omega_0+C\rho e^{i\theta}e^{-i\tau\omega_0}=0.
\label{eq:stabilite}
\end{equation}
Since $\beta_k$ belongs to the disk centered at -1 of radius 1, we can choose $\theta\in[\pi/2,3\pi/2]$ and $\rho\le2|cos\theta|$.
Let $\epsilon\in\{-1,1\}$ denote the sign of $\omega_0$. Then \eqref{eq:stabilite} reads
\begin{equation}
\epsilon\omega_0=C\rho\quad {\rm and} \quad e^{i(\theta-\tau\omega_0)}=\epsilon i.
\label{eq:stab1}
\end{equation}
Thanks to the second equation (\ref{eq:stab1}), there exists an integer $m\in{\mathbb Z}$ such that $\theta=\tau\omega_0+\epsilon \pi/2+ 2m\pi$,  and with the first equation (\ref{eq:stab1}), this leads to $\theta=\epsilon (C \rho\tau+\pi/2)+2m\pi$, or $C \rho\tau=-\pi/2 + \epsilon \theta + 2m\pi$. Then, we distinguish the two cases $\epsilon = 1$ and $\epsilon = -1$: 

\begin{itemize}
\item[] Case 1: $\epsilon=1$. Then, we have: 
\begin{equation}
\tau^\star = \frac{\theta-\pi/2}{\rho C} \geq \frac{\theta-\pi/2}{2 C|\cos\theta|} \geq \frac{1}{2C}.
\label{eq:stab3}
\end{equation}

\item[] Case 2: $\epsilon=-1$. Then: 
$$\tau^\star = \frac{3\pi/2-\theta}{\rho C} \geq \frac{3\pi/2-\theta}{2 C|\cos\theta|} \geq \frac{1}{2C}. $$
\end{itemize}

In both cases, we find that $\tau^*$ is bounded from below by $\frac{1}{2C}$, which ends the proof. \endproof

\medskip
In the case of a relaxation to the global mean velocity (i.e. when all the $b_\ell$'s are equal), we can prove some refined bounds on the critical delay $\tau^\star$, as shown in the following.

\begin{proposition} 
\label{prop:bounds}
Let us assume that $b_\ell=1/N$ for $\ell=0, \ldots, N-1$. 
If $N$ is even, then the critical delay satisfies
$$\frac{\max \big( 1, \arccos (1-\alpha) \big) }{(2-\alpha) \, C} \leq \tau^\star \leq \frac{\pi}{2(2-\alpha)\, C}.$$
The lower bound is also valid if $N$ is odd. It improves that given in Prop. \ref{prop:delay1} if $\alpha \leq 1-\cos(1) \sim 0.46$.
\end{proposition}

\medskip
\noindent
{\bf Proof.} By elementary properties of $N$-th roots of unity, we have $\sum_{\ell=0}^{N-1} b_\ell \nu^{k \ell} =\frac{1}{N} \sum_{\ell=0}^{N-1}\nu^{k \ell} $ $= 0$. Then, from (\ref{eq:stab2}), we deduce that $\beta_k=-1+(1-\alpha)\nu^k$ belongs to the circle of center $-1$ and radius $1-\alpha$. If we write $\beta_k=\rho e^{i\theta}$ (and assume by symmetry that $\theta\in[0,\pi]$) then simple geometric considerations show that $\rho\le 2-\alpha$ and $\sin(\pi-\theta) \leq 1-\alpha$. Hence  $\theta-\pi/2 \geq \arccos(1-\alpha)$. Therefore 
$$\tau^\star=\frac{\theta-\pi/2}{\rho C} \geq \frac{\arccos(1-\alpha)}{(2-\alpha) C}.$$

Another possibility is to note that all the eigenvalues $\beta_k$ lie in the disk of center $(1-\alpha/2)$ and of radius $1-\alpha/2$. Hence as above, if $\beta_k=\rho e^{i\theta}$ then $\rho \leq (2-\alpha) |\cos \theta|$. This provides the following lower bound:
$$\tau^\star \geq \frac{1}{(2-\alpha) C}.$$

If $N$ is even, then $-2+\alpha$ is an eigenvalue of $A$ and is such that $\rho=2-\alpha$ and $\theta=\pi$. Then, by (\ref{eq:stab3}), the delay at which the corresponding eigenvalue of the delay problem reaches the imaginary axis is equal to $\frac{\pi}{2(2-\alpha)}$. Therefore, we deduce that $\tau^\star \leq \frac{\pi}{2(2-\alpha)}$. This ends the proof. \endproof

\begin{figure}
  \centering
  \includegraphics[scale=0.5]{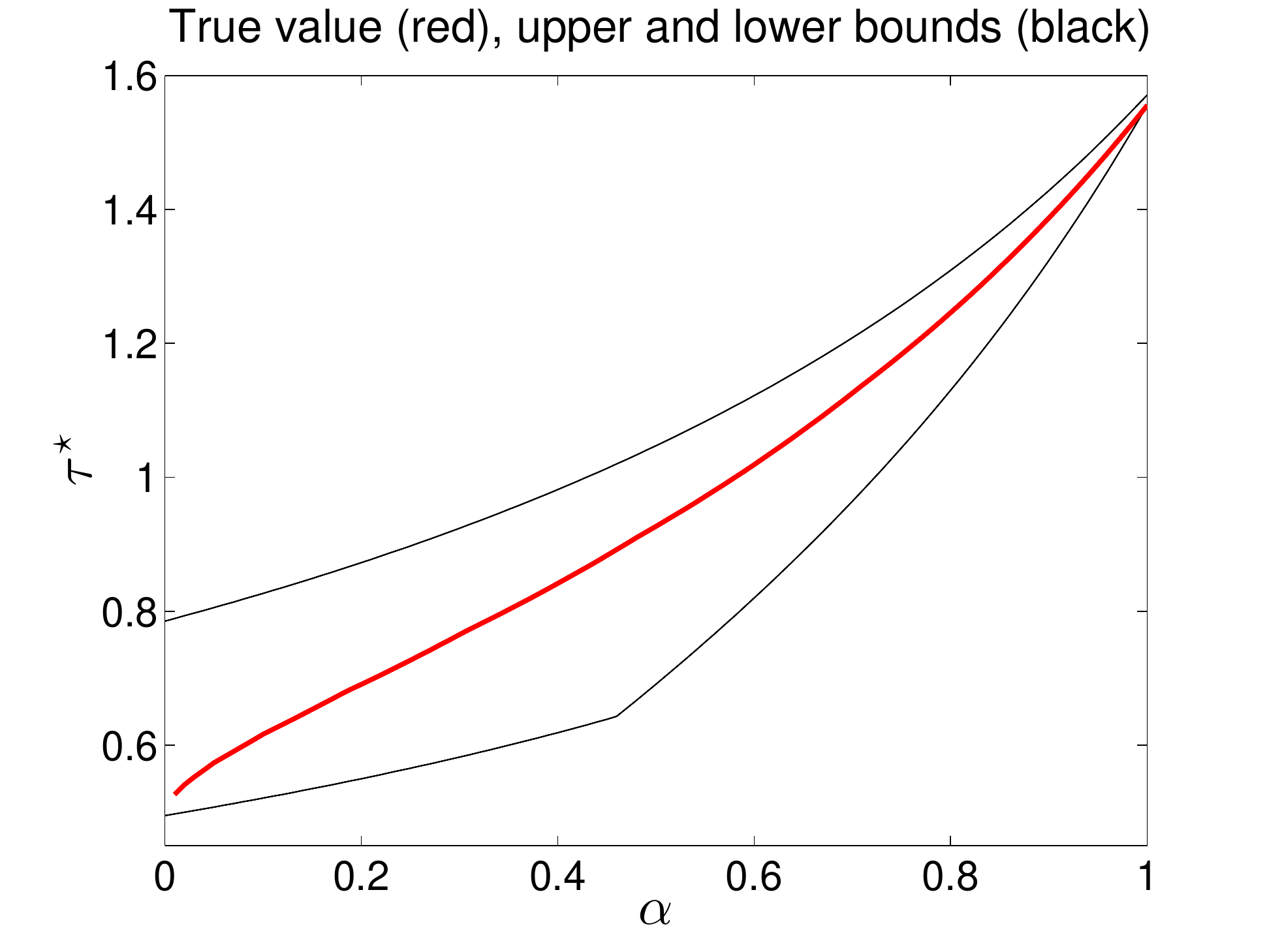}
  \caption{The maximal delay $\tau^*$ which maintains the stability of the delay differential system, plotted as a function of  $\alpha$ in the case of a relaxation to the global average velocity ($b_\ell=1/N$ for all $\ell$). Here, we have chosen $C=1.01$. The upper and lower bounds given by Prop. \ref{prop:bounds} are plotted in black for comparison.}
  \label{fig:3}
\end{figure}

\medskip
The precise value of $\tau^\star$ can also be computed numerically since all the eigenvalues $\beta_k$ are known. The result is presented in Figure \ref{fig:3}. The quantity $\tau^\star$ is plotted as a function of $\alpha$ (in red). By comparisons, the lower and upper bounds given by Prop. \ref{prop:bounds} are displayed in black color.

\subsection{Model calibration}
\label{sec:model_calibration}

In order to estimate the model parameters, i.e. the delay $\tau$ and the reaction constant $C$ from experimental data we proceed by cross correlation. However we will consider small sub-windows since the parameters are not strictly constant over time. The procedure for the estimation of the delay is as follows:

We consider a time-window $I$. For each pedestrian $i$, we estimate the delay $\tau$ that accounts the best for the measurements in this time window. This delay maximizes
   \begin{gather*}
            \tau^{\rm obs}=\text{argmax}_{\tau\in
       [\tau_{\min},\tau_{\max}]}\dfrac{\langle
   \dot{\omega}_i(\cdot+\tau),(\omega_{i+1}-\omega_i)(\cdot)\rangle_{L^2_{I}}}{\norm{\dot{\omega}_i(\cdot+\tau)}_{L^2_{I}}}, 
   \end{gather*}
where for two functions of time $f(t)$ and $g(t)$ defined on $I$, we denote by $\langle f,g\rangle_{L^2_{I}}$ and $\|f\|^2_{L^2_{I}}$ the $L^2$ inner product of $f$ and $g$ and the squared $L^2$ norm of $f$, respectively defined by 
$$ \langle f,g\rangle_{L^2_{I}} = \int_{t \in I} f(t) \, g(t) \, dt, \qquad \|f\|^2_{L^2_{I}} = \int_{t \in I} |f(t)|^2 \, dt. $$
In this work, the delay $\tau^{\rm obs}$ is searched within the interval $[\tau_{\min},\tau_{\max}]=[-2 \,\mbox{s},3 \, \mbox{s}]$.

For the same given pedestrian and the same time window, the best constant $C$ is then determined by least-squares minimization, which reads
 \begin{gather*}
   C=\frac{\langle
   \dot{\omega}_i(\cdot+\tau^{\rm obs}),(\omega_{i+1}-\omega_i)(\cdot)\rangle_{L^2_I}}{\norm{(\omega_{i+1}-\omega_i)}_{L^2_I}^2}.
 \end{gather*}
Associated to these quantities, the correlation coefficient is a  quantity which belongs to $[-1,1]$ and which provides a reliability measurement. It is defined by:
 \begin{equation*}
   \epsilon=\dfrac{\langle  \dot{\omega}_i(\cdot+\tau),(\omega_{i+1}-\omega_i)(\cdot)\rangle_{L^2_{I}}}{\norm{\dot{\omega}_i(\cdot+\tau)}_{L^2_{I}}\norm{\omega_{i+1}-\omega_i}_{L^2_{I}}}.
 \end{equation*}

We say that the model appropriately describes the data from a given pedestrian and a given time-window $I$ (or that "the data from $I$ are compliant with the model") if the following two constraints are simultaneously satisfied: 
\begin{itemize}
\item[(i)] the correlation $\epsilon$ is close to $1$, i.e. is larger than a given threshold $\epsilon_t$. We choose $\epsilon_t=0.6$ unless stated otherwise.
\item[(ii)] the resulting time-delay $\tau$ satisfies $0 \leq \tau \leq \tau_{\max}-0.05$, where $[\tau_{\min},\tau_{\max}]$ is the interval where $\tau$ is sought.
\end{itemize}
We also discard all the data from one given pedestrian, if there are less than $1/3$ compliant data for this pedestrian, collected on all the given time windows. These outliers originate either from intrinsic differences in the subject's behavior or more likely from incorrect data reconstruction. They are discarded to avoid pollution of the calibrated parameters by incorrect data. 

Thanks to this procedure, for each experiment, we collect samples which consist of all the values of $\tau$ (or of $C$) for all the considered time-windows and all the pedestrians. In our data processing, a set of windows is defined by shifting by steps equal to $50\Delta t=5/12s$.

\subsection{Numerical approximation}
\label{subsec:implementation}

We now describe the numerical scheme that was implemented to solve the
model \eqref{eq:generalODE}. This is a delay differential equation. A
4th order Runge-Kutta method was used to solve this problem. Due to the delay, an interval of initial condition is required and the resolution on a time interval $[t_1,t_2]$ requires the knowledge of the data in the interval $[t_1-\tau,t_1]$.


We present below simulations with different values of the delay, or with density-dependent delay, however the delays are always less than $10s$. 
In all our simulations, we therefore used the observations on the interval $[0,10s]$ as initial condition and performed a simulation on the interval $[10s,80s]$.

\setcounter{equation}{0}
\section{Results}
\label{sec:results}

\subsection{Model calibration}
\label{subsec:resul_model_calibration}

\subsubsection{Calibration of model parameters: case of constant parameters}
\label{subsec:resul_estimation}

In a first step, we suppose that the model parameters $\tau$ and $C$ are constant and in particular independent of the local pedestrian density. In a forthcoming section, we will see that the calibration is improved by making the parameters $\tau$ and $C$ dependent of the local density.  

\begin{figure}
\centering
    \subfigure[High density ($28$ pedestrians walking on the inner circle: average density $1.86$ ped m$^{-1}$), mean=0.82, std=0.51.]{\includegraphics[scale=0.31]{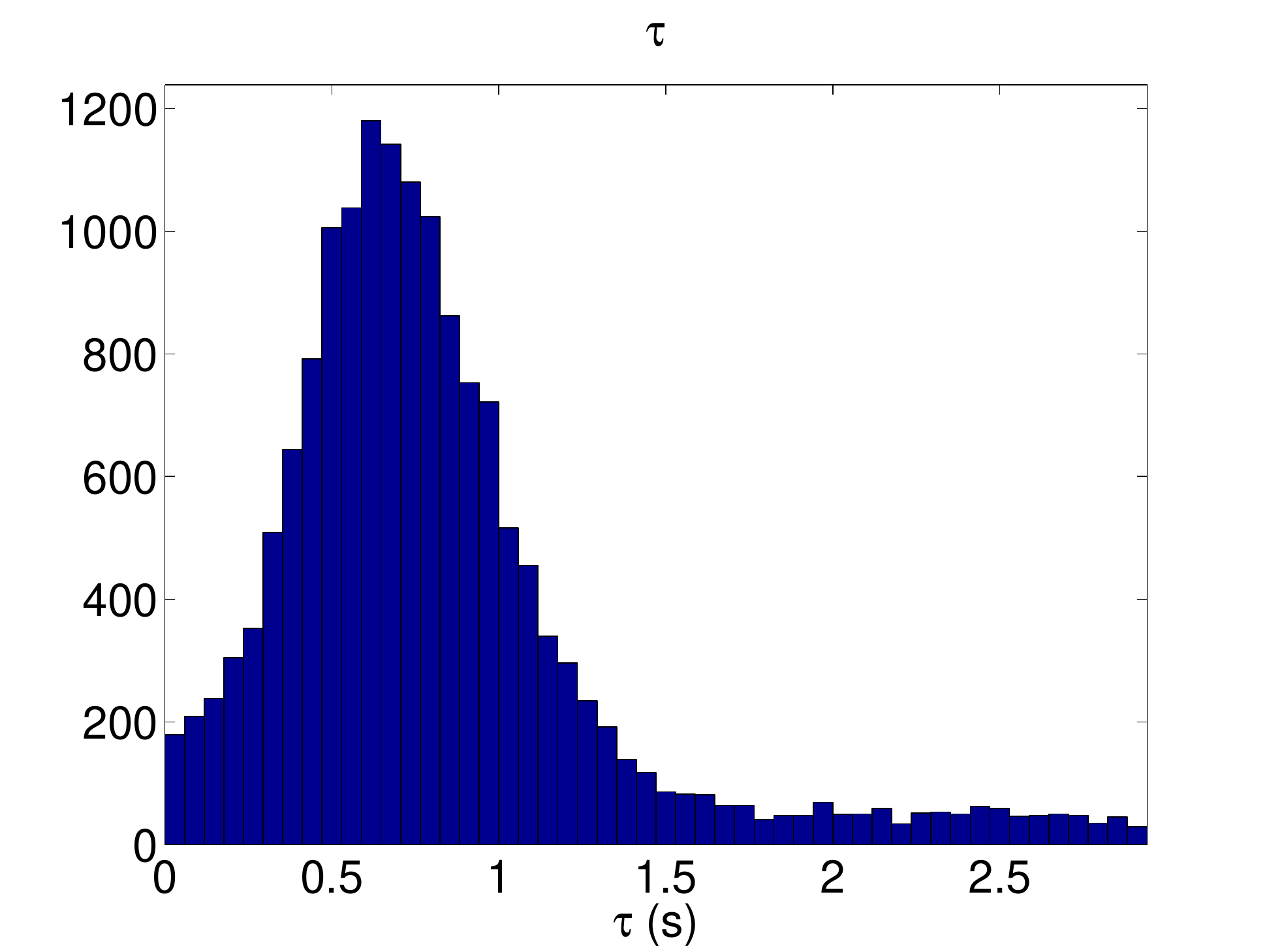}} \hspace{0.2cm}
    \subfigure[medium high density case (24 pedestrians walking on the inner circle: average density $1.59$ ped m$^{-1}$), mean=0.71, std=0.45.]{\includegraphics[scale=0.31]{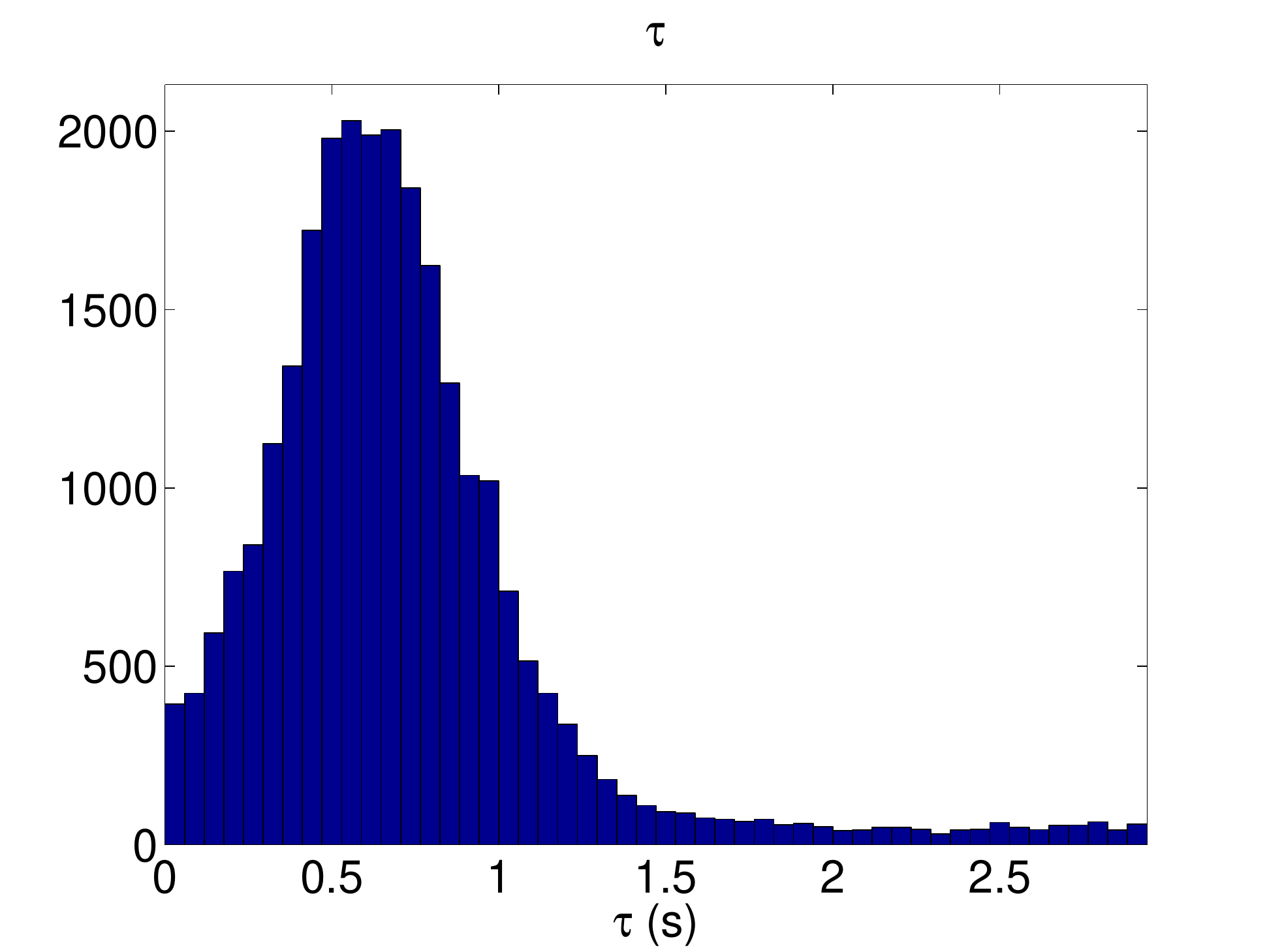}}\\
      \subfigure[Medium low density case ($24$ pedestrians walking on the outer circle: average density $0.93$ ped m$^{-1}$), mean=0.8, std=0.42.]{\includegraphics[scale=0.31]{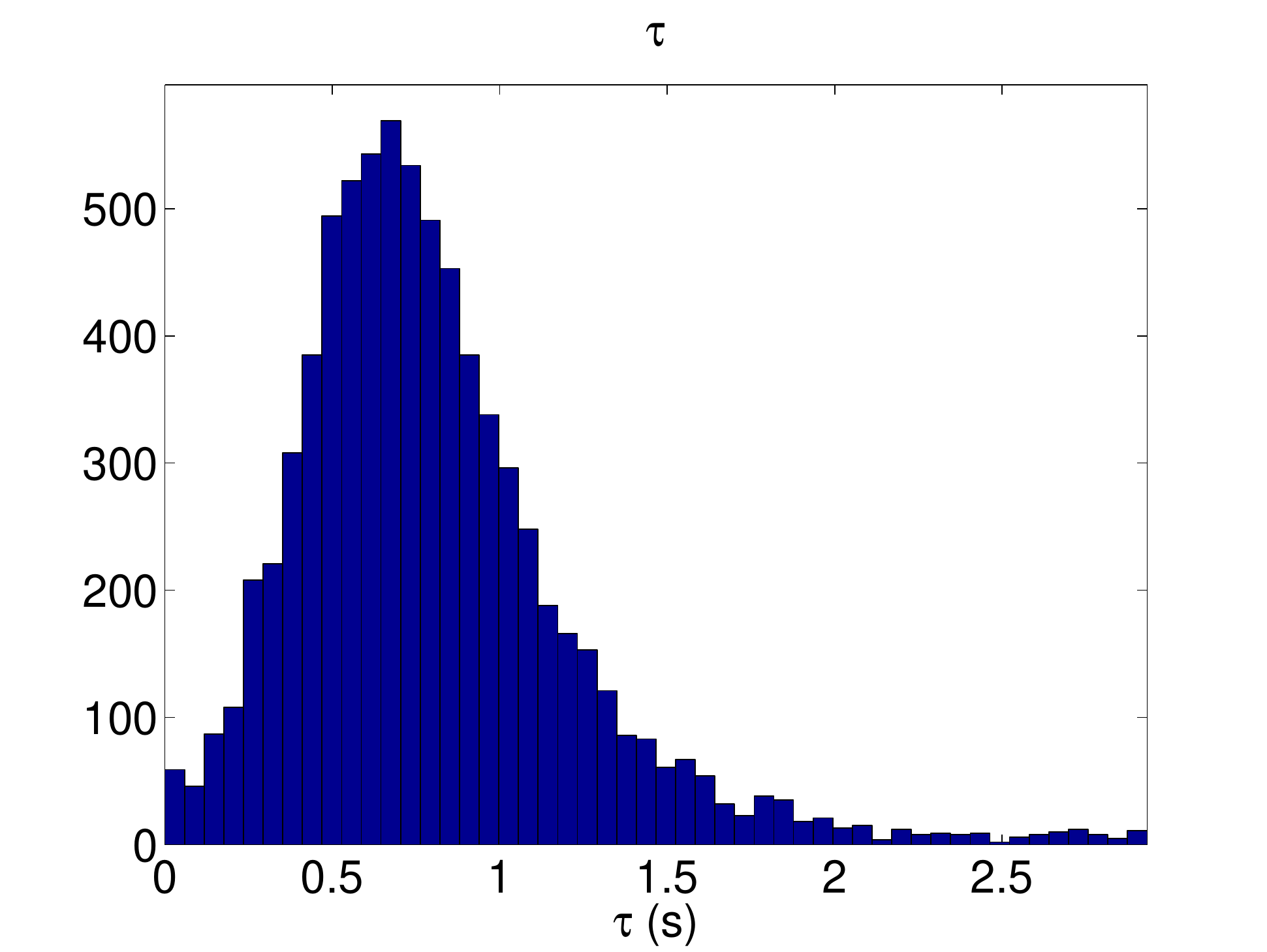}}\hspace{0.2cm}
   \subfigure[Low density case ($8$ pedestrians: average density $0.31$ ped m$^{-1}$), mean=1.04, std=0.58.]{\includegraphics[scale=0.31]{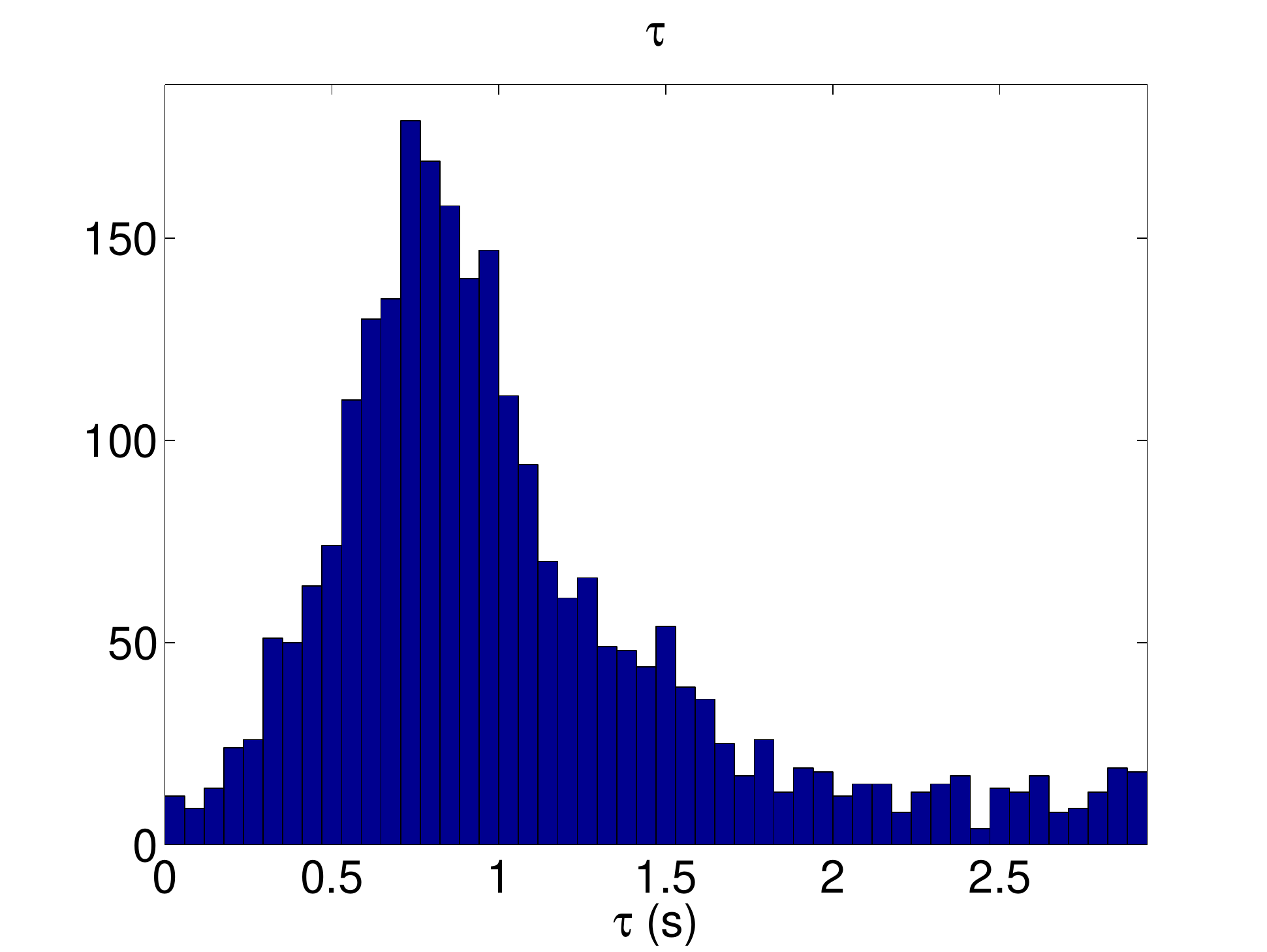}}
     \caption{Histograms of the time delay $\tau$ for different average densities $\rho_{\mbox{\scriptsize av}}$:  (a)~$\rho_{\mbox{\scriptsize av}} = 1.86 \mbox{\, ped \, m}^{-1}$ ; (b) $\rho_{\mbox{\scriptsize av}} = 1.59 \mbox{\, ped \, m}^{-1}$ ; (c) $\rho_{\mbox{\scriptsize av}} = 0.93 \mbox{\, ped \, m}^{-1}$ ; (d)~$\rho_{\mbox{\scriptsize av}} = 0.31 \mbox{\, ped \, m}^{-1}$ (with ``ped'' standing for ``pedestrian'').  The samples are defined as being pairs (pedestrian, time window) and the samples range through all experiments with the same average pedestrian density. For each case we indicate the mean value and the standard deviation of $\tau$.  For each pedestrian, the time delay is estimated within the range $[\tau_{\min},\tau_{\max}]=[-2 \,\mbox{s},3 \, \mbox{s}]$. The time windows are shifted by steps equal to $50\Delta t=0.417s$.}\label{fig:tauexp}
  \end{figure}

\begin{figure}
\centering
      \subfigure[High density case ($28$ pedestrians walking along the inner circle: average density $1.86$ ped m$^{-1}$), mean=0.95, std=0.4.]{\includegraphics[scale=0.31]{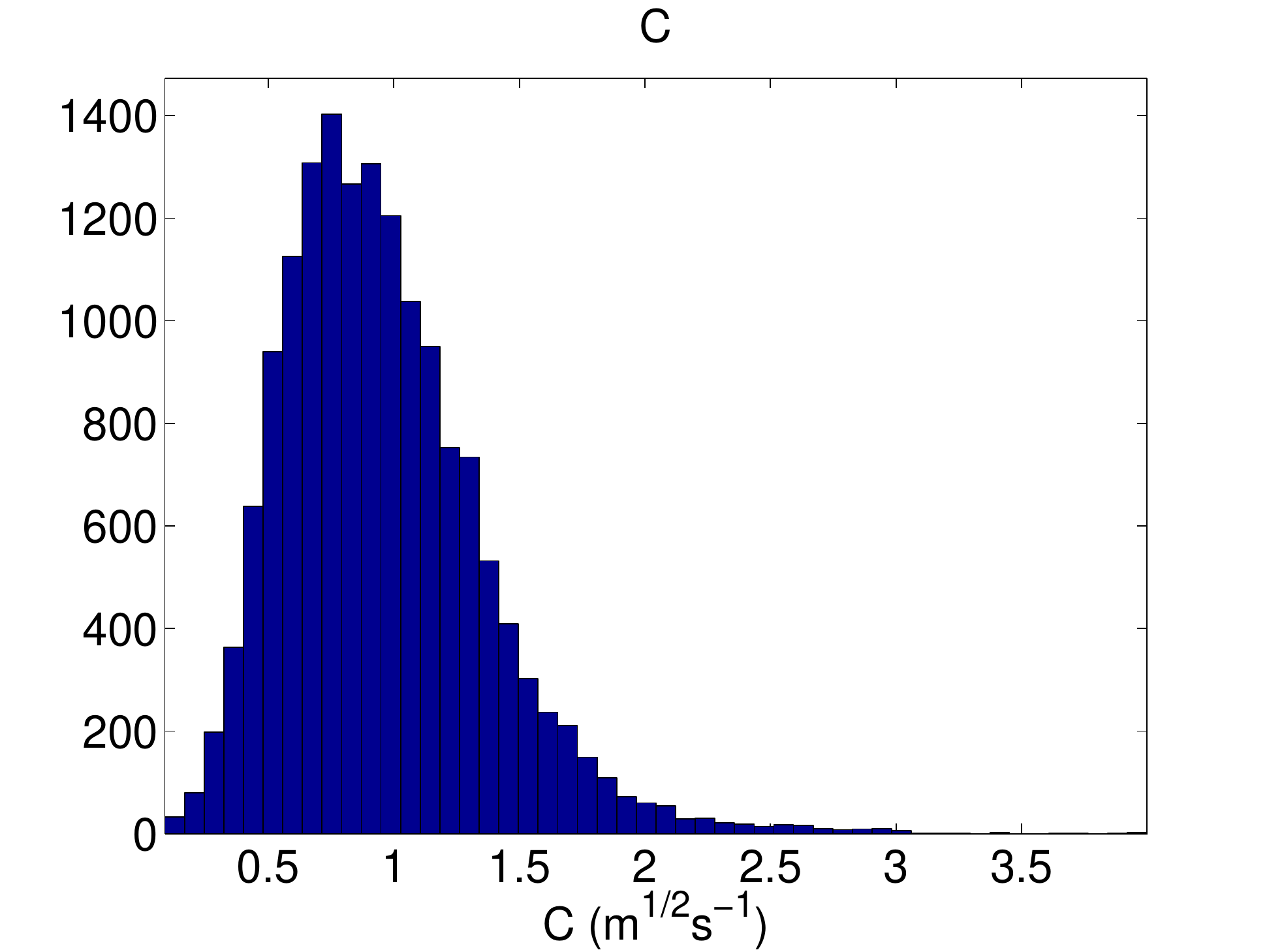}}  \hspace{0.2cm} 
      \subfigure[Medium high density case ($24$ pedestrians walking along the inner circle: average density $1.59$ ped m$^{-1}$), mean=1.09, std=0.44.]{\includegraphics[scale=0.31]{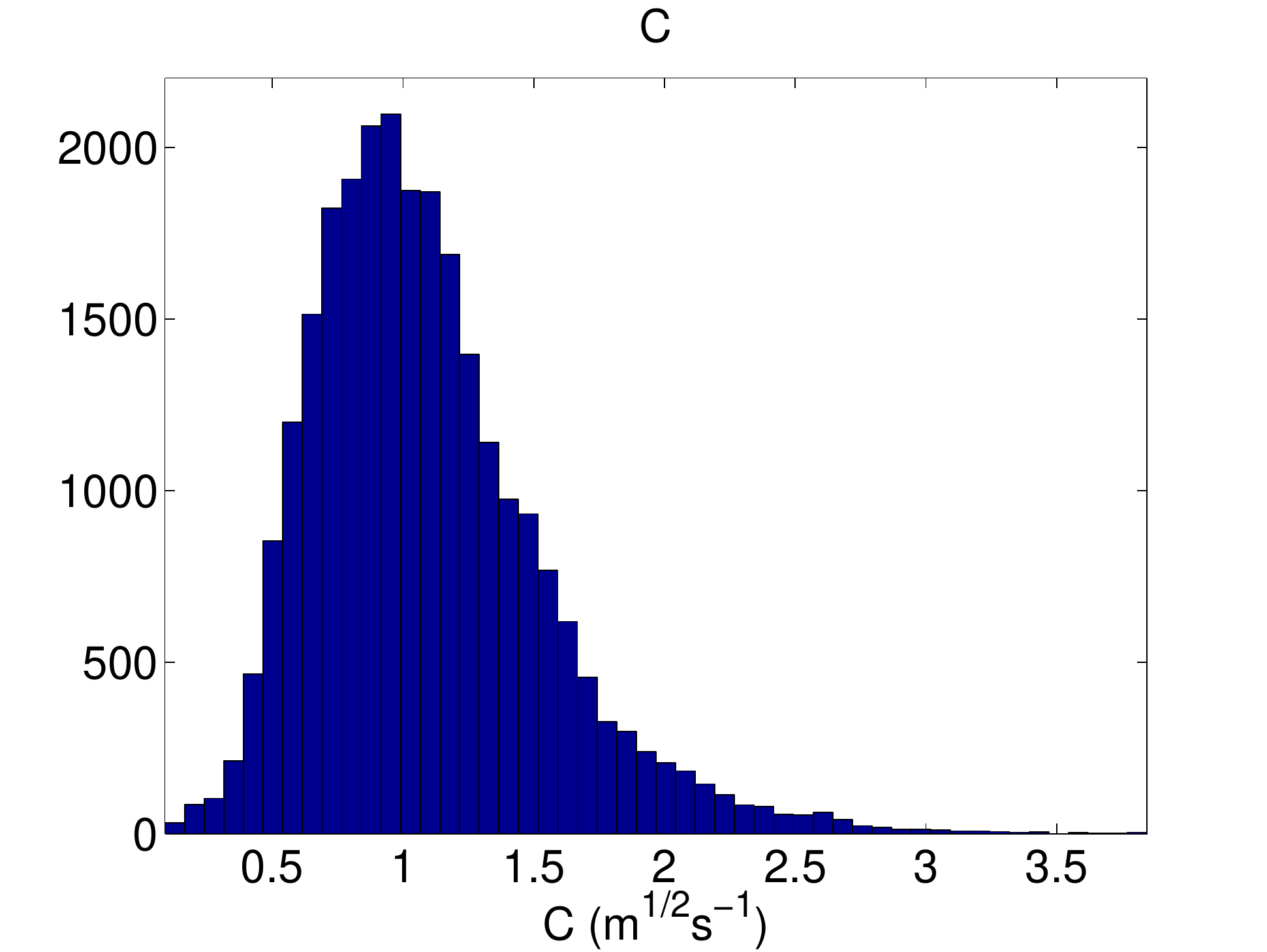}}\\
      \subfigure[Medium low density case($24$ pedestrians walking along the outer circle: average density $0.93$ ped m$^{-1}$), mean=0.84, std=0.36.]{\includegraphics[scale=0.31]{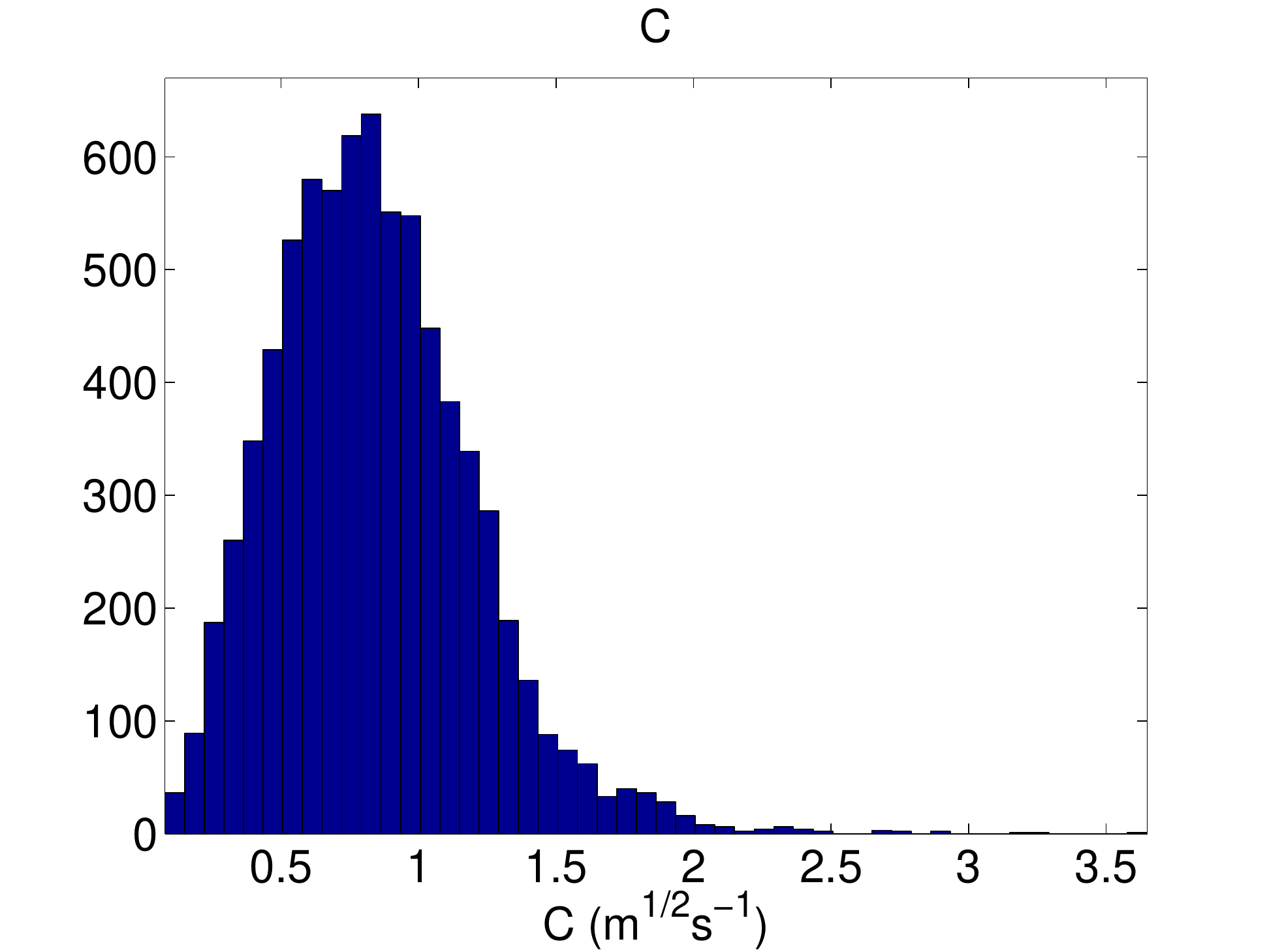}}  \hspace{0.2cm}
        \subfigure[Low density case ($8$ pedestrians: average density $0.31$ ped m$^{-1}$), mean=0.62, std=0.37.]{\includegraphics[scale=0.31]{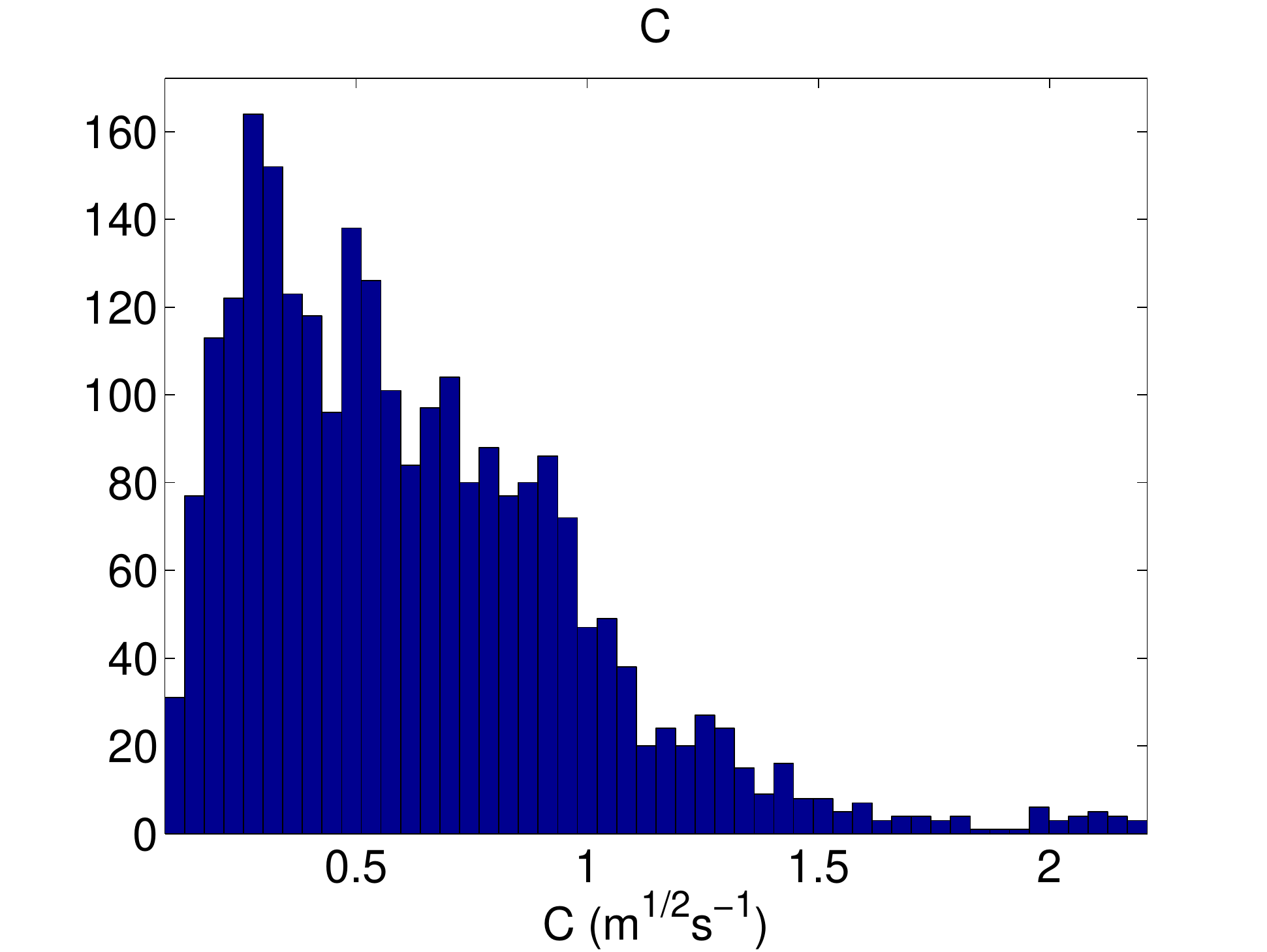}}
     \caption{Histogram of the reaction constant $C$ for different average densities $\rho_{\mbox{\scriptsize av}}$:  (a)~$\rho_{\mbox{\scriptsize av}} = 1.86 \mbox{\, ped \, m}^{-1}$ ; (b) $\rho_{\mbox{\scriptsize av}} = 1.59 \mbox{\, ped \, m}^{-1}$ ; (c) $\rho_{\mbox{\scriptsize av}} = 0.93 \mbox{\, ped \, m}^{-1}$ ; (d)~$\rho_{\mbox{\scriptsize av}} = 0.31 \mbox{\, ped \, m}^{-1}$ (with ``ped'' standing for ``pedestrian'').  The samples are defined as being pairs (pedestrian, time window) and the samples range through all experiments with the same average pedestrian density. For each case we indicate the mean value and the standard deviation of $C$. For each pedestrian, the values of $C$ is estimated within the range $[\tau_{\min},\tau_{\max}]=[-2 \,\mbox{s},3 \, \mbox{s}]$. The time windows are shifted by steps equal to $50\Delta t=0.417s$.}\label{fig:Cexp}
  \end{figure}

The estimated values of the model constants $\tau$ and $C$ from the experimental data are presented in Figs. \ref{fig:tauexp} and \ref{fig:Cexp}  respectively. More precisely, Fig. \ref{fig:tauexp} and \ref{fig:Cexp} show histograms of the estimated values of $\tau$ and $C$ where the samples are defined as being pairs (pedestrian, time window) and the samples range through all experiments with the same average pedestrian density. The average density $\rho_{\mbox{\scriptsize av}}$ is defined as the total number of pedestrian involved in a given experiment divided by the average walking radius for that experiment. 

Four different histograms, corresponding to four different values of the average pedestrian density $\rho_{\mbox{\scriptsize av}}$ are shown by order of decreasing density:  (a) $\rho_{\mbox{\scriptsize av}} = 1.86 \mbox{\, ped \, m}^{-1}$ ; (b) $\rho_{\mbox{\scriptsize av}} = 1.59 \mbox{\, ped \, m}^{-1}$ ; (c) $\rho_{\mbox{\scriptsize av}} = 0.93 \mbox{\, ped \, m}^{-1}$ ; (d) $\rho_{\mbox{\scriptsize av}} = 0.31 \mbox{\, ped \, m}^{-1}$ (with ``ped'' standing for ``pedestrian''). In these histograms, only the time windows where the data are compliant with the model are retained. For these Figures, the following parameters were used: cut-off frequency $f_c = 0.5$ Hz ; window width $w_w = 6.67$ s, correlation threshold $\varepsilon_t = 0.6$; model parameter $\gamma = 1$.

In Table \ref{tab:percent}, we summarize  the proportion of compliant data as a function of the total number of pedestrians (we do not distinguish between the inner or outer circles). The proportion of compliant data is small for the low average density cases (for the experiments with $8$ pedestrians, the average density is $0.31$ ped m$^{-1}$) but quite large above $16$ pedestrians (average density of $0.6$ ped m$^{-1}$). This  indicates that the delay-differential model \eqref{eq:modelcst} is well-adapted for densities above $0.6$ ped m$^{-1}$ but has poorer match with the data for lower densities.

\begin{table}[htbp]
  \centering
  \begin{tabular}[bc]{|c|c|}
\hline
    Number of pedestrians & percentage of compliant data\\
    \hline
    $8$ & $44.96$ \%\\
    $16$ & $74.91$\%\\
    $20$ & $82.34$\%\\
    $21$ & $84.96$\%\\
    $24$ & $79.75$\%\\
    $28$ & $77.44$\%\\
\hline
  \end{tabular}
  \caption{Proportion of compliant data as
a function of the total number of pedestrians.}\label{tab:percent}
\end{table}

For all the experiments where the model is relevant (all cases but low average density)
the delay $\tau$ has a distribution around a mean value of the order of $0.8$ s, with a standard deviation of the order of $0.45$ s, see Figs. \ref{fig:tauexp} (a), (b) and (c). We can conclude that there is no fixed value for the delay $\tau$ but rather a range of values. Similarly, for the experiments other than the low average density experiments, the constant $C$ has a distribution around a mean value of the order of $1$ m$^{1/2}$s$^{-1}$, with a standard deviation of the order of $0.4$ m$^{1/2}$s$^{-1}$, see Figs. \ref{fig:Cexp} (a), (b) and (c).
The choice of the delay model \eqref{eq:modelcst} or its stabilized version \eqref{eq:generalODE} is comforted by the fact that a large amount of data fits the model (see Table~\ref{tab:percent}).

\subsubsection{Sensitivity with respect to the processing parameters}
\label{subsubsec:sensitivity}

The calibration procedure of the model from the data depends on several processing parameters, namely the correlation threshold $\epsilon_t$ for a sample defined by a (pedestrian, time-window) pair to be compliant with the model, the time window width $w_w$, the cut-off frequency $f_c$. We have tested the influence of each parameter on the estimation of $\tau$ and $C$. The detailed analysis is presented in the appendix B and  summarized in Table \ref{tab:sensitivity}.

As a summary, the following observations can be made: 
\begin{enumerate}
\item The parameters $\epsilon_t$ and $w_w$ have a little influence on the results, and therefore our analysis is stable with respect to these parameters. 
\item The parameter $f_c$ has a strong influence on the estimated time-delay $\tau$ and reaction constant $C$. Therefore the choice of this parameter is of importance, and its value must be chosen according to 'physics-based' criteria. In the rest of the work we use $f_c = 0.5$ Hz, which is of the order of magnitude of the stepping frequency. Therefore, this value allows  to smooth out the oscillations due to the stepping of the pedestrians without perturbing phenomena occuring at longer time-scales.
\end{enumerate}

\begin{center}
\begin{table}[htb]
  \begin{tabular}[bc]{|c|c|c|c|c|}
\hline
    parameter & range & relative  & relative varia- & relative varia\\
     &  &  range & tion of $\tau$ & tion of $C$\\
    \hline
    \hline
    $\epsilon_t$ & $0.6$ -- $0.8$ & 29 \, \% & 1.6\, \% & 5.5\, \%\\
    \hline
    $w_w$ & $5$ -- $8$ s & 46\, \% & 2.5\, \% & 5.3\, \% \\
    \hline
    $f_c$ & $0.2$ -- $1.2$ Hz & 143 \, \% & 90\, \% & 45\, \% \\
	 \hline
  \end{tabular}
  \caption{Sensitivity of $\tau$ and $C$ with respect to the correlation cut-off $\epsilon_t$, time-window width $w_w$ and cut-off frequency $f_c$. For each parameter, we indicate the range, i.e. the interval of values where this parameter was tested, the relative range (in \%), i.e. the interval length divided by the median value, and the associated relative variations of the two parameters $\tau$ and  $C$ (in \%)}
  \label{tab:sensitivity}
\end{table}
\end{center}

\subsubsection{Calibration of local density-dependent model parameters}
\label{subsubsec:density_dep}

In Figures \ref{fig:tauexp}-\ref{fig:Cexp}, we observe that the distribution of $\tau$ and $C$ depend on the average density of the pedestrians. The value of $\tau$ is decreasing with respect to the average density while that of $C$ is increasing. Both quantities tend to be only mildly varying upon the average density when this density is large. On the other hand, we observe a strong variability of $\tau$ and $C$ among the different samples. This suggests that these quantities could actually depend on the local density at the location of pedestrian $i$ defined as $\rho_i =1/d_{i,i+1}$, with $d_{i,i+1} = R_{\mbox{\scriptsize av}} (\theta_{i+1}-\theta_i)$ and $R_{\mbox{\scriptsize av}}$ being the average radius of the walking trajectory. 

In order to investigate this hypothesis, we use density-dependent fits of $\tau$ and $C$, using the compliant data collected during the complete set of experiments. Here, following the general form of the FTL model (\ref{eq:1.0}), (\ref{eq:2.0}), we use piecewise power laws. More precisely, we use the  following forms of $\tau$ and $C$: 
\begin{equation}
  \tau(\rho)=
  \begin{cases}
   \alpha_1(\frac{\rho}{\rho_\tau})^{\beta_1} & \text{ for }\rho\leq\rho_\tau\\
   \alpha_1(\frac{\rho}{\rho_\tau})^{\beta_2} & \text{ for }\rho>\rho_\tau
  \end{cases} \, \, , \qquad 
 C(\rho)=
  \begin{cases}
   \alpha_2(\frac{\rho}{\rho_C})^{\beta_3} & \text{ for }\rho\leq\rho_C\\
   \alpha_2(\frac{\rho}{\rho_C})^{\beta_4} & \text{ for }\rho>\rho_C
  \end{cases}
	\, \, .
	\label{eq:5}
\end{equation}
The functions of $\tau$ and $C$ are required to be continuous with respect to $\rho$.  This is a nonlinear regression problem since we want to determine the thresholds $\rho_{\tau}$ and $\rho_C$. To estimate $\alpha_{1}$, $\alpha_{2}$, $\beta_{1}$, $\beta_{2}$, $\rho_{\tau}$, $\rho_{C}$, we performed a least squares regression after taking the log. For comparison, we performed a robust regression in $L^1$ norm, which reduces the influence of outliers. We also fitted a single power law for $C$ and $\tau$, i.e.
\begin{gather}
 \tau=\alpha_1\rho^{\beta_1}, C=\alpha_2\rho^{\beta_2},
\label{eq:9}
\end{gather}
using robust regression. We finally estimated constants $\tau$ and $C$ by taking the median values of their distributions.

\begin{figure}[htb]
    \centering
    \subfigure[Horizontal: $\log_{10} \rho$ , vertical: $\log_{10} \tau$]{\includegraphics[scale=0.31]{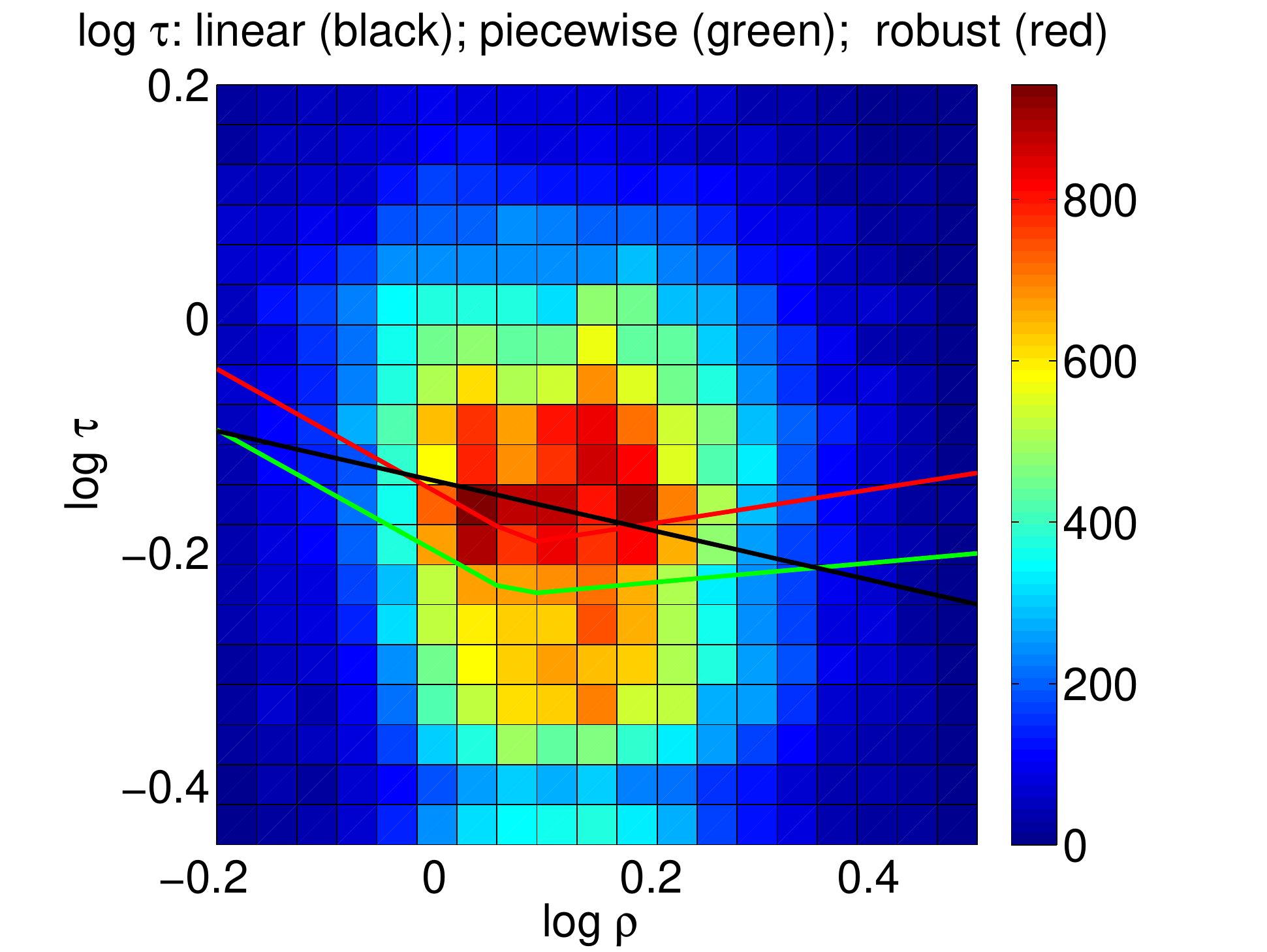}}
    \subfigure[Horizontal: $\log_{10} \rho$ , vertical: $\log_{10}
    C$]{\includegraphics[scale=0.31]{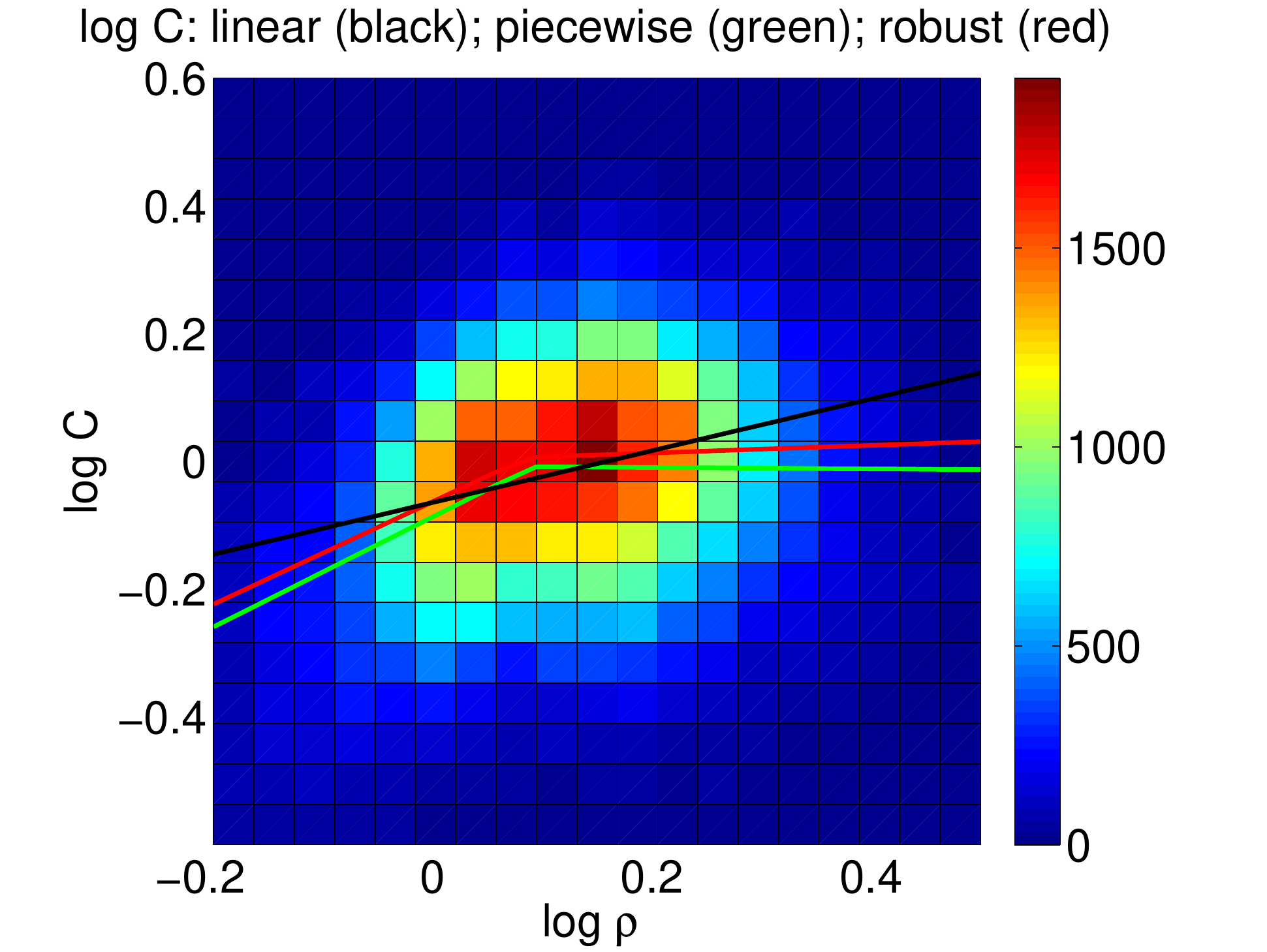}}
    \caption{Two-dimensional color coded histograms of the (density $\rho$, time-delay $\tau$) pairs (left) and (density $\rho$, reaction constant $C$) pairs (right) and associated fitted curves. The fitted curves for piecewise laws \eqref{eq:5} using the standard and robust regression are represented by the green and red color broken lines respectively. The fitted curves for a single power law (\ref{eq:9}) are represented by the black lines. 
}\label{fig:24}
  \end{figure}

The results are presented in Fig.~\ref{fig:24} for the estimation of the time-delay $\tau$ (left) and of the reaction constant $C$ (right). We present two-dimensional color-coded histograms of of the (density $\rho$, time-delay $\tau$) pairs (left) and (density $\rho$, reaction constant $C$) pairs (right) and associated fitted curves. The fitted curves for piecewise laws \eqref{eq:5} using the standard and robust regression are represented by the green and red color broken lines respectively. The fitted curves for a single power law (\ref{eq:9}) are represented by the black lines. 

From Fig.~\ref{fig:24}, it seems that the robust regression to the dual power law  (\ref{eq:5}) provides a better match than the standard regression  to the dual power law or the robust regression to the single power law (\ref{eq:9}). However, the differences between these three calibrations as read from Fig.~\ref{fig:24} is not striking. In order to better assess the quality of each of these calibrations, in the next section, we perform numerical simulations of all these models and compare them with the experimental data. More precisely, we select the three different sets of the parameters for $\tau$ and $C$ given by the robust fit to the dual power law (\ref{eq:5}), the robust linear fit to the single power law (\ref{eq:9}) and the constant values defined by the median of the data. These choices give rise to the following model parameters:

\begin{enumerate}
\item Piecewise power laws (Pm1):
\begin{equation}
       \tau=
  \begin{cases}
     0.712 \, \rho^{-0.522}, & \rho \le 1.22\\
     0.625 \, \rho^{0.145}, & \rho > 1.22
  \end{cases} 
	\, \, , \qquad 
 C=
  \begin{cases}
     0.864 \, \rho^{0.803}, & \rho\le1.22\\
     1.000 \, \rho^{0.06}, & \rho>1.22
  \end{cases}
	\, \, . 
	\label{eq:PM1}
\end{equation}

\item Single power law (Pm2):
\begin{equation}
\tau= 0.726\rho^{-0.212},
  \qquad
 C=0.862\rho^{0.405},
\label{eq:PM2}
\end{equation}

\item Constant parameters (Pm3):
\begin{equation}
\tau= 0.643,\qquad C=1.01.
\label{eq:PM3}
\end{equation}

\end{enumerate}

It follows from the analysis of Section \ref{subsec:stab_ana} that the model with $\alpha=0$ is unstable for the constant values (\ref{eq:PM3}) of the parameters since the stability condition is $2C\tau\le 1$, and this condition is not fulfilled by the constant parameters \eqref{eq:PM3}. The other fits (\ref{eq:PM1}) and (\ref{eq:PM2}) appear numerically unstable. For this reason,  we need to add a relaxation term.  To this effect, we consider $\alpha>0$ and the relaxation model \eqref{eq:generalODE}. The value we chose for the relaxation velocity $\hat{\omega_i}$ is the average of the velocities of the $n'$ pedestrians in front of the considered pedestrian $i$. Our trials with $n'$ equal to 1/4 of the total number of pedestrians showed a reasonnable agreement with the experimental data and this value $n'=1/4$ is the one retained in all the numerical simulations.

\subsection{Comparisons between simulations and experimental data}
\label{subsec:simulations}

In this section we present comparisons of the simulation results with the experimental  data. We consider model \eqref{eq:generalODE} where $C,\tau$ are given by (Pm1), (Pm2) or (Pm3) and $\alpha=0.2$, $0.25$ or $0.3$.  These comparisons are first presented in detail for an experiment (called experiment (A)) corresponding to a large average density. We then briefly present comparisons for two other experiments, one also for a large average density (called experiment (B)) and one for a smaller average density (called experiment (C)) in order to document the versatility of the model and its range of validity.

We first consider experiment (A). This experiment involved $24$ pedestrians walking on the inner circle with an average walking radius of $2.35$ m. Initially, the pedestrians were instructed to form a compact group where each subject (except the leader) was almost in contact with his predecessor. We simulated the model taking as initial values the data extracted from the experiment, i.e. the initial data were taken as the observations on the time interval $[0 \, s, \, 10\, s]$. Indeed, for a delay differential system, we need initial conditions on a whole time interval. The simulations were run on the time interval $[10s,80s]$, which corresponded to the duration of the experiments.

\begin{figure}
\centering
\subfigure[Experimental data]{\includegraphics[scale=0.24,angle=-90]{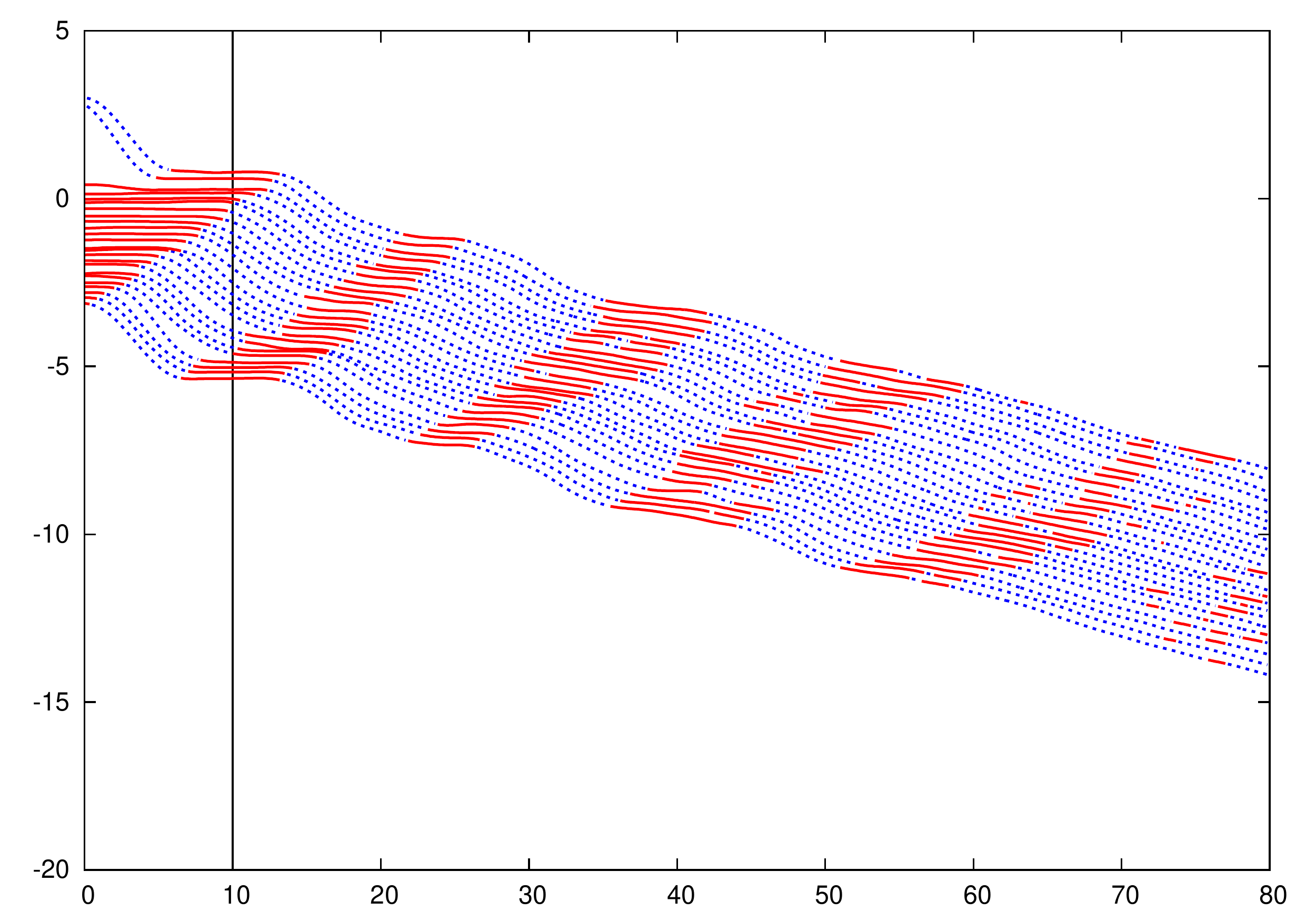}}
\subfigure[Parameters (Pm1) and $\alpha=0.3$]{\includegraphics[scale=0.24,angle=-90]{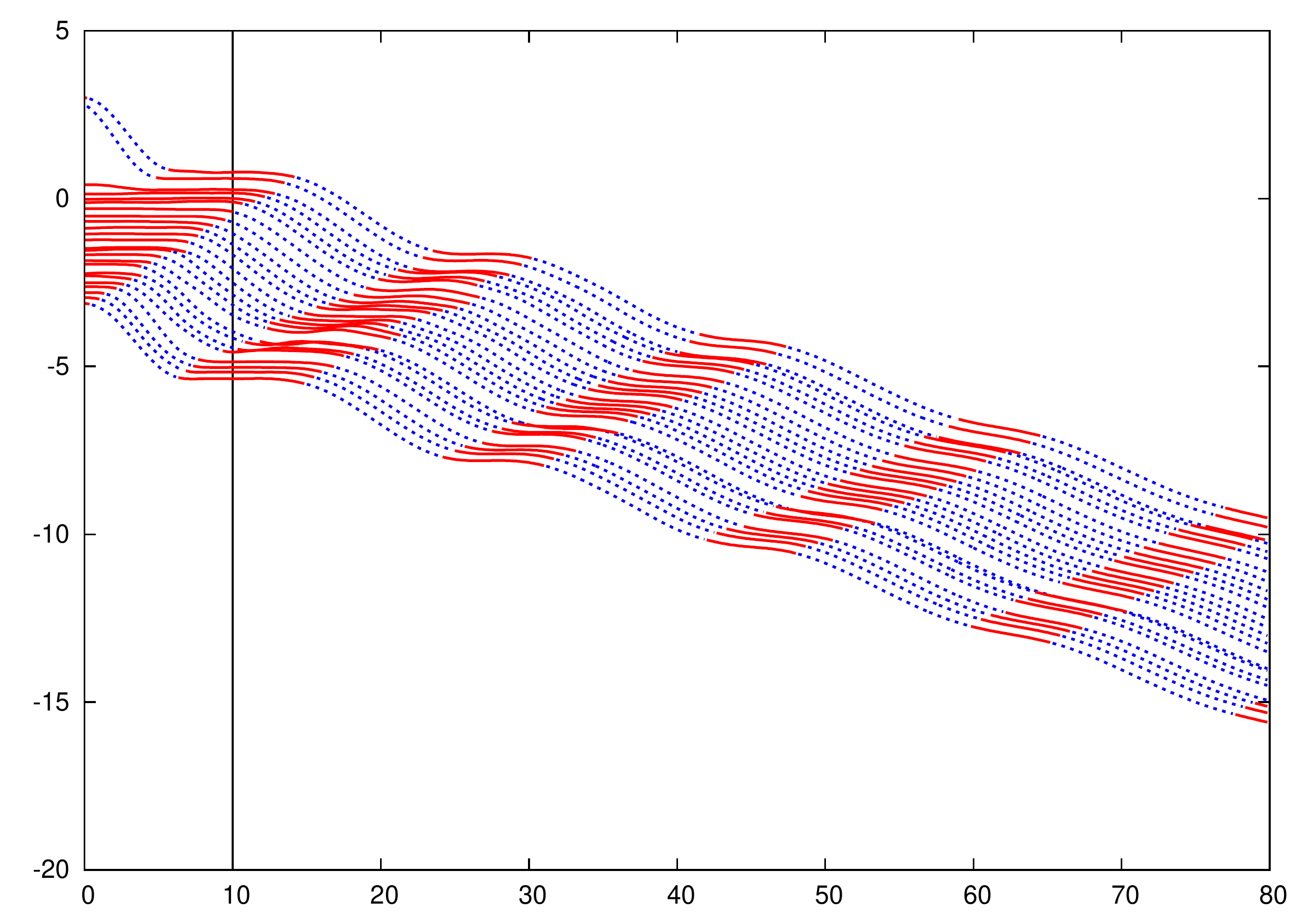}}
 \subfigure[Parameters (Pm2) and $\alpha=0.25$]{\includegraphics[scale=0.24,angle=-90]{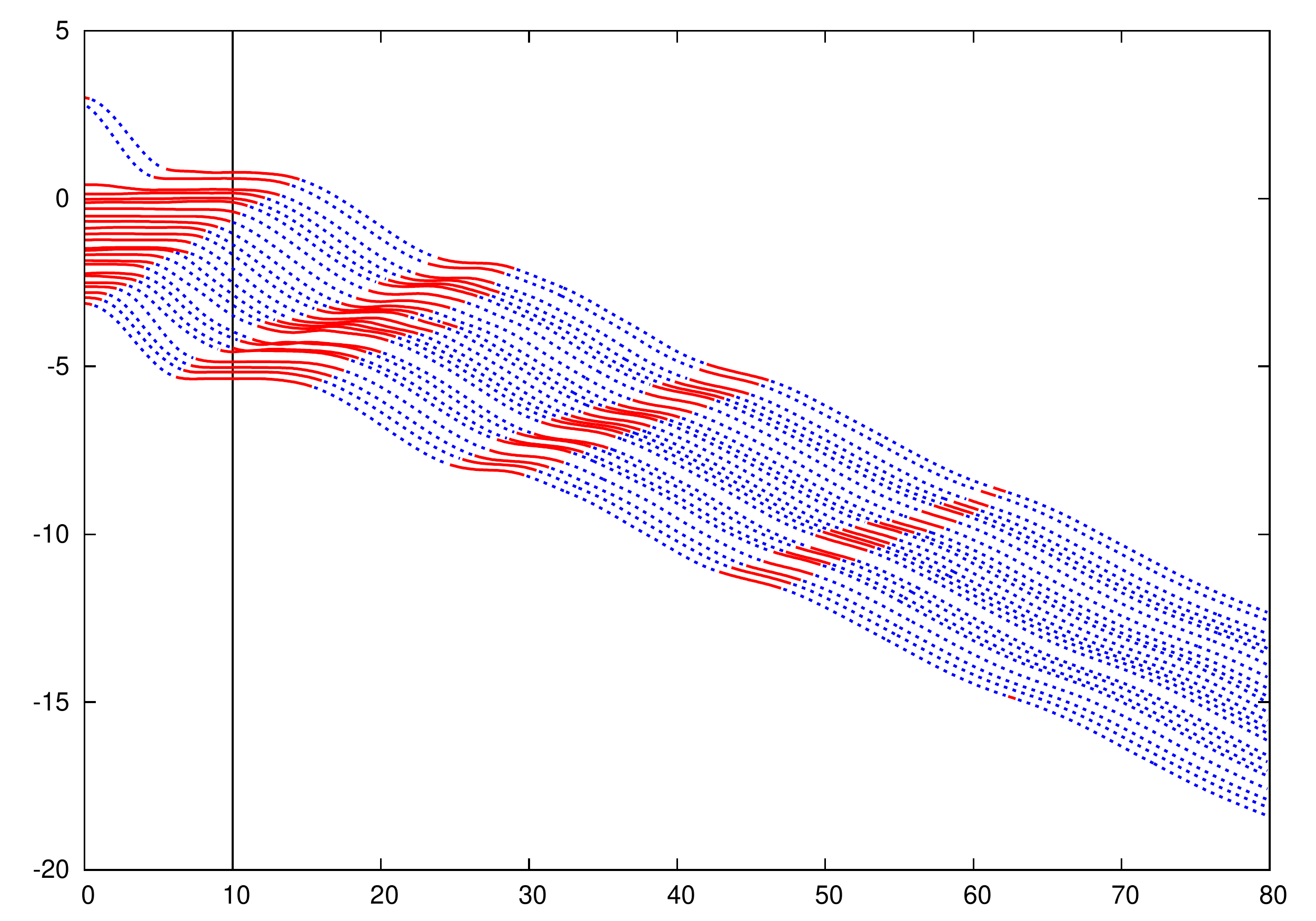}}
\subfigure[Parameters (Pm3) and $\alpha=0.3$]{\includegraphics[scale=0.24,angle=-90]{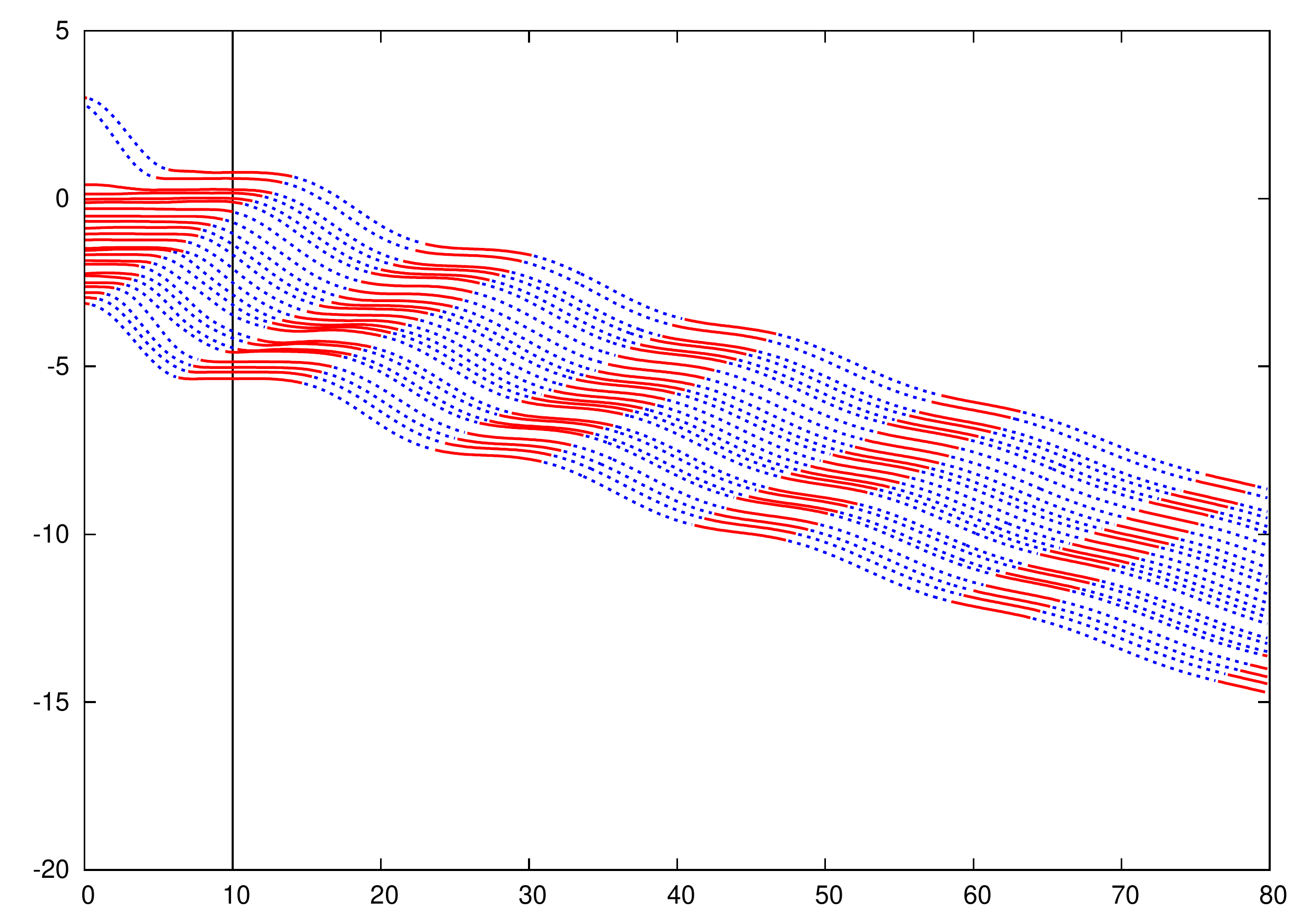}}
\caption{Positions of the pedestrians (in radians, on the circle) as functions of time (in~s) in experiment (A). Experimental data (top left) and models with model parameters (Pm1) (top right), (Pm2) (bottom left) and (Pm3) (bottom right). Portions of a pedestrian trajectory corresponding to a jam are highlighted in red (a jam is defined as a set of pedestrians whose velocity is less than $0.8$ times the average velocity). During the time interval between 0 and 10s (left-hand side of the vertical bar) the delay system is initialized by the experimental data. }
\label{fig:position}
\end{figure}

Fig.  \ref{fig:position}  shows the positions of the pedestrians (in units of radians on their circular path) as a function of time (in s) in experiment (A). Since several laps were performed during one experiment, the position angle is incremented by $2 \pi$ times the number of laps performed at the corresponding time so as to make the trajectories continuous. Each of the pedestrians gives rise to a different trajectory represented by a different curve. The portions of the trajectory of a given pedestrian where he is caught in a jam are highlighted in red, where a jam is defined as a connected set of pedestrians whose velocity is less than $0.8$ times the average velocity. Fig.  \ref{fig:position}  shows the experimental data (top left) and the simulation with model parameters (Pm1) (top right),  (Pm2) (bottom left) and (Pm3) (bottom right).  For each set of parameters we display the results obtained with the value of $\alpha$ that gives the closest similarity between the simulations and the experimental data. Model parameters (Pm1) (Fig. \ref{fig:position} top right) and (Pm3) (Fig. \ref{fig:position} bottom right) seem to give the best match to the experimental data (compare with Fig. \ref{fig:position} (top left)). Indeed, the number of jams is about the same in these two simulations and in the experiments. By contrast, model parameters (Pm2) (Fig. \ref{fig:position} bottom left) seem to lead to an excessive damping of the jams and a too fast convergence towards a state where the pedestrians are equidistant and move with the same velocity.  

\begin{figure}
\centering
 \subfigure[Experimental data]{\includegraphics[scale=0.24,angle=-90]{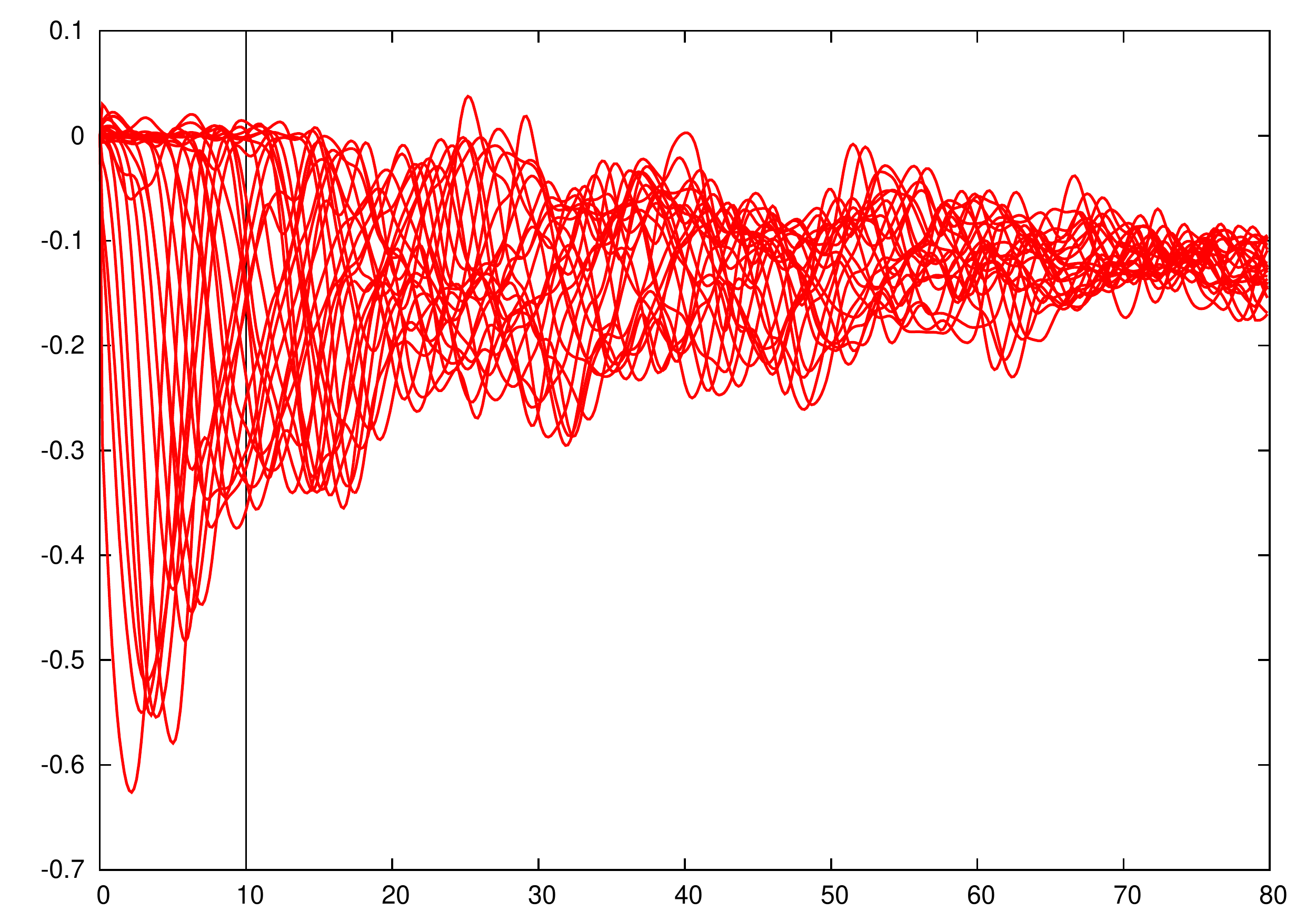}}
\subfigure[Parameters (Pm1) and $\alpha=0.3$]{\includegraphics[scale=0.24,angle=-90]{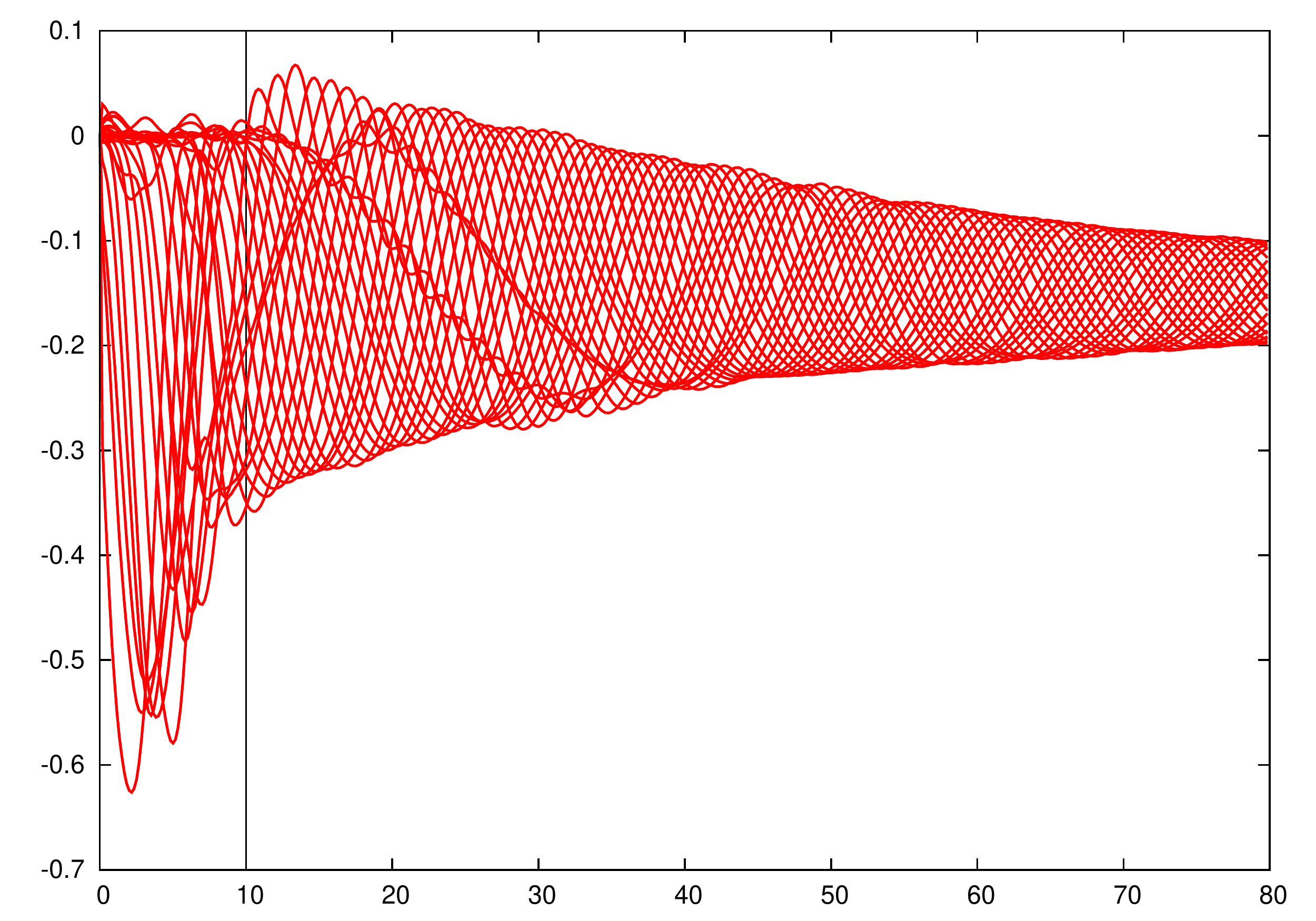}}
\subfigure[Parameters (Pm2) and $\alpha=0.25$]{\includegraphics[scale=0.24,angle=-90]{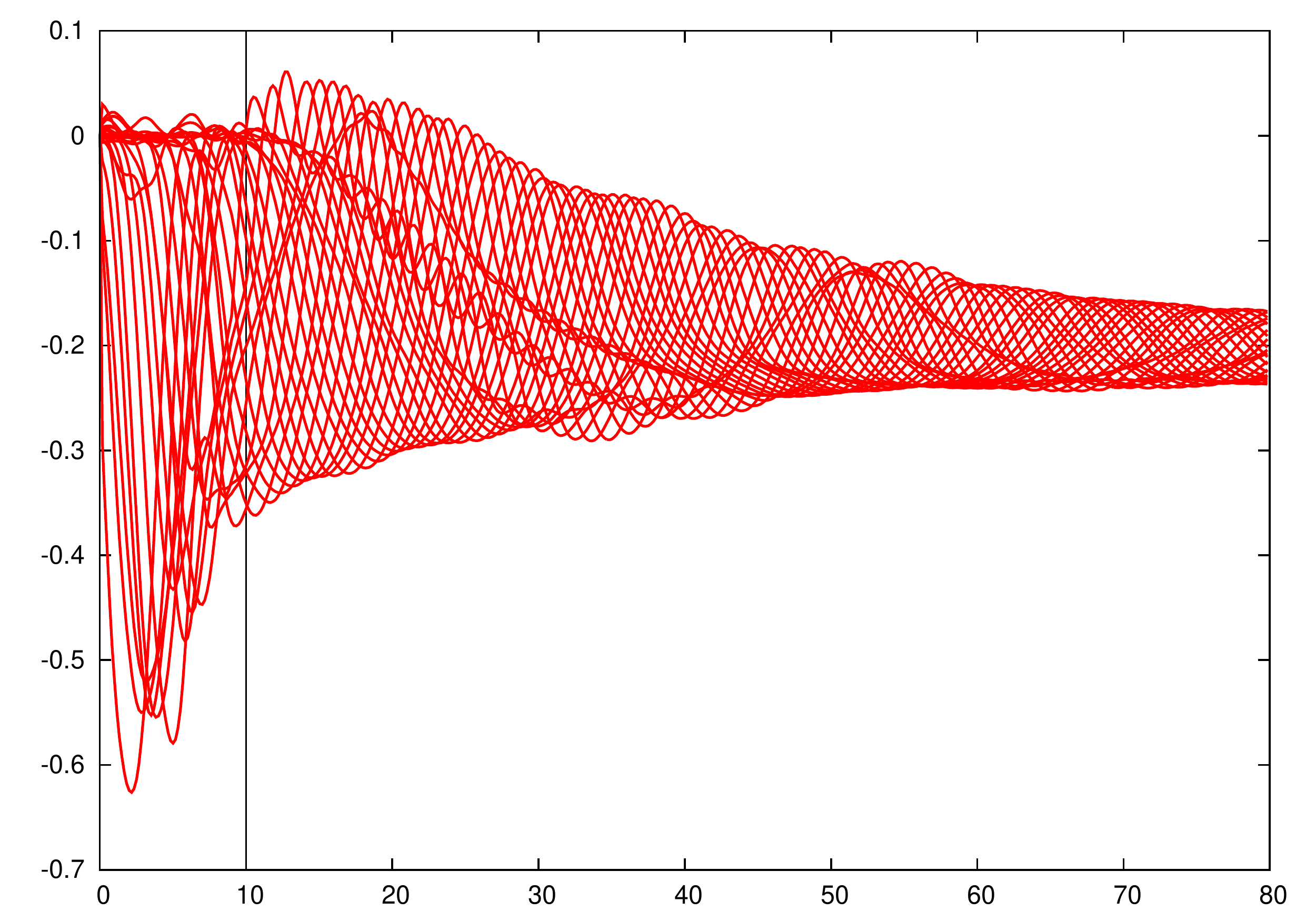}}
\subfigure[Parameters (Pm3) and $\alpha=0.3$]{\includegraphics[scale=0.24,angle=-90]{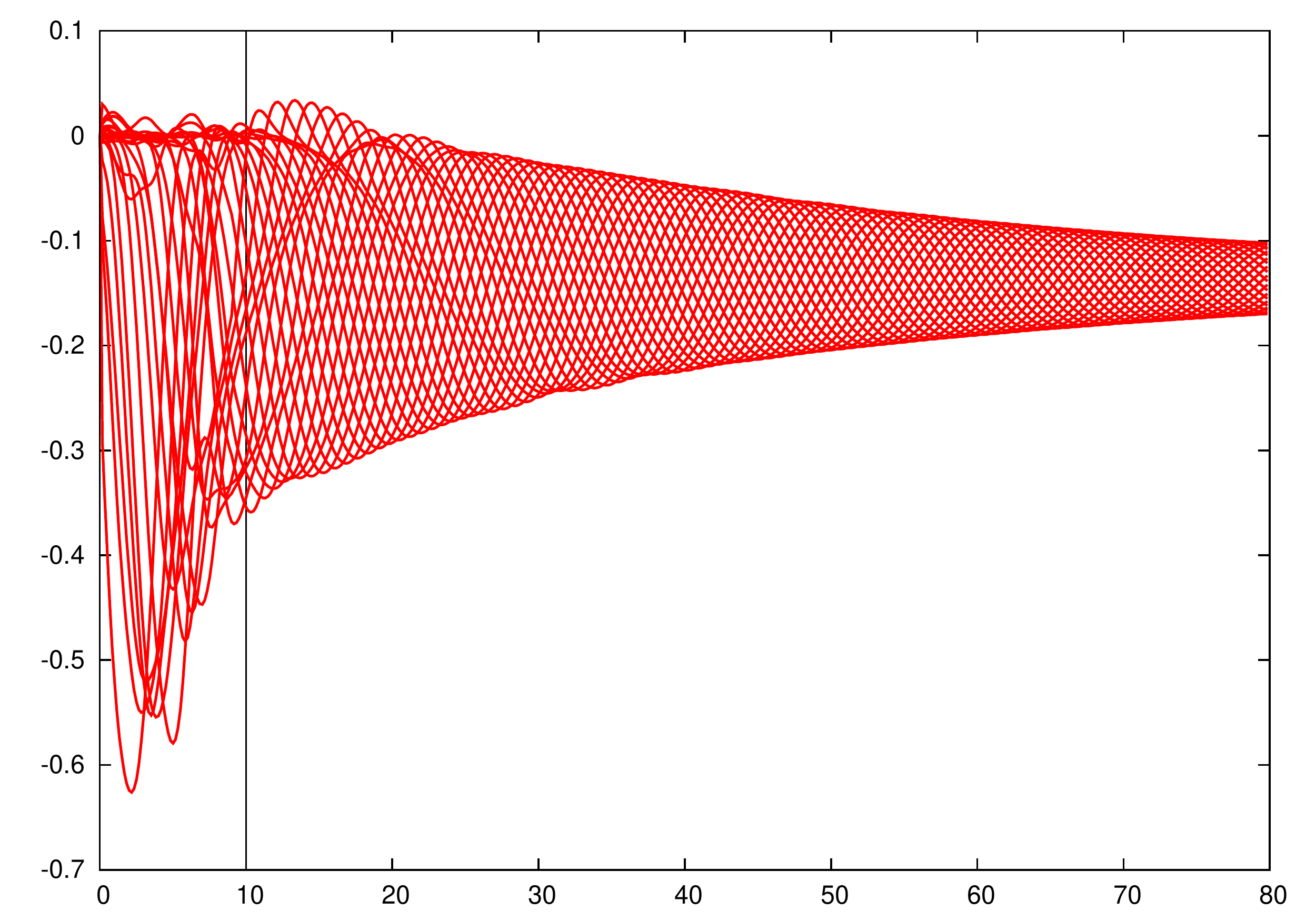}}
 \caption{Velocities of the pedestrians in experiment (A). Experimental data (top left) and  model parameters (Pm1) (top right), (Pm2) (bottom left) and (Pm3) (bottom right). During the time interval between 0 and 10s (left-hand side of the vertical bar) the delay system is initialized by the experimental data. }
\label{fig:velocity}
\end{figure}

Likewise, Fig.  \ref{fig:velocity} displays the velocities of the pedestrians (in ms$^{-1}$) as a function of time (in s) in experiment (A) (Fig.  \ref{fig:velocity} (top left)) and for model parameters (Pm1) (Fig.  \ref{fig:velocity} (top right)),  (Pm2) (Fig.  \ref{fig:velocity} (bottom left)) and (Pm3) (Fig.  \ref{fig:velocity} (bottom right)). These figures confirm that model parameters (Pm1) and (Pm3) give better results that model parameters (Pm2). Indeed, in the latter, pedestrian velocities are too much damped. We observe however, that model parameters (Pm1) and (Pm3) also produce too much velocity damping, but to a much lesser extent than model parameters (Pm2). 

In order to assess the quality of the model, we test its ability to reproduce macroscopic features of the system, such as the dynamics of jams. Indeed, from the definition of a jam as being a connected set of pedestrians whose velocity is less than $0.8$ times the average velocity, we can retrieve descriptors of the jam dynamics such as the velocity of the jam head, the average velocity of the pedestrians in jams or the number of pedestrian in jams. 

\begin{figure}
\centering
\subfigure[Jam head velocity. ]{\includegraphics[scale=0.24,angle=-90]{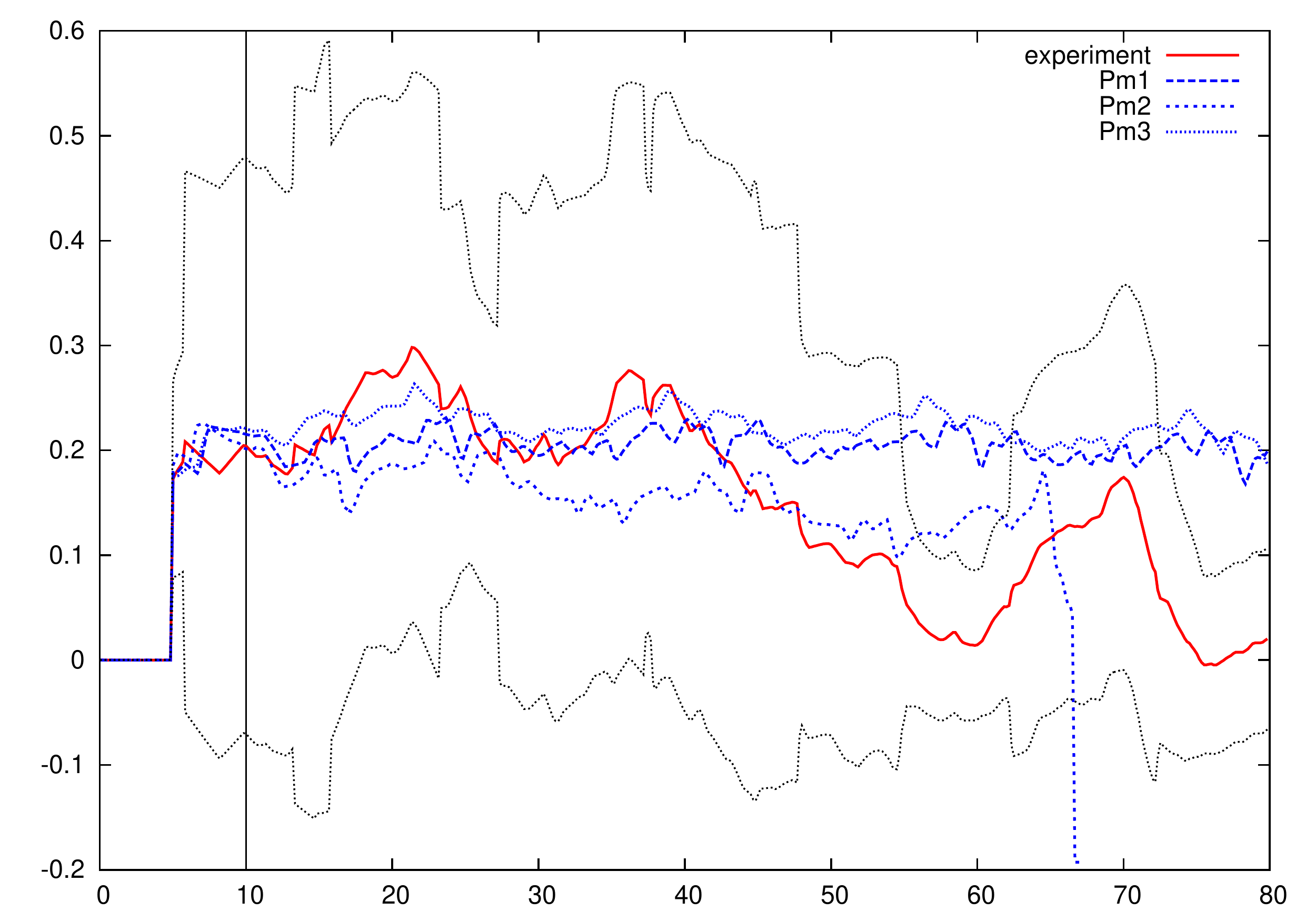}} 
\subfigure[Average velocity in jams. ]{\includegraphics[scale=0.24,angle=-90]{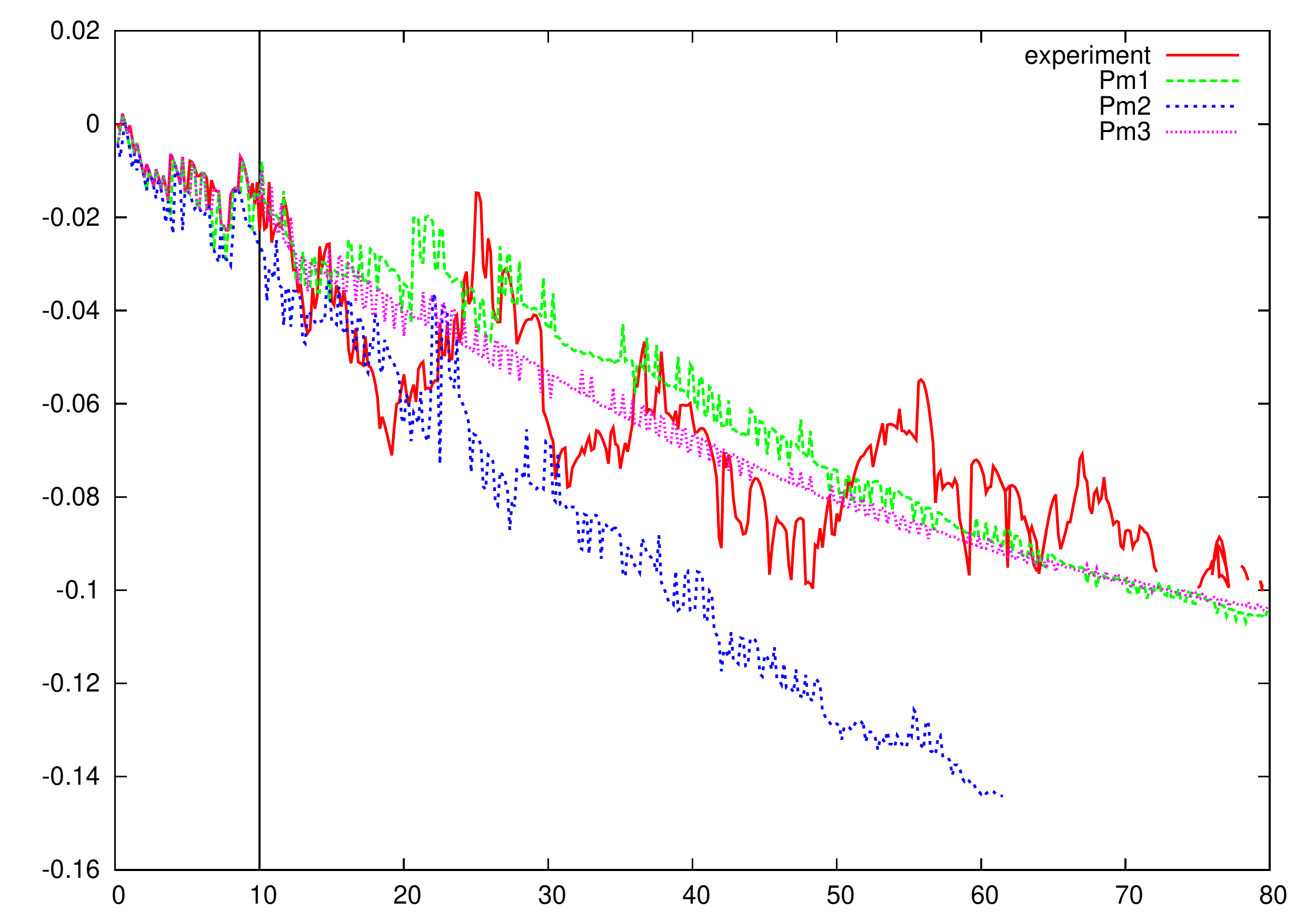} }
 \caption{{\em Left:} Average Jam head velocity in ms$^{-1}$ as a function of time in s in experiment (A). Experimental values are displayed in solid red line and the standard deviation is displayed in black dotted line. Simulations with model parameters (Pm1) to (Pm3) are displayed with dotted blue lines. Thick dots are for (Pm1), medium thick dots for (Pm2) and light dots for (Pm3). The three models provide average jam head velocities that are consistent with the uncertainties of the measurements but there are no jams for (Pm2) after time $70$ s, by contrast with experimental values or models (Pm1) and (Pm3).  
{\em Right:} Average velocity of pedestrians in jams in ms$^{-1}$  as a function of time in s in experiment (A). Experimental values are displayed in thick red line, while simulation results for models (Pm1), (Pm2) and (Pm3) are displayed in dotted green, blue and magenta lines respectively. Experimental values have a greater dispersion than simulated ones but values obtained by models (Pm1) and (Pm3) are within the range of uncertainties of the experimental data. By contrast, values obtained by model (Pm2) are off by almost $100$ \%. 
}
\label{fig:jamvel}
\end{figure}

Fig. \ref{fig:jamvel} (top) displays the average jam head velocity in ms$^{-1}$ as a function of time in~s in experiment (A). Experimental values are shown in solid red line and the standard deviation is shown in black dotted line. Simulations with model parameters (Pm1) to (Pm3) are displayed with dotted blue lines (thick dots for (Pm1), medium thick dots for (Pm2) and light dots for (Pm3)). The three models provide average jam head velocities that are consistent with the uncertainties of the measurements but there are no jams for (Pm2) after time $70$ s, by contrast with experimental values or models (Pm1) and (Pm3).

Fig. \ref{fig:jamvel} (bottom) displays the average velocity of pedestrians in jams in ms$^{-1}$ as a function of time in s in experiment (A). Experimental values are shown in thick red line, while simulation results for models (Pm1), (Pm2) and (Pm3) are shown with dotted green, blue and magenta lines respectively.  Experimental values for the average velocity in jams have a greater dispersion than simulated ones. On the other hand, values obtained by models (Pm1) and (Pm3) are within the range of uncertainties of the experimental data. By contrast, model (Pm2) is off by almost $100$ \%. 

\begin{figure}
\centering
\includegraphics[scale=0.4]{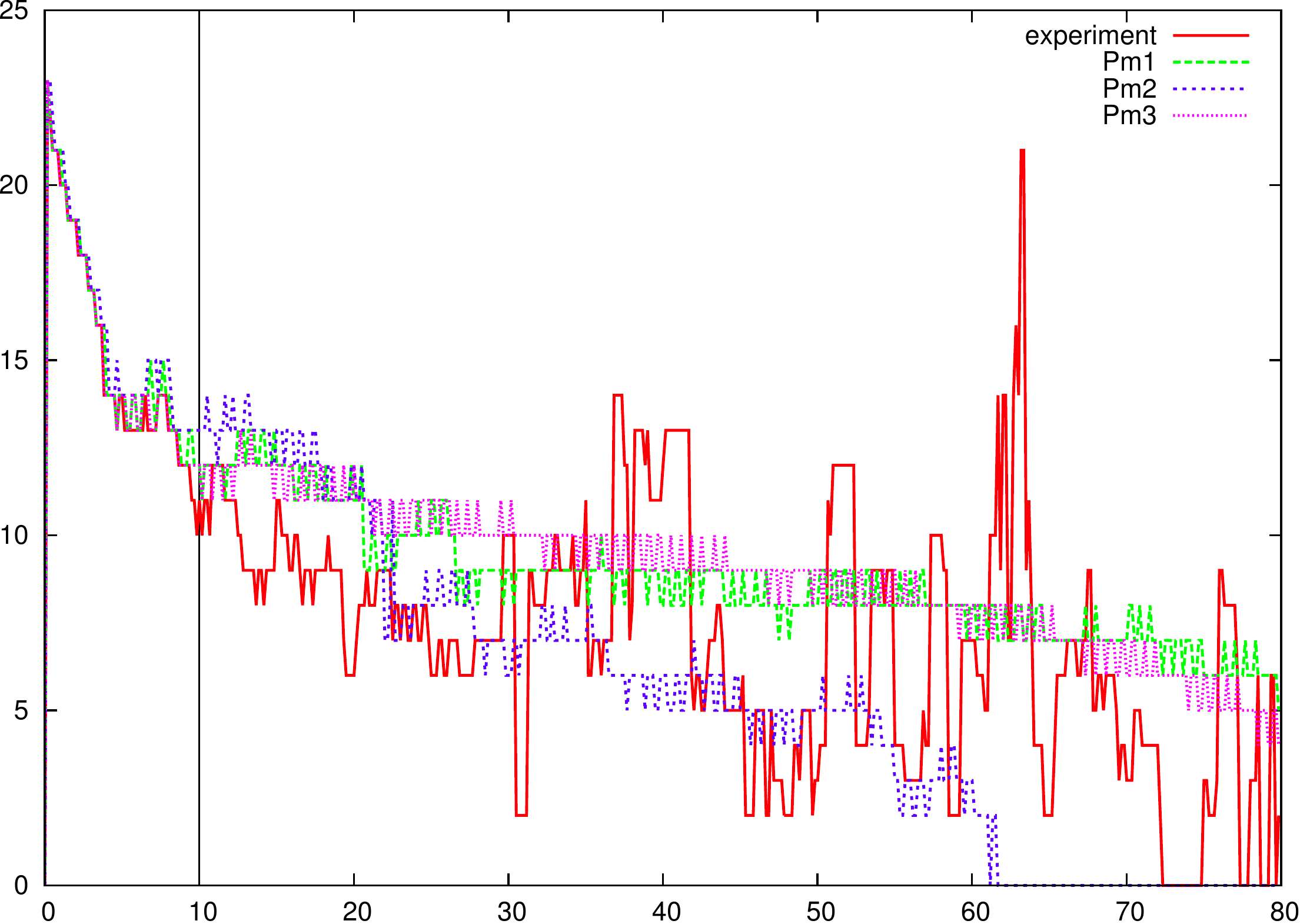}
\caption{Number of pedestrians in jams as a function of time in s in experiment (A). Experimental values are displayed with the thick red line, while simulation results for models (Pm1), (Pm2) and (Pm3) are displayed with dotted green, blue and magenta lines respectively. Again, the values provided by models (Pm1) and (Pm3) are within the uncertainties of the measurements, with a slight tendency to overestimate this number. By contrast, model (Pm2) has too few pedestrians in jams.  }
\label{fig:number} 
\end{figure}

Fig. \ref{fig:number} shows the number of pedestrians in jams as a function of time in s in experiment (A). Experimental values are shown in thick red line, while simulation results for models (Pm1), (Pm2) and (Pm3) are shown with dotted green, blue and magenta lines respectively. We observe that models (Pm1) and (Pm3) provide values within the uncertainties of the measurements. Measurements have more variability and the two models tend to slightly  overestimate the average number of pedestrian in jams. By contrast, model (Pm2) confirms the tendency to overdamp jams and displays fewer pedestrians in jams than experimental measurements. 

\begin{figure}
\centering
\subfigure[Experimental data.]{\includegraphics[height=0.32\textwidth,angle=-90]{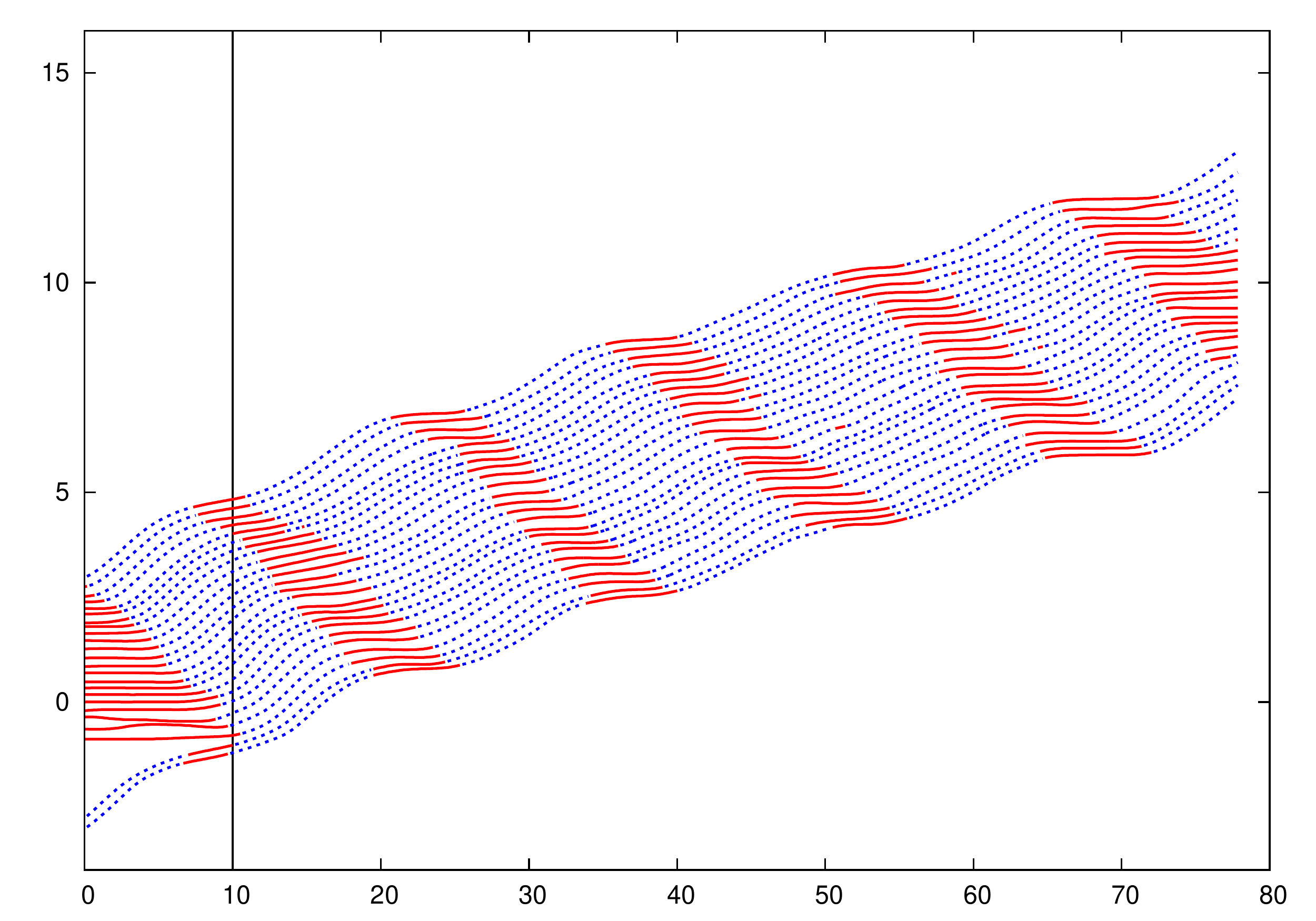}}
\subfigure[Parameters (Pm1).]{\includegraphics[height=0.32\textwidth,angle=-90]{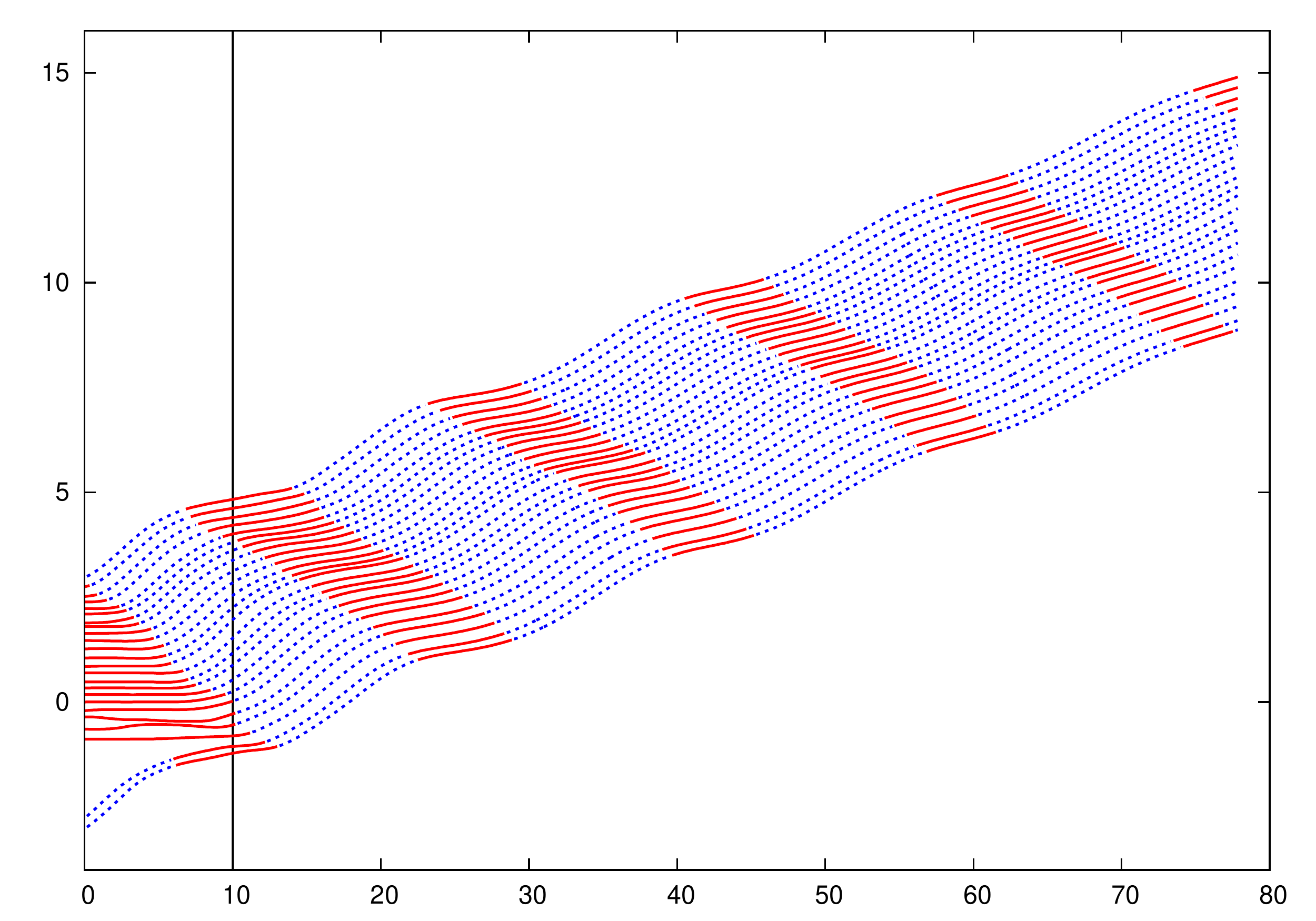} }
\subfigure[Parameters (Pm3).]{\includegraphics[height=0.32\textwidth,angle=-90]{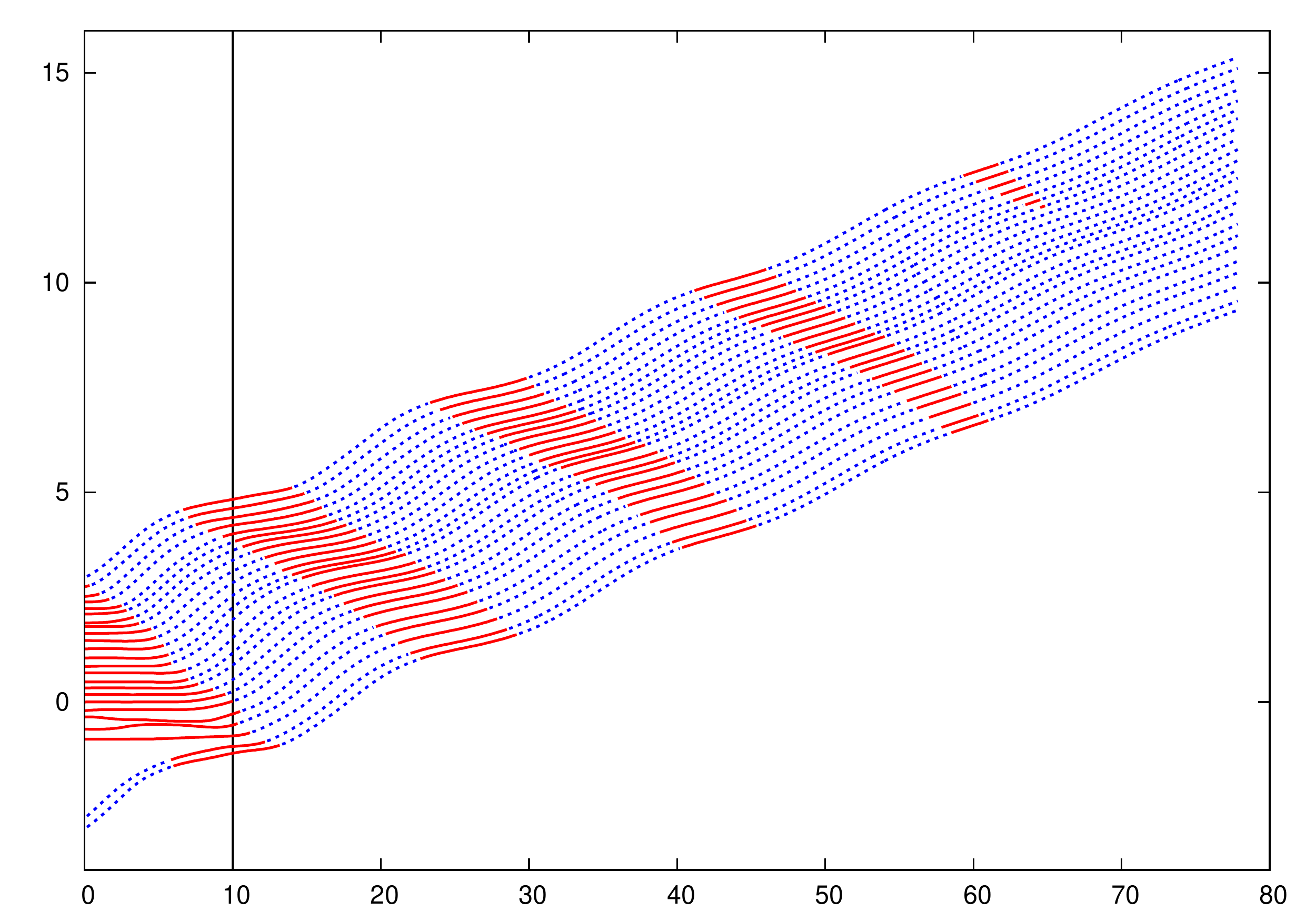} }
 \caption{Positions of the pedestrians (in radians, on the circle) as functions of time (in~s) in experiment (B). Experimental data (left), model (Pm1) (middle) and model (Pm3) (right). Portions of a pedestrian trajectory corresponding to a jam are highlighted in red. During the time interval between 0 and 10s (left-hand side of the vertical bar) the delay system is initialized by the experimental data.  Parameter $\alpha$ was set to $0.3$ in both (Pm1) and (Pm3). Here model (Pm1) provides a better match than model (Pm3). }
\label{fig:exp25}
\end{figure}

We now briefly present comparisons between experimental data and simulations of models (Pm1) and (Pm3) (we discard model (Pm2) on the basis of the poor matches it provided to the previous experiment) in two other experiments. The first one, called experiment (B) is, like the previous one, a large density experiment, with $24$ pedestrians walking along the inner circle with average walking radius of $2.50$ m and initially arranged in a compact group. Fig. \ref{fig:exp25} provides the positions of the pedestrians (in radians, on the circle) as functions of time (in s). Experimental data are shown in the left figure, while models (Pm1) and (Pm3) are shown in the middle and right figures respectively. Parameter $\alpha$ was set to $0.3$ in both (Pm1) and (Pm3). Here model (Pm1) provides a better match than model (Pm3). Model (Pm3) exhibits an excessive damping of the jams, with jams disappearing after $70$ s of time. By contrast, jams are maintained with model (Pm1) but they seem to travel more quickly than in the experimental data. This is correlated with the observation that the pedestrians seem also to travel faster than in the experiment. In the experiment, pedestrians seem to slow down after a time of about $10$-$20$~s. This could explain the observed discrepancy with the numerical simulation since the initialization of the delay differential system is set by using the data over the first interval of $10$ s duration. This change of pace of the pedestrians could be due to the transition from a compact group to an equally spaced group. This tends to indicate that the model is not precise enough for the very large densities that prevail in the compact group. It is not very surprising since very few experimental data were available in that density range, and so, the calibration does simply not take it into account.  

\begin{figure}
\centering
\subfigure[Experimental data.]{\includegraphics[height=0.32\textwidth,angle=-90]{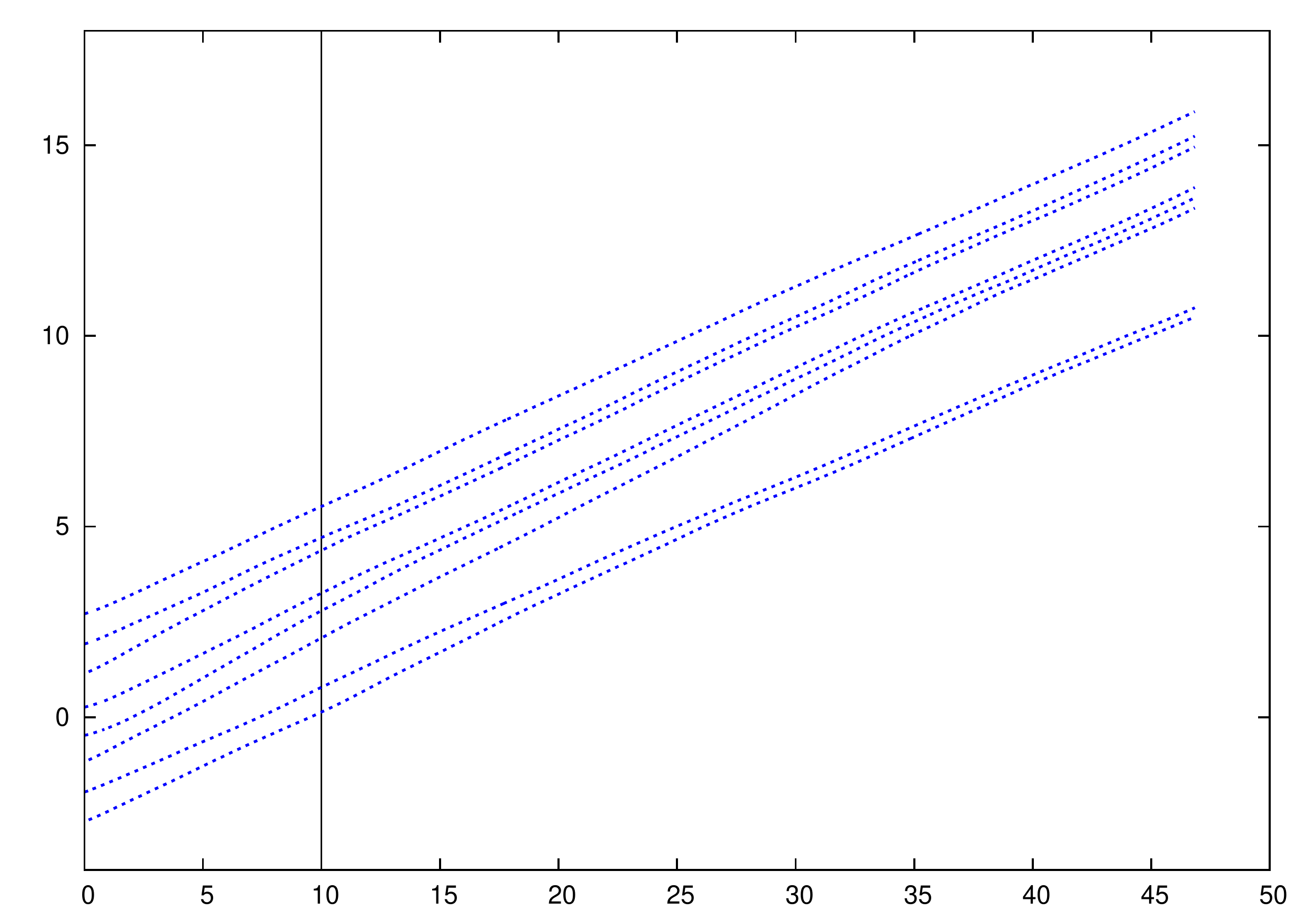}}
\subfigure[Parameters (Pm1).]{\includegraphics[height=0.32\textwidth,angle=-90]{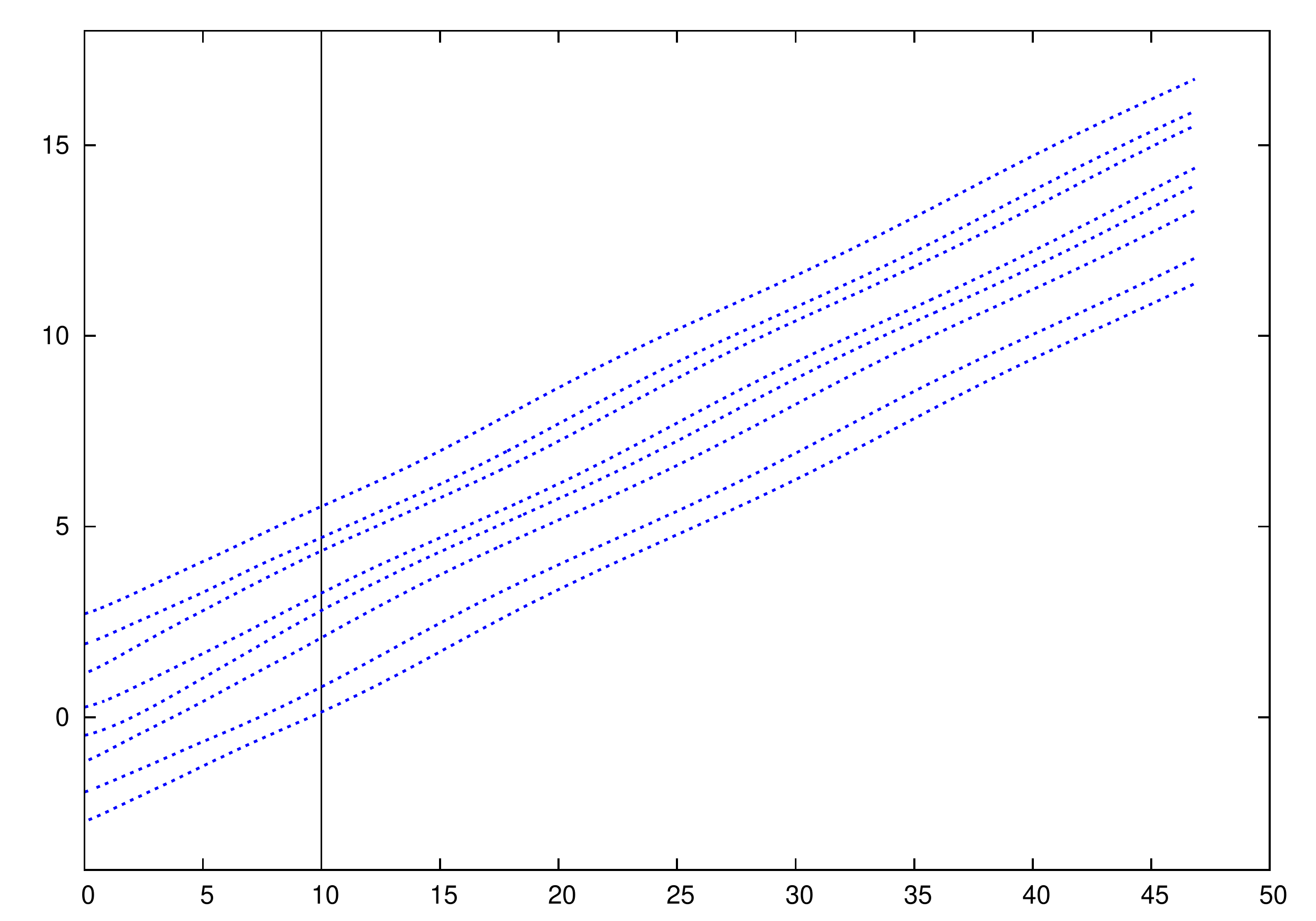} }
\subfigure[Parameters (Pm3).]{\includegraphics[height=0.32\textwidth,angle=-90]{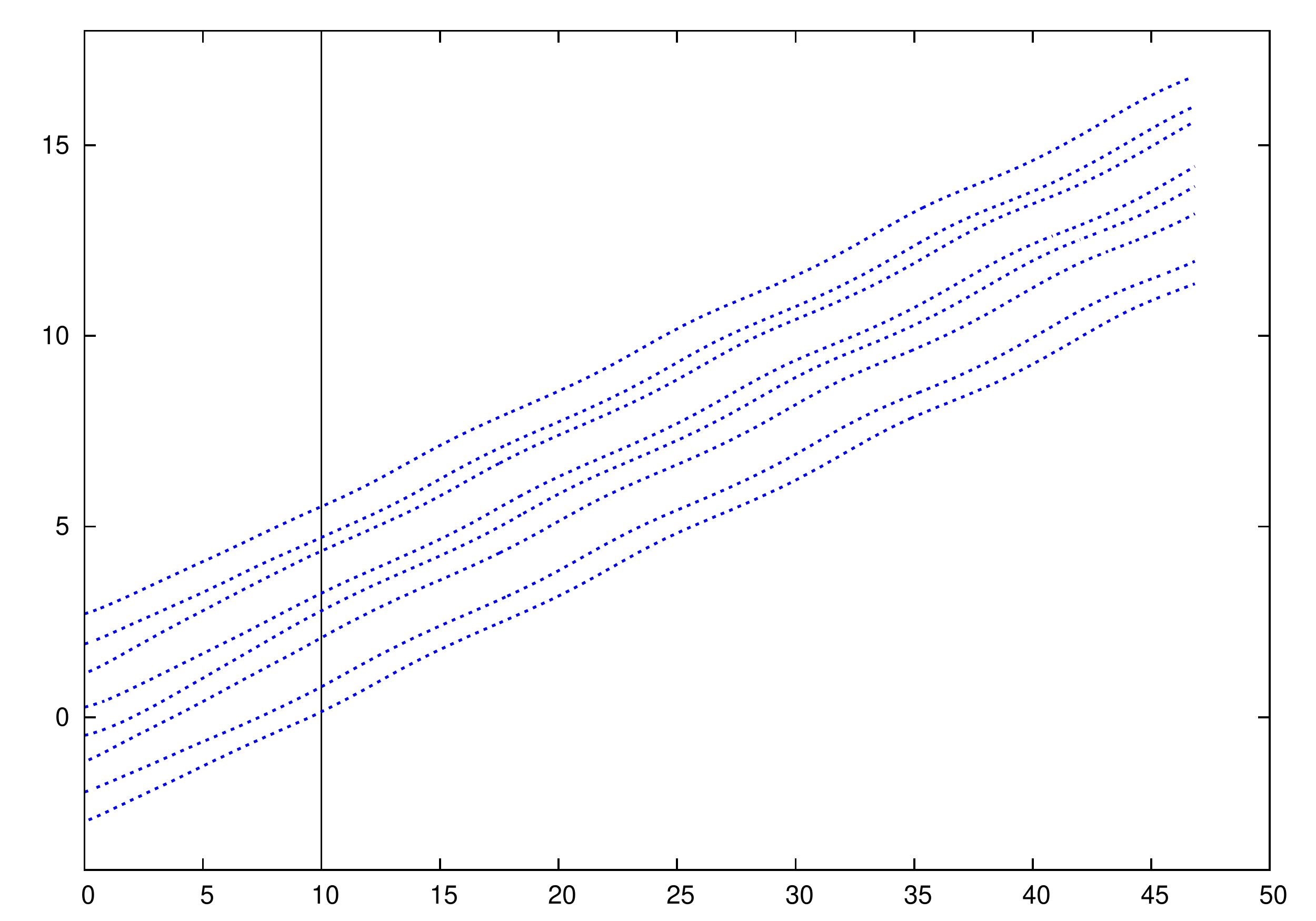} }
 \caption{Positions of the pedestrians (in radians, on the circle) as functions of time (in~s) in experiment (C). Experimental data (left), model (Pm1) (middle) and model (Pm3) (right). During the time interval between 0 and 10s (left-hand side of the vertical bar) the delay system is initialized by the experimental data. Parameter $\alpha$ was set to $0.3$ in both (Pm1) and (Pm3). Here both models lead to slightly too fast pedestrian velocities. }
\label{fig:exp51}
\end{figure}

The final experiment, called experiment (C) is a small density experiment, with $8$ pedestrians walking along the outer circle with an average walking radius of $4.15$ m and initially arranged in an equidistant manner. Fig. \ref{fig:exp51} provides the positions of the pedestrians (in radians, on the circle) as functions of time (in s). Experimental data are shown in the left figure, while models (Pm1) and (Pm3) are shown in the middle and right figures respectively. Parameter $\alpha$ was set to $0.3$ in both (Pm1) and (Pm3). Here, the density is too low and no jam forms, so we cannot use jams to assess the quality of the model. But we easily notice  that simulations provide too fast pedestrian velocities compared to the experiments. Also, some clustering of the pedestrians emerges (though not affecting their velocity and thus, not qualifying as jams according to our previous definition). Indeed, there are essentially four clusters at the end of the experiments, with respectively $1$, $2$, $3$, and $2$ pedestrians separated by bigger gaps. This clustering is midly reproduced in the two models (a little bit more accurately by model (Pm3)) but does not look as sharp as in the experiments. Therefore, different observables than those defined above should be set up for assessing the validity of the models in the low density case. 

\setcounter{equation}{0}
\section{Discussion}
\label{sec:discussion}

The present study of pedestrians walking in line has demonstrated the existence of a significant time-delay, i.e.  that the acceleration of a pedestrian at a given time is determined by the relative position and velocity of this pedestrian with respect to his predecessor at an earlier time. This time delay is about  $0.6$ s. Given that finding, we have developed three different time-delayed Follow-the-Leader models. The first model (Pm3) considers that both the time delay $\tau$ and the reaction constant $C$ (i.e. the acceleration intensity) are constant. The second model (Pm2) considers that $\tau$ and $C$ are power laws of the local density. Finally, the third model (Pm1) supposes that there are two density regimes and that the power law dependences of $\tau$ and $C$ are different at low and high local densities, with a larger power at small density than at large density. 

In all cases, the models needed to be stabilized by the addition of a relaxation of each pedestrian's velocity to the average velocity of a certain number of his predecessors. In our simulation, the number of predecessors was equal to $25$ \% of the total number of pedestrians involved in the experiment and about $30$ \% of the reaction of a given pedestrian was triggered by this relaxation and $70$ \% by the leader following behavior itself. 

Our finding is that the model that matches the experimental results with the highest degree of accuracy is model (Pm1), i.e. the model where the power law dependences of $\tau$ and $C$ are different at low and high local densities. This model was carefully assessed by investigating the ability of the simulation to reproduce large-scale dynamical features of the experimental data such as jam formation and dynamics. 

According to this model, the leader-following behavior of the pedestrians is different at low and high local densities. Refering to (\ref{eq:PM1}), we notice that $\tau$ decreases like $1/\sqrt{\rho}$ as long as the density is lower than a crossover value of about $1.2$ ped m$^{-1}$ and then stays approximately constant. A possible interpretation of this behavior is that pedestrians become increasingly aware of their leader's behavior when their distance to him decreases. But once this distance exceeds a certain value, there is no further decay of this time-delay as information processing and decision-making take an incompressible amount of time. In a similar way, the reaction constant increases almost linearly with the density at low density but saturates to a constant once the crossover value of the density if reached. 

It is instructive to remember that the worst model has been shown to be model (Pm2), where $\tau$ and $C$ are given by a single power law throughout the whole range of values of the local density. Even if, in that case, the decay of $\tau$ and the increase of $C$ at small density is less pronounced that in model (Pm1), it seems that keeping the same law above the critical density leads to a significant detoriation of the result. Therefore, it seems essential to take into account the saturation of the time delay and response intensity at large densities. This is confirmed by the fact that model (Pm3) which keeps $\tau$ and $C$ constant and independent of $\rho$ performs better than (Pm2), as if the request that $\tau$ and $C$ should be constant at large density overrid the necessity of making them $\rho$-dependent at low density. To some extent, the model with constant $\tau$ and $C$ is the simplest, and offers a very attractive cost-benefit ratio for large-scale simulations. In \cite{Lemercier_etal_CompGraphForum12}, it has been shown to compare favorably to other models in the literature such as \cite{Helbing_Molnar_PRE95, Reynolds_ProcGameDev99}. 

To assess the model, we compared the simulated jam dynamics with the experimentally observed one. This assessment methodology is restricted to the large average density case. Indeed, in the small average density case, no jam is formed. In this case, different assessment methods need to be developed. During calibration, we observed that, in the low density case, there were significantly less samples which were compliant with the model than in the large density case. This seems to indicate that the Follow-the-Leader model alone is unable to correctly account for the low-density observations. New theoretical models need to be developped in that case. 

This model, either in the form (Pm3), or in the form (Pm1), can be used as a building block for two-dimensional models. Indeed, it can account for the speed adjustments of the pedestrians due to the presence of other pedestrians walking in front of them in the same direction. It can be complemented by a model describing how the direction of motion is changed to account for the presence of obstacles or other pedestrians moving in the opposite direction, in the spirit of \cite{Degond_etal_JSP13, Degond_etal_KRM13, Moussaid_etal_PNAS11}. Such an approach has been already outlined in \cite{Lemercier_etal_CompGraphForum12} and will be pursued in the future.

\setcounter{equation}{0}
\section*{Appendix}
\label{sec:appendix_B}

The pre-processing of the experimental data depends on several parameters, namely the cutoff frequency $f_c$, the window width $w_w$, the model parameter $\gamma$ and the correlation threshold $\epsilon_t$ below which a sample (pedestrian, time window) is discarded. We study the influence of each of these parameters on the retrieved time delay $\tau$ and constant $C$. The data are gathered by local density and the influence of each parameter is studied by letting the other ones fixed.

\subsection*{Influence of the parameter $f_c$}
\label{subsec:parameterfc}

\begin{figure}[htb]
\centering
  \subfigure{\includegraphics[scale=0.31]{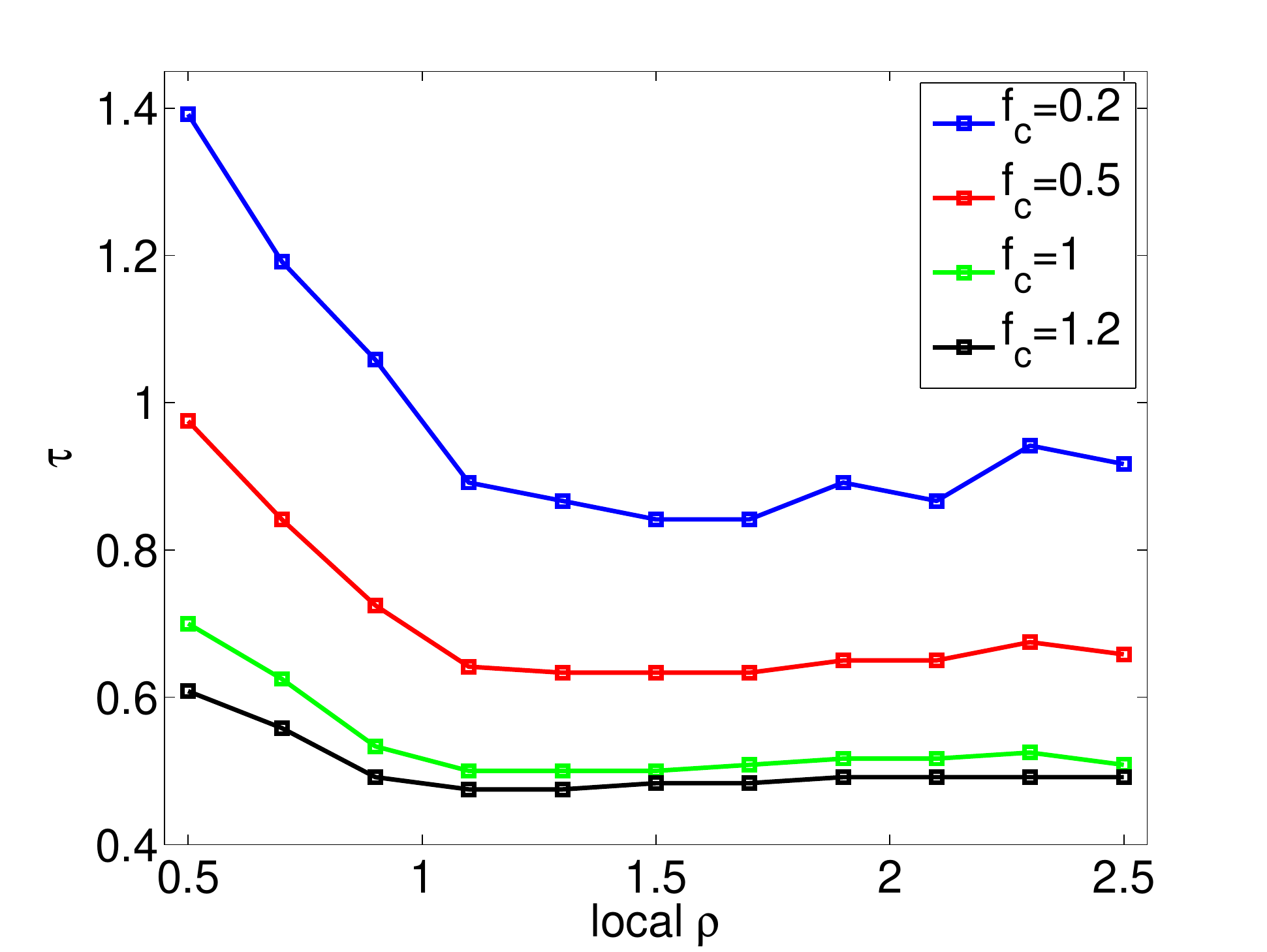}}
  \subfigure{\includegraphics[scale=0.31]{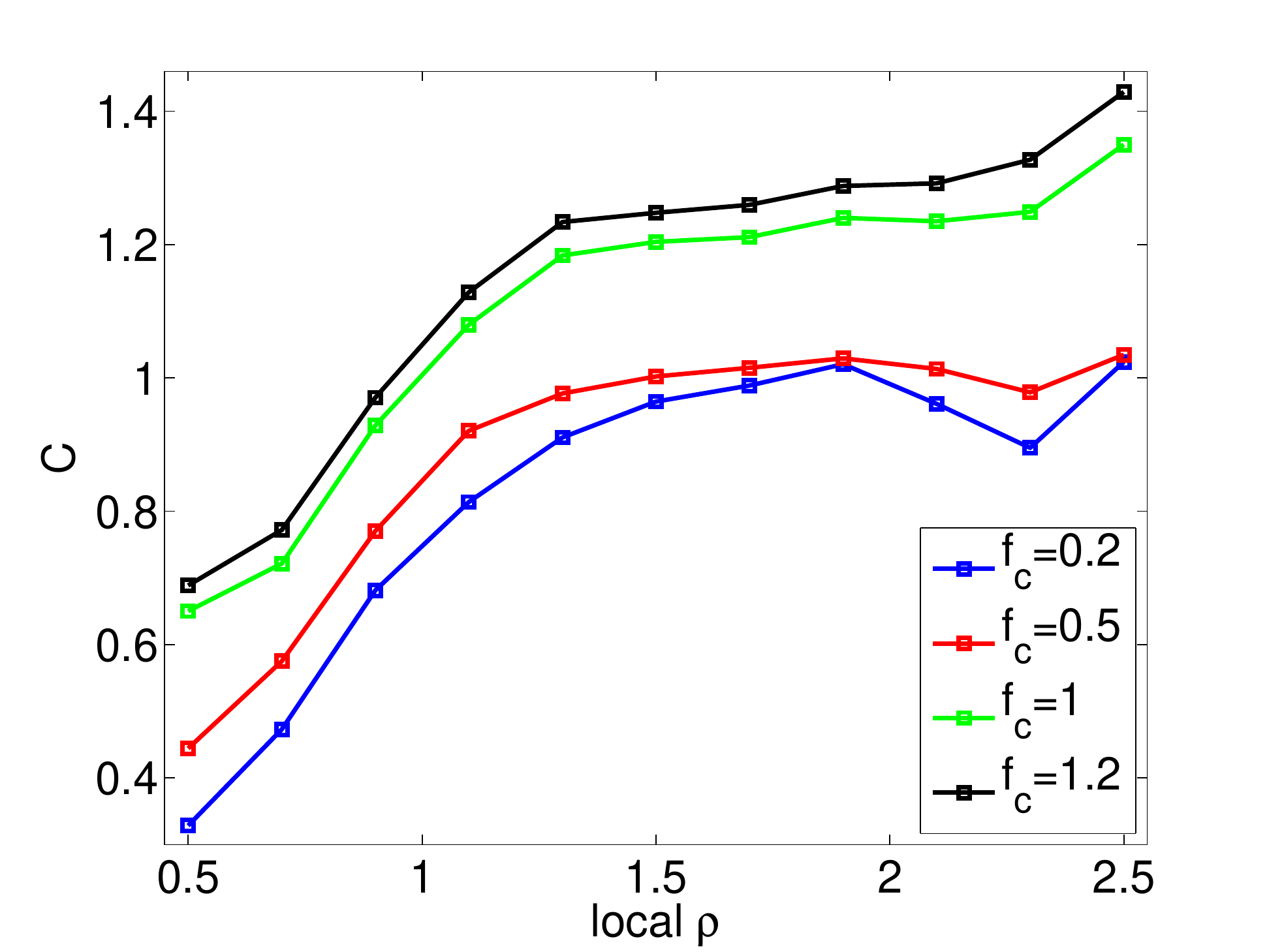}}
\caption{Median of estimated delay $\tau$ in s (left) and constant $C$ in m$^{1/2}$s$^{-1}$ (right) over the set of samples consisting of (pedestrian, time window) pairs ranging through the whole set of experiments as functions of the local density $\rho$ in ped m$^{-1}$, for different values of the cutoff frequency~$f_c$. The curves were obtained with $f_c\in\{0.2, \, 0.5, \, 1, \, 1.2\}$ Hz and are respectively displayed in blue, red, green and black, while the other parameters  $\epsilon_t=0.6$, $w_w=6.67$s, $\gamma=-1$ are fixed.}\label{fig:35}
\end{figure}

To test the effect of the parameter $f_c$, we process the data of all experiments with the following values: $\epsilon_t=0.6$, $w_w=6.67$ s, $\gamma=-1$. The cutoff frequency $f_c$ is given the following  values $f_c\in\{0.2, \, 0.5, \, 1, \, 1.2\}$ Hz. The medians of the estimated values of $\tau$ and $C$ over the set of samples consisting of (pedestrian, time window) pairs ranging through the whole set of experiments  are presented in Fig. \ref{fig:35} as functions of the local density $\rho$ (see section \ref{subsubsec:density_dep} for the definition of the local density).

From Fig. \ref{fig:35}, we notice that as $f_c$ increases, the median of $\tau$ decreases  while the median of $C$ increases. The medians of $\tau$ and $C$ are only slightly influenced by $f_c$ when $f_c \geq 1$ Hz and in this range of cut-off frequencies, the dependence of these medians with respect to the local density $\rho$ is mild. By contrast, the medians of $\tau$ and $C$ depend much more strongly on $f_c$ for $f_c \leq 1$ Hz and their dependence on the local density if stiffer. We have also noticed (not illustrated by a figure) that the percentage of compliant data increases as $f_c$ decreases until reaching the value $f_c=0.2$ Hz. In this last case, the time-delay found from the calibration is too large for the interval in which it is searched for and the percentage of compliant data then drops dramatically. Given these observations, we choose a cut-off frequency $f_c = 0.5$ Hz as it roughly corresponds to the stepping frequency of the pedestrians. This choice allows us to retain all phenomena occuring at a frequency larger that $0.5$ Hz.

\subsection*{Influence of the window width $w_w$}
\label{subsec:influ_wd}

\begin{figure}[htb]
\centering
\subfigure{\includegraphics[scale=0.31]{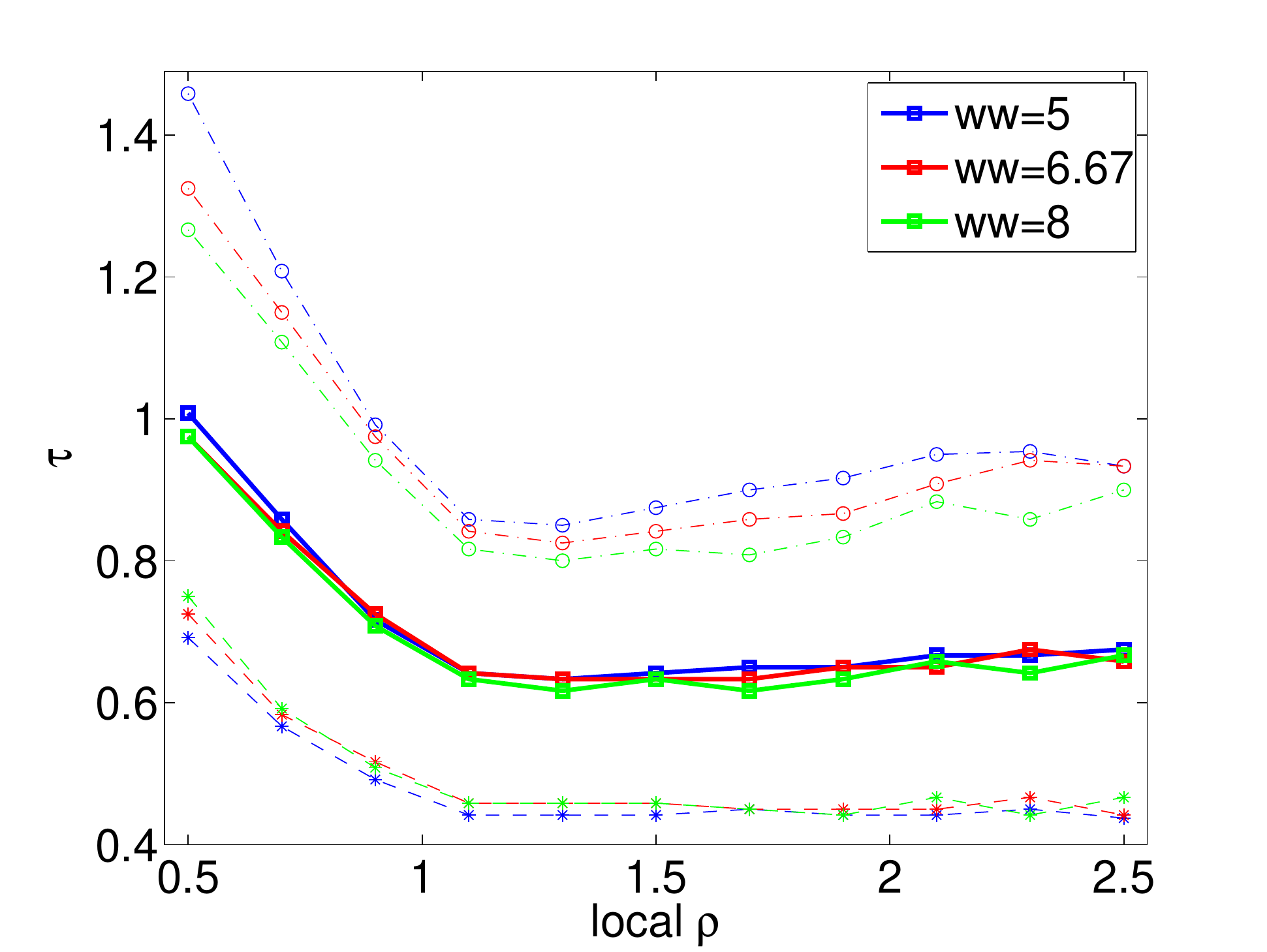}}
\subfigure{\includegraphics[scale=0.31]{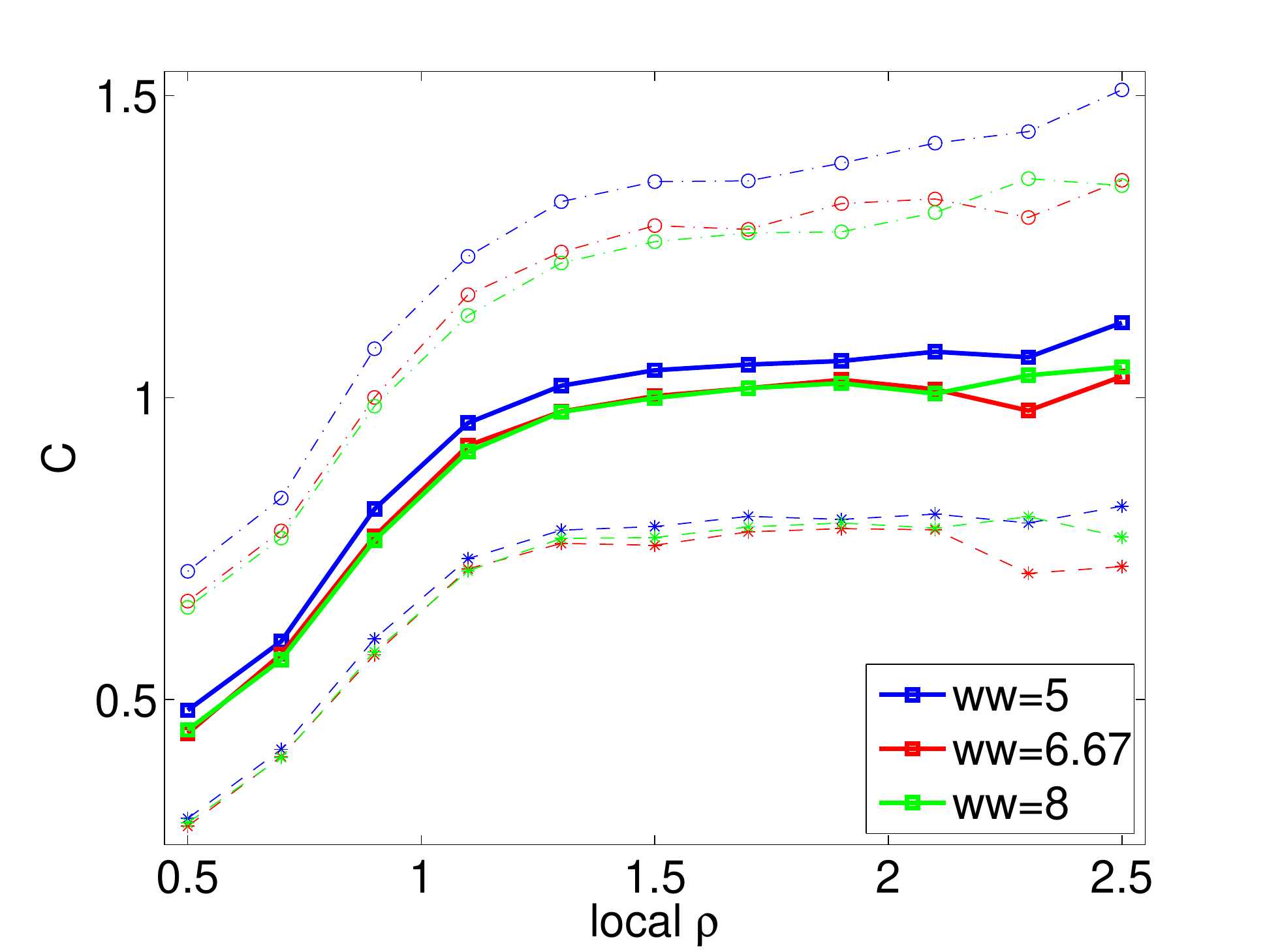}}
\caption{Median and quartiles of estimated delay $\tau$ in s (left) and constant $C$ in m$^{1/2}$s$^{-1}$ (right) over the set of samples consisting of (pedestrian, time window) pairs ranging through the whole set of experiments as functions of the local density $\rho$ in ped m$^{-1}$,  for different values of the window width $w_w$. The different curves were obtained with $w_w=5$ s, $6.67$ s, $8$ s and are respectively displayed in blue, red and green while the other parameters $\epsilon_t=0.6$, $f_c=0.5$ Hz, $\gamma=-1$ are fixed. Medians are displayed in solid lines while quartiles are shown in dotted lines of the corresponding color. }\label{fig:33}
 \end{figure}

To test the effect of the parameter $w_w$, we process the data of all experiments with
$\epsilon_t=0.6$, $f_c=0.5$ Hz,  $\gamma=-1$ and $w_w$ is given the following values: $w_w\in\{5, \, 6.67, \, 8\}$~s.  The median and quartiles of the estimated $\tau$ and $C$ over the set of samples consisting of (pedestrian, time window) pairs ranging through the whole set of experiments  are presented in Fig. \ref{fig:33} (left and right respectively) as functions of the local density $\rho$ (see section \ref{subsubsec:density_dep} for the definition of the local density).

The estimated values of of $\tau$, $C$ are only very midly dependent of the window width. The variations of $\tau$ as a function of $w_w$ are of the order of $2.5$ \%, and the variations of $C$ are of the order of $5.3$ \%.  The distributions of $\tau$ and $C$ are slightly more concentrated as the window length $w_w$ becomes larger (as shown by the interval between two quartiles becoming narrower) but this effect is really small. These findings show that the dependence of the results on the window width is negligible.

\subsection*{Influence of the model parameter $\gamma$}
\label{subsec:influ_gamma}

\begin{figure}[htb]
\centering
 \subfigure{\includegraphics[scale=0.31]{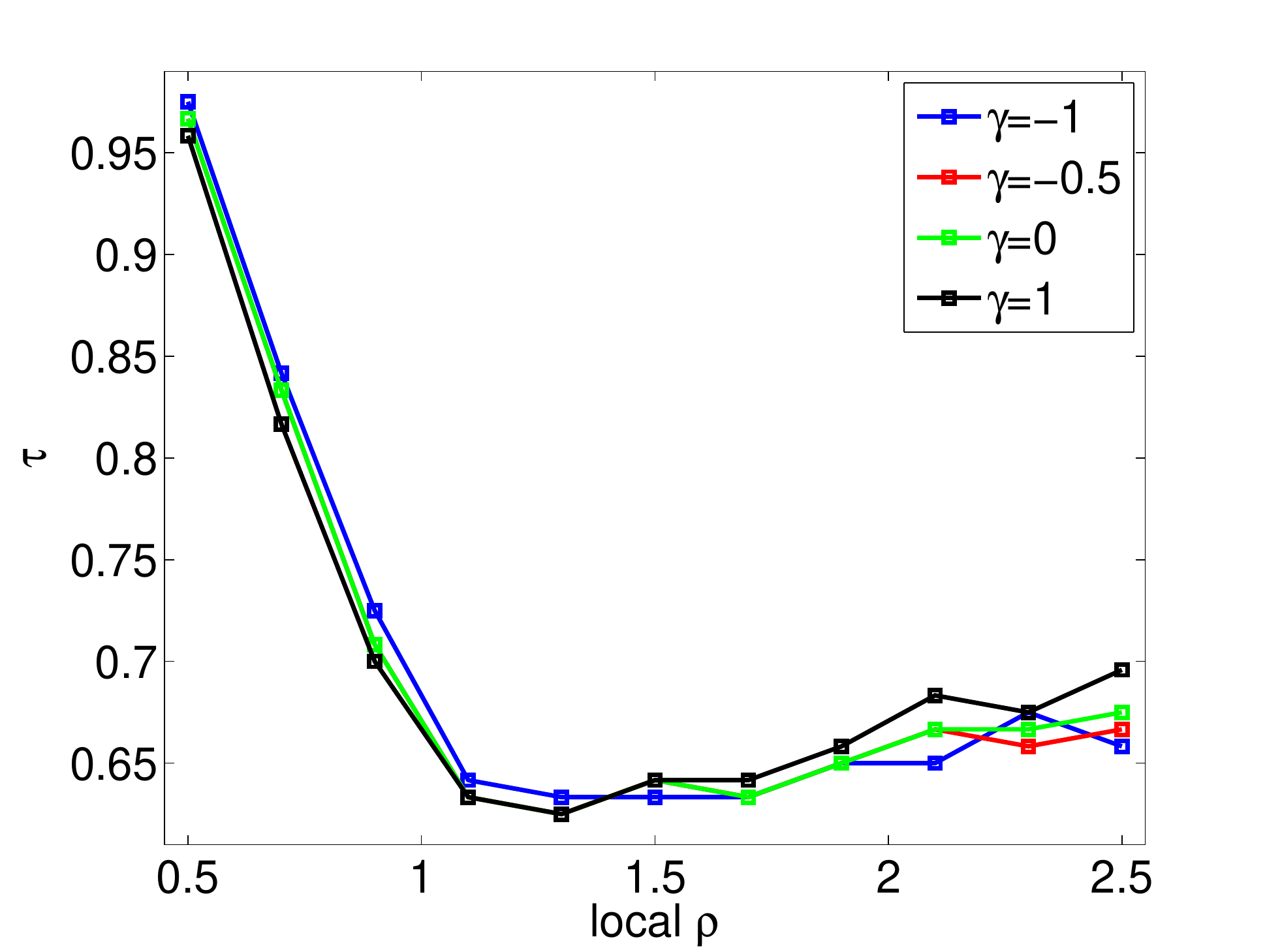}}
\caption{Median of estimated delay $\tau$ over the set of samples consisting of (pedestrian, time window) pairs ranging through the whole set of experiments  for different values of the parameter $\gamma$. The different curves are obtained with the values $\gamma=-1,\, -0.5,\, 0, \, 1$ and represented in blue, red, green and black respectively, while the other parameters $\epsilon_t=0.6$, $w_w=6.67$ s, $f_c=0.5$ Hz are fixed.}\label{fig:38}
\end{figure}

To test the effect of the model parameter $\gamma$, we process the data of  all experiments with $\epsilon_t=0.6$, $w_w=6.67$ s, $f_c=0.5$ Hz, and $\gamma$ is given the following values: $\gamma\in\{-1, \,-0.5,\, 0,\, 1\}$. The median of the estimated delay $\tau$ over the set of samples consisting of (pedestrian, time window) pairs ranging through the whole set of experiments  are presented in Fig. \ref{fig:38} as a function of the local density $\rho$ (see section \ref{subsubsec:density_dep} for the definition of the local density).

From Fig. \ref{fig:38}, it can be observed that there is no significant dependence of the values of the median of the time delay $\tau$ on the model parameter $\gamma$.  Indeed, the variations of the median of $\tau$ are of the order $2.6$ \%, which is almost negligible. The values of $C$ are actually of different physical dimensions for different values of $\gamma$, which makes their simple comparison not meaningful. Since all cases seem to perform well, we keep the value $\gamma=-1$ when we deal with model (Pm3) (i.e. when we deal with local-density independent values of $\tau$ and $C$).

\subsection*{Influence of the correlation threshold $\epsilon_t$}
\label{subsec:influ_threshold}

\begin{figure}[htb]
\centering
 \subfigure{\includegraphics[scale=0.31]{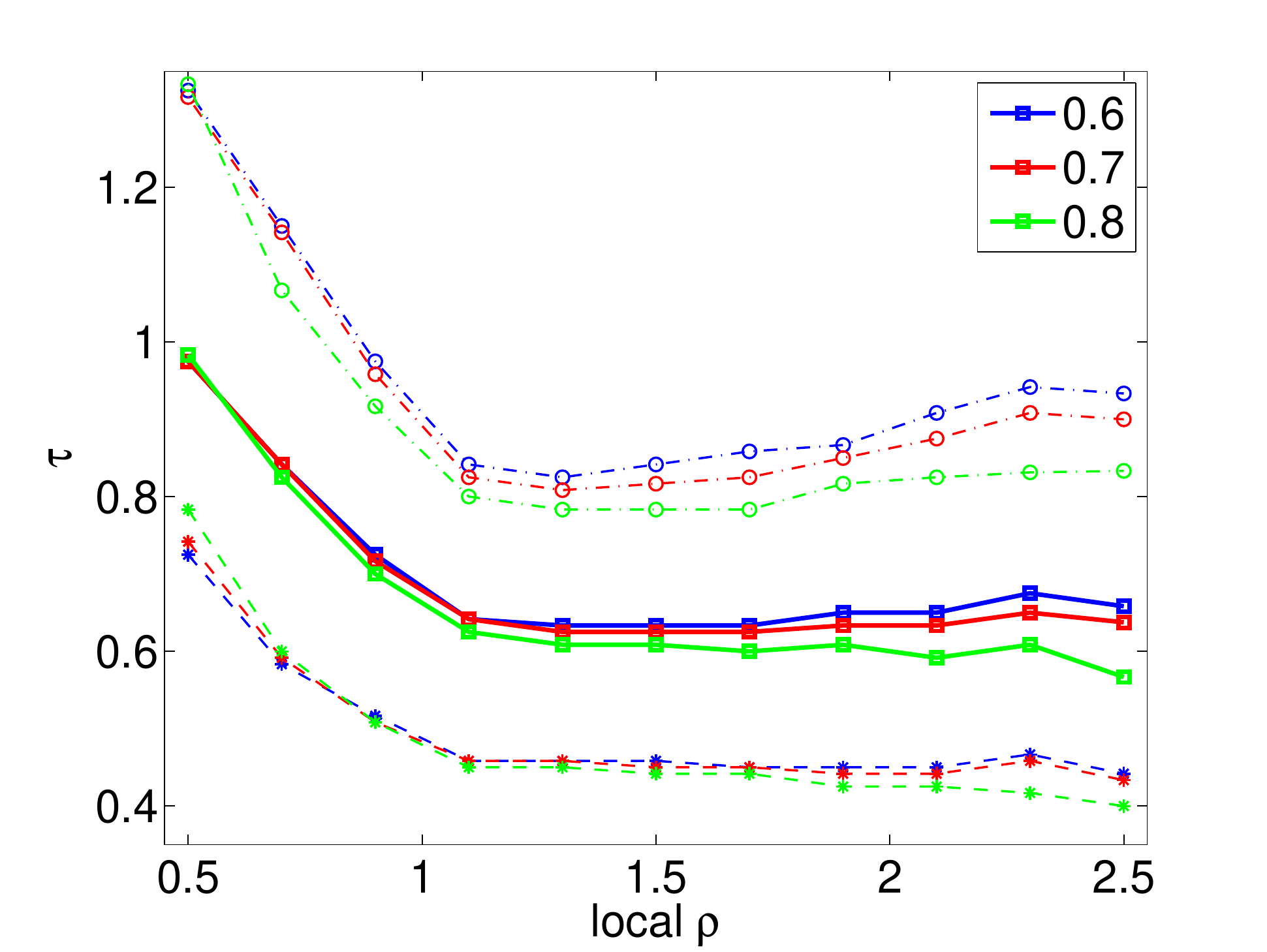}}
  \subfigure{\includegraphics[scale=0.31]{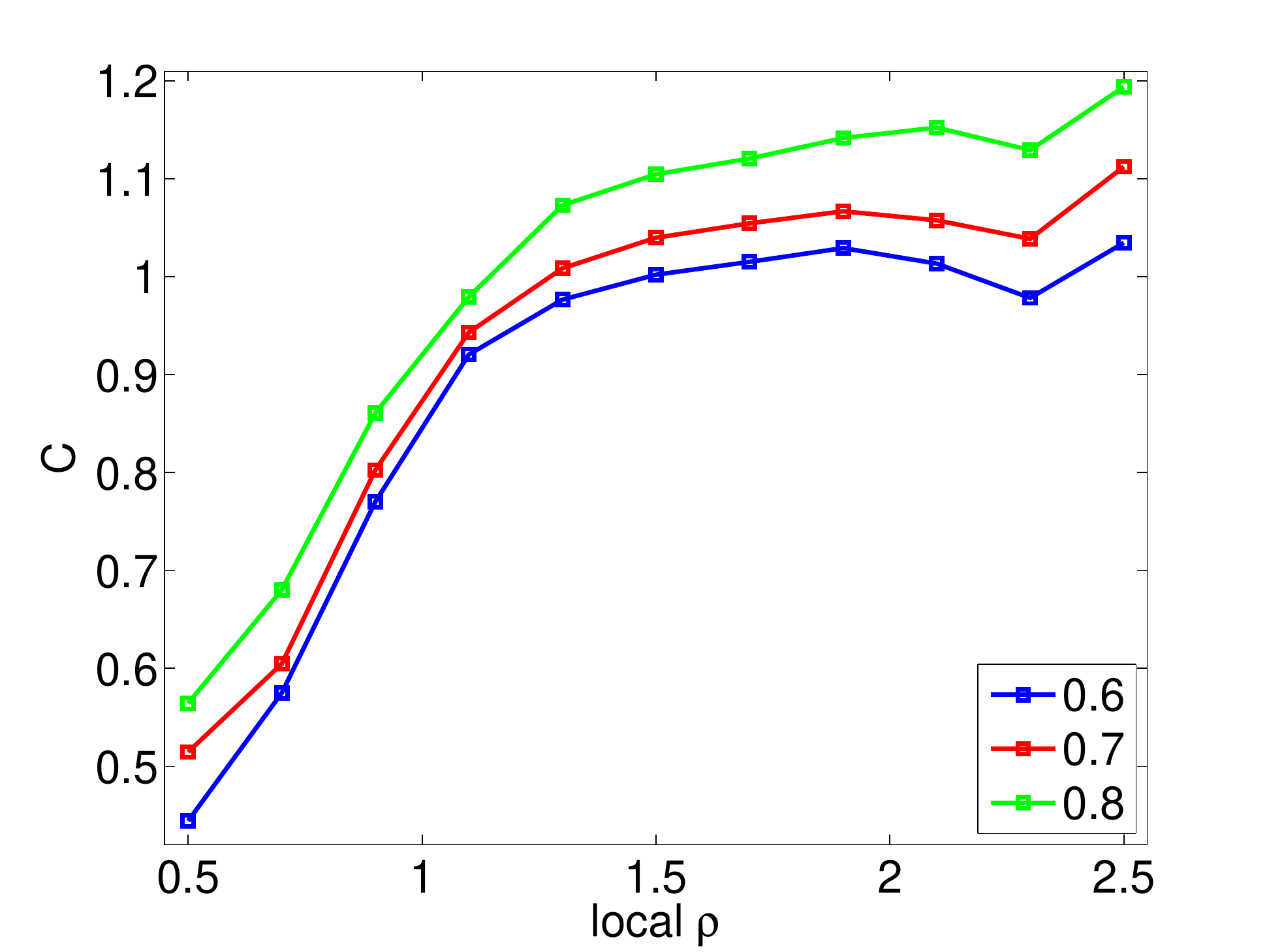}}
\caption{Median and quartile of the estimated delay $\tau$ (left) and median of the estimated constant $C$ (right) over the set of samples consisting of (pedestrian, time window) pairs ranging through the whole set of experiments, for different values of the threshold $\epsilon_t$. The curves were obtained with $\epsilon_t=0.6, \, 0.7, \, 0.8$ and correspond to the blue, red and green curves respectively, while the other parameters are fixed to the values  $w_w=6.67$ s, $f_c=0.5$ Hz, $\gamma=-1$. Medians are displayed in solid lines while quartiles are shown in dotted lines of the corresponding color.
}\label{fig:31}
\end{figure}

We remind that the correlation threshold is used to discard samples which are not compliant with the model. For a given sample consisting of a pair (pedestrian, time window), we compute the correlation parameter (see section \ref{sec:model_calibration}) and if this parameter is less that $\epsilon_t$, we discard this sample as being 'not compliant with the model'. To test the effect of this parameter, we process all experiments with $w_w=6.67$ s, $f_c=0.5$~Hz, $\gamma=-1$.  We test the following values:  $\epsilon_t\in\{0.6, \, 0.7, \, 0.8\}$. The median and quartile of the estimated delay $\tau$ and the median of the estimated constant $C$ over the set of samples consisting of (pedestrian, time window) pairs ranging through the whole set of experiments  are presented in Fig. \ref{fig:31} (left and right respectively) as functions of the local density $\rho$ (see section \ref{subsubsec:density_dep} for the definition of the local density).

We observe that the parameter $\epsilon_t$ has little influence on the medians and quartiles of the estimated values of  $\tau$ and on the medians of the estimated values of $C$.   The relative variations of $\tau$ are of the order of $1.6$ \%, and the relative variations of $C$ are of the order of $5.5$ \%.


\end{document}